\documentclass[a4paper,11pt]{article}
\pdfoutput=1
\usepackage{jcappub}
\usepackage[T1]{fontenc} 
\usepackage{hyperref}
\usepackage{graphics}
\usepackage{amssymb, amsmath}
\usepackage[dvipsnames]{xcolor}
\usepackage{hhline}
\usepackage{multirow}
\usepackage{placeins}
\usepackage{afterpage}

\author[b,c]{Alex Krolewski}
\author[b]{Simone Ferraro}
\author[b]{Edward F. Schlafly}
\author[a,b,c]{Martin White}

\affiliation[a]{Department of Physics, University of California, Berkeley, CA 94720}
\affiliation[b]{Physics Division, Lawrence Berkeley National Laboratory, Berkeley, CA}
\affiliation[c]{Department of Astronomy, University of California, Berkeley, CA 94720}
\emailAdd{krolewski@berkeley.edu}
\emailAdd{sferraro@lbl.gov}
\emailAdd{eschlafly@lbl.gov}
\emailAdd{mwhite@berkeley.edu}

\title{unWISE tomography of Planck CMB lensing}

\abstract{CMB lensing tomography, or the cross-correlation between CMB lensing maps and large-scale structure tracers over a well-defined redshift range, has the potential to map the amplitude and growth of structure over cosmic time, provide some of the most stringent tests of gravity, and break important degeneracies between cosmological parameters. In this work, we use the unWISE galaxy catalog to provide three samples at median redshifts $z \sim 0.6, 1.1$ and 1.5, fully spanning the Dark Energy dominated era, together with the most recent Planck CMB lensing maps. We obtain a combined cross-correlation significance $S/N = 79.3$ over the range of scales
$100 < \ell < 1000$.
We measure the redshift distribution of unWISE sources by a combination of cross-matching with the COSMOS photometric catalog and cross-correlation with BOSS galaxies and quasars and eBOSS quasars. We also show that magnification bias must be included in our analysis and perform a number of null tests.
In a companion paper, we explore the derived cosmological parameters by modeling the non-linearities and propagating the redshift distribution uncertainties.}
\arxivnumber{19MM.NNNNN}

\begin{document}
\maketitle
\flushbottom






\section{Introduction}

As they travel from the surface of last scattering to the Earth, Cosmic Microwave Background (CMB) photons are deflected by the gravitational potentials associated with large-scale structure (LSS), providing a probe of late-time physics directly in the CMB sky (see Refs.~\cite{Lewis06,Hanson10} for reviews).
The lensing effect is dominated by structures on Mpc scales over a very broad range of redshifts from $z<1$ to $z\sim 10$.  By cross-correlating the lensing map with another tracer of large-scale structure which spans a narrower range in redshift, we can simultaneously increase the signal-to-noise ratio and isolate particular redshifts of interest. Doing this on multiple lens redshift planes (``CMB lensing tomography'') breaks important degeneracies between the expansion history and the growth of perturbations, as well as providing greater control over systematics \cite{Hu:1999ek, Hu:2002rm}. 
The first detections of CMB lensing were obtained in cross-correlation between galaxy samples and WMAP data \cite{Smith2007, Hirata2008}, and some of the early work employing cross-correlations with ACT, SPT and Planck are presented in refs.~\cite{Sherwin2012, Bleem2012, PlanckLens13} respectively. Since then, there have been a large number of cross-correlation analysis with a wide variety of samples (see for example refs.~\cite{Omori2015,Allison2015,Bianchini2015,Baxter2016,Giannantonio2016,Omori2018b, Marques:2019aug}).

In this work, we use galaxies from the unWISE catalog \cite{Schlafly19}, containing angular positions and magnitudes of over two billion objects observed by the Wide-field Infrared Survey Explorer (WISE, \cite{Wright10}) mission. The unWISE catalog builds upon earlier WISE-based catalogs by  including additional data from the post-hibernation NEOWISE mission, and is the largest full-sky galaxy catalog currently available \cite{Schlafly19}, containing over half a billion galaxies across the full sky. We further divide the catalog based on magnitude and color and reject stars based on Gaia data \cite{Gaia16}, creating three samples, referred here as ``blue'', ``green'' and ``red,'' at median redshifts $\sim 0.6, 1.1$ and 1.5, respectively, allowing a tomographic analysis of the amplitude of fluctuations in the Dark Energy dominated era. Previous cross-correlations between WISE-derived catalogs and CMB lensing were presented in refs.~\cite{PlanckLens13, Ferraro15,Ferraro2016,Hill2016,Shajib2016,Peacock2018,Marques:2019aug}.

In this paper, we present the auto correlation of the galaxy samples and their cross-correlation with the Planck CMB lensing maps \cite{PlanckLens18}.
We also measure the redshift distribution of the unWISE galaxies, which is crucial for the cosmological interpretation of the signal. While obtaining photometric redshifts from the two WISE colors alone is 
not feasible, cross-matching sources with the COSMOS photometric catalog as well as cross correlation with a number of spectroscopic surveys allows us to determine the ensemble redshift distribution of our samples, together with an estimate of its uncertainty.

The outline of the paper is as follows: In Section \ref{sec:data} we summarize the data used and in Section \ref{sec:model} we describe our modelling.  In Section  \ref{sec:angular_clustering} we discuss the auto and cross correlation measurements and in Section \ref{sec:dndz}, we measure the redshift distribution of the unWISE sample and characterize its uncertainties.
The results are presented in Section \ref{sec:results}. Possible systematics and null tests are explored in Section \ref{sec:systematics}, and in Section \ref{sec:conclusions} we summarize our results.
This paper is focused on the measurement of the cross-correlation.
In a companion paper \cite{paper2}, we will extract cosmological information by modeling the non-linearities in the signal and marginalizing over uncertainties in the stellar contamination fraction and the galaxy redshift distribution.

Where necessary we assume a fiducial $\Lambda$CDM cosmology with the Planck 2018 maximum likelihood parameters (the final column in Table 2 in ref.~\cite{2018arXiv180706209P}).
We quote magnitudes in the Vega system, noting that we can easily convert these to AB magnitudes with AB = Vega + 2.699, 3.339 in W1, W2, respectively.

\section{The data}
\label{sec:data}

\begin{figure}
    \centering

    \resizebox{\columnwidth}{!}{\includegraphics{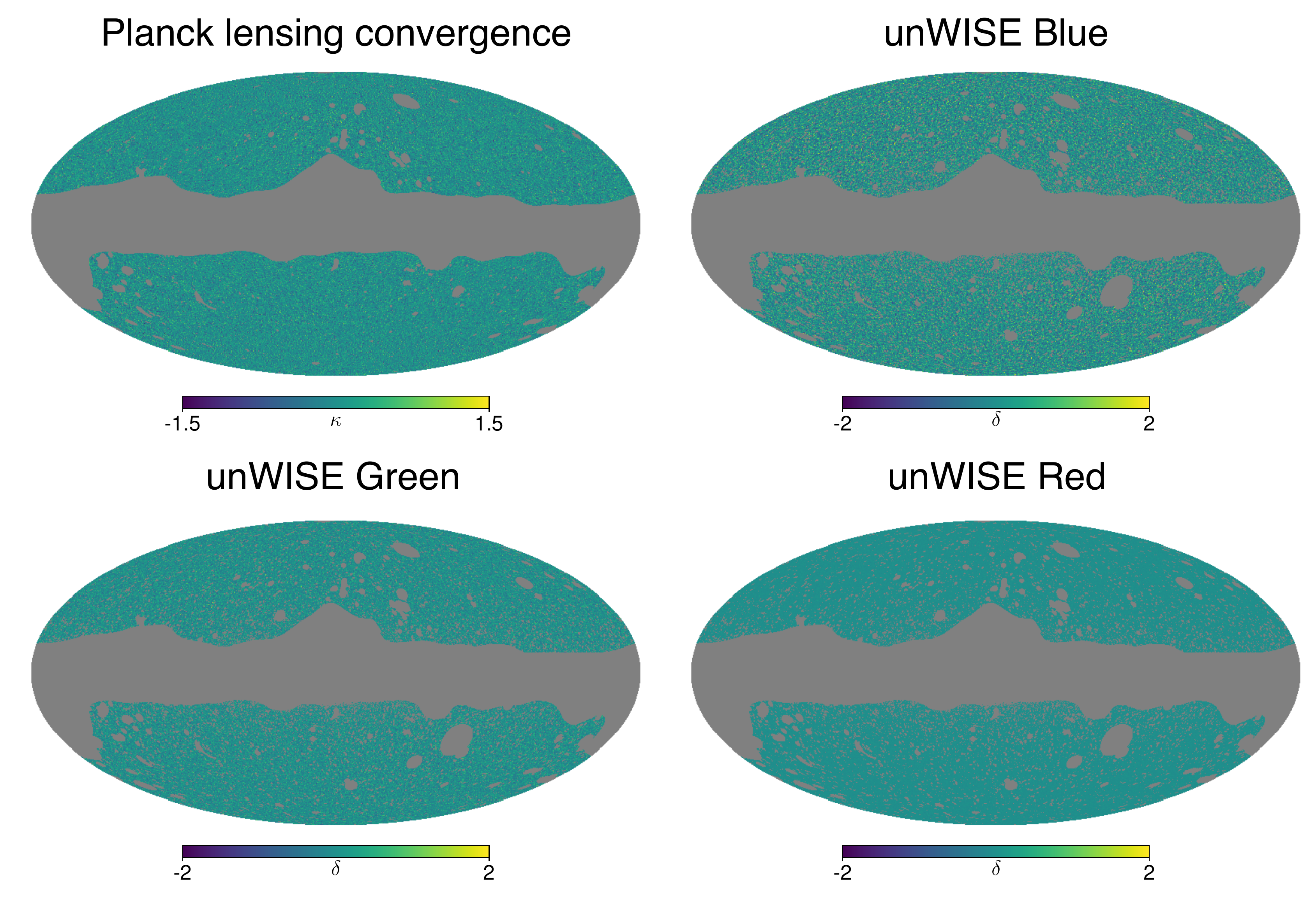}}
    \caption{Plot of the maps used in the analysis ($\kappa$ for Planck lensing convergence and density contrast $\delta$ for the galaxy samples). The maps have been filtered to only contain the range of scales used in this analysis, i.e.\ $\ell_{\rm min} = 100$ and $\ell_{\rm max} = 1000$, and this explains the lack of large-scale power.
    }
    \label{fig:sky_maps}
\end{figure}

\subsection{Planck CMB lensing maps}

Gravitational lensing of the CMB remaps the temperature and polarization fields, altering their statistics in a well-defined way \cite{Lewis06}.  By searching for these statistical patterns it is possible to reconstruct the lensing convergence, $\kappa$, from quadratic combinations of the foreground-cleaned maps \cite{HuOkamoto02}.
We use the latest CMB lensing maps from the Planck 2018 release \cite{PlanckLens18} and their associated masks, downloaded from the Planck Legacy Archive.\footnote{PLA: \url{https://pla.esac.esa.int/}}
These maps are provided as spherical harmonic coefficients of the convergence, $\kappa_{\ell m}$, in HEALPix format \citep{Gorski05} and with $\ell_{\rm max} = 4096$.  In particular, for our fiducial analysis we use the minimum-variance (MV) estimate obtained from both temperature and polarization, based on the \texttt{SMICA} foreground-reduced CMB map. Since the MV reconstruction is dominated by temperature, residual galactic and extragalactic foregrounds may contaminate the signal. Extensive testing has been performed by the Planck team, indicating no significant problems at the current statistical level.  Nonetheless, thermal Sunyaev-Zel'dovich (tSZ) contamination has been shown to be one of the largest potential contaminants to cross correlations with tracers of large-scale structure in other analyses \cite{Schaan:2018tup,Madhavacheril:2018bxi,vanEngelen:2013rla,Osborne:2013nna}. For this reason, as a test, we shall repeat the analysis with a lensing reconstruction on \texttt{SMICA} foreground-reduced maps where tSZ has been explicitly deprojected \cite{PlanckLens18}, and we will refer to this analysis as ``tSZ-free.'' Possible foreground contamination is discussed more in detail in Section \ref{subsec:foregrounds}.

\subsection{unWISE}

The WISE mission mapped the entire sky at $3.4$, $4.6$, $12$, and $22\,\mu$m (W1, W2, W3, and W4) with angular resolutions of $6.1''$, $6.4''$, $6.5''$ and $12''$, respectively \cite{Wright10}. 
The AllWISE data release encompassed the full WISE cryogenic mission as well as the initial NEOWISE post-cryogenic mission, from 2010 January to 2011 February, after which the instrument was placed into hibernation \cite{Mainzer11, Cutri2013}.  The W1 and W2 bands do not require cryogen to operate efficiently, motivating reactivation of WISE in December 2013 \cite{Mainzer14}.  Observations from the continuing NEOWISE mission have been incorporated into increasingly deep ``unWISE'' coadded images of the sky \cite{fulldepth_neo1, fulldepth_neo2, fulldepth_neo3}, which now feature more than $4\times$ longer exposure times than were available for the AllWISE data release.
In the future, at least another two years of NEOWISE data will be available
(NEO5 and NEO6), which would further increase the depth by ${\sim}0.2$ magnitudes.

The deeper imaging coupled with the $\sim6.5''$ angular resolution leads to crowded images with many overlapping sources, requiring a new approach to the analysis of the WISE coadded images.  The \texttt{crowdsource} crowded field photometry pipeline \cite{Schlafly18}, originally designed for surveys of the Galactic plane, was employed to generate a new catalog based on the deep unWISE coadded images \cite{Schlafly19}.  The resulting catalog provides a sample of $>500$ million galaxies with $0 < z < 2$ and improves the uniformity of the depth and photometric calibration of the WISE survey.

\subsection{Galaxy selection}
\label{subsec:galaxyselection}

Galaxies are selected on the basis of their WISE W1 and W2 magnitudes.  Inspection of the average colors of galaxies detected in WISE as a function of redshift shows a clear trend in which fainter and redder galaxies tend to be at higher redshift.  Accordingly we made three selections of galaxies in $\mathrm{W1} - \mathrm{W2}$ color, with a sliding cut on color with magnitude reflecting that fainter galaxies tend to be at higher redshifts.  Table \ref{tab:galaxyselection} gives the adopted color selection for the three samples considered in this work, which we term the blue, green, and red samples \cite{Schlafly19}.  
Table \ref{tab:galaxyselection} also summarizes
important properties of each sample including the redshift distribution,
the number density, and the response of number density to galaxy magnification $s \equiv d\log_{10}N/dm$. We measure $s$ using galaxies with ecliptic latitude $|\lambda| >60^{\circ}$, where the WISE depth of coverage
is greater and thus the measurement of $s$ is less affected
by incompleteness.  We describe the measurement of $s$ in Appendix~\ref{sec:magbias_slope}.

\begin{table}[]
\centering
\begin{tabular}{c|ccccccc}
Label & $\mathrm{W1}-\mathrm{W2} > x$ & $\mathrm{W1}-\mathrm{W2} < x$ & $\mathrm{W2} < x$ & $\bar{z}$ & $\delta z$ & $\bar{n}$ & $s$\\
\hline
Blue & & $(17-\mathrm{W2})/4+0.3$ & 16.7 & 0.6 & 0.3 & 3409 & 0.455 \\
Green & $(17-\mathrm{W2})/4 + 0.3$ & $(17-\mathrm{W2})/4 + 0.8$ & 16.7 & 1.1 & 0.4 & 1846 & 0.648 \\
Red & $(17 - \mathrm{W2})/4 + 0.8$ & & 16.2 & 1.5 & 0.4 & 144  & 0.842\\
\end{tabular}
\caption{Color and magnitude cuts for selecting galaxies of different redshifts, together with the mean redshift, $\bar{z}$, and the width of the redshift distribution, $\delta z$ (as measured by matching to objects with photometric redshifts on the COSMOS field \cite{Laigle16}), number density per deg${}^2$ within the unWISE mask, $\bar{n}$, and response of the number density to magnification, $s \equiv d\log_{10}N/dm$.  Galaxies are additionally required to have $\mathrm{W2} > 15.5$, to be undetected or not pointlike in Gaia (see \textsection\ref{subsec:galaxyselection}), and to not be flagged as diffraction spikes, latents or ghosts. $s$ is measured using galaxies at ecliptic latitude $|\lambda|>60^{\circ}$, where WISE reaches fainter limiting magnitudes due to increased depth of coverage (see Appendix~\ref{sec:magbias_slope}).
}
\label{tab:galaxyselection}
\end{table}

We require that the blue and green samples have $15.5 < \mathrm{W2} < 16.7$, and the red sample has $15.5 < \mathrm{W2} < 16.2$. If we allow the red sample to include sources with $16.2 < \mathrm{W2} < 16.7$, we find that the red-blue cross-correlation
is inconsistent at the 2-3$\sigma$ level with the expected
cross-correlation given the bias measured from the
CMB-cross spectra (Appendix~\ref{sec:galaxy-galaxy}).  The fainter red samples also exhibit
a decrease in number density closer to the Galactic plane,
and may have more angular variation in $dN/dz$.
As a result, we suspect that the fainter red sample is more affected by stellar
contamination or systematics-driven fluctuations, and exclude it from our
fiducial definition of the red sample.

Each of the samples is required to be either undetected or not pointlike in Gaia.  Here a source is taken as ``pointlike'' if 
\begin{equation}
\mathrm{pointlike}(G, A) = \begin{cases}
\log_{10} A < 0.5 & \text{if $G < 19.25$} \\
\log_{10} A < 0.5 + \frac{5}{16} (G-19.25) & \text{otherwise} \, ,
\end{cases}
\end{equation}
where G is the Gaia G band magnitude and $A$ is \texttt{astrometric\_excess\_noise} from Gaia DR2 \cite{Gaia18}.  A source is considered ``undetected'' in Gaia if there is no Gaia DR2 source within $2.75''$ of the location of the WISE source.  High \texttt{astrometric\_excess\_noise} indicates that the Gaia astrometry of a source was more uncertain than typical for resolved sources; this cut essentially takes advantage of the $0.1''$ angular resolution of Gaia to morphologically separate point sources from galaxies.
We additionally remove sources classified as diffraction spikes, first or second latents, or ghosts in either W1 or W2,
corresponding to ``unWISE flags'' 1, 2, 3, 4 or 7.\footnote{See Table 5 here: \url{http://catalog.unwise.me/files/unwise_bitmask_writeup-03Dec2018.pdf}}

\subsection{Masks}
\label{sec:masking}

For the lensing map, we use the official 2018 Planck lensing mask, provided together with the other data products \cite{PlanckLens18}. This is created using a combination of the \texttt{SMICA} 70\% Galactic mask, retaining the cleanest 70\% of the sky, together with the 143 and 217 GHz point source masks and the tSZ-detected clusters with $S/N$ > 5 \cite{PlanckLens18}. We additionally masked a small region of the sky with $|b| < 10^{\circ}$ that was unmasked in the Planck map.  Overall, this leaves an unmasked sky fraction $f_{\rm sky} = 0.670$ for the lensing map.
As a test of Galactic contamination we also use the 60 and 40\% temperature masks from the Planck 2018 data release\footnote{Always multiplied by the original lensing mask.}, masking an increasing fraction of the Galactic plane. The impact on the results is discussed in Section \ref{subsec:gal_mask}.

For WISE, we found it convenient to use the Planck lensing mask as an effective galactic mask, to avoid excessive stellar contamination close to the galactic plane. We additionally mask stars, galaxies, planetary nebulae, and \texttt{NSIDE} = 2048
HEALPix
pixels with substantial area lost due to sub-pixel unWISE masks
(e.g.\ for diffraction spikes from bright stars).

We mask the 6678 brightest stars
in the infrared sky
(6156 at $|b| > 20^{\circ}$)
 with $\mathrm{W1} < 2.5$ or $\mathrm{W2} < 2$ or $\mathrm{K} < 2$ (where the WISE magnitude is the brighter of the
AllWISE or unWISE magnitudes). 
We use the bright star list provided by the CatWISE team (Eisenhardt et al., in prep)\footnote{\url{catwise.github.io}} for these objects.\footnote{We additionally
add the carbon star IRC+20326, which had problematic photometry in both AllWISE and unWISE.}
We find that a disk of radius $0.5^\circ$ is adequate to prevent contamination due to spurious detections around the majority of these bright sources.  
For the very brightest stars, diffraction spikes extend beyond the $\sim 1^{\circ}$ extent of the diffraction spike mask;
we therefore use a $1.5^\circ$ radius around 32 stars with $-3 < \mathrm{W2} < -2$ and a $3^\circ$ radius around 11 stars with $\mathrm{W2} < -3$.  
Finally, we mask $0.2^\circ$ around 6212 stars with $2 < \mathrm{W2} < 2.5$, $\mathrm{W1} > 2.5$, and $\mathrm{K} > 2$, where we find that in rare cases, the unWISE PSF model does not extend far enough into the wings of the star, leading to spurious sources at the edge of the modeled region.

We also mask bright galaxies using the LSLGA catalog\footnote{\url{https://github.com/moustakas/LSLGA}}, selecting 715 galaxies from Hyperleda \cite{Makarov14} with magnitudes $< 13$ (almost always in the $B$ filter, though in rare cases the $K$ or $I$ filter), diameter $D_{25} > 3$ arcmin, and surface brightness within $D_{25}$ of $< 26$ mag/arcsec$^2$.
Using the position angle and ellipticity in the catalog we mask ellipses around each galaxy out to $1.5 R_{25}$, and we visually confirm that this radius removes the impact of galaxies on our samples.

We also find that planetary nebulae can contaminate our samples, particularly
the red sample.
We mask 1143 planetary nebulae \citep{Acker92}, masking out to twice the optical radius of
each planetary nebula.

In all three cases (stars, galaxies, and planetary nebulae) we create a binary mask on an \texttt{NSIDE} = 2048 HEALPix map, masking all pixels within the specified distance of the source. For the planetary nebulae, we additionally use the ``inclusive=True'' option in the HEALPix {\sc{query\_disc}} command since the pixels in our map are often larger than the mask radius.

Finally, we correct for area lost in each (\texttt{NSIDE} = 2048) HEALPix pixel
from sub-pixel masking.  Sub-pixel masking arises from two sources:
foreground Galactic stars from Gaia, which will mask any unWISE source within $2.75''$
due to our Gaia point-source exclusion,
and unWISE masking of diffraction spikes, latents and ghosts around bright stars\footnote{\url{http://catalog.unwise.me/files/unwise_bitmask_writeup-03Dec2018.pdf}}.
We apply a binary mask to remove all pixels with more than 20\% area lost due to sub-pixel masking, and
we correct the density in the remaining pixels by dividing by the fractional
unmasked area of each pixel.

We apodize the Planck lensing mask (with additional exclusion of $|b| < 10^{\circ}$)
with a 1$^{\circ}$ FWHM Gaussian. We do not apodize the stellar, large galaxy,
planetary nebulae or area lost masks.  We use the apodized Planck lensing mask
for the CMB lensing map and the product of the apodized lensing mask and the unapodized stellar, large galaxy, planetary nebulae and area-lost masks for the unWISE galaxy map.  This yields $f_{\rm sky} = 0.586$ for the unWISE galaxy map.

\section{Model}
\label{sec:model}

\subsection{Angular Clustering}

Both the CMB lensing convergence $\kappa$ and the unWISE projected galaxy density are projections of 3D density fields.  We define the projection through kernels $W(\chi)$, where $\chi$ is the line-of-sight comoving distance.  Given two such fields $X, Y$ on the sky their angular cross-power spectrum is
\begin{equation}
  C_\ell^{XY} = \frac{2}{\pi}\int_0^\infty
  d\chi_1\,d\chi_2\ W^X(\chi_1)W^Y(\chi_2) \int_0^\infty k^2\,dk
  \ P_{XY}(k;z_1,z_2)j_\ell(k\chi_1)j_\ell(k\chi_2)
  \quad .
\end{equation}
On small angular scales (high $\ell$) one may make the Limber approximation \cite{Limber53}, under which $C_\ell$ reduces to a single integral of the equal-time, real-space power spectrum:
\begin{equation}
  C_\ell^{XY} = \int d\chi\ \frac{W^X(\chi)W^Y(\chi)}{\chi^2} 
  \ P_{XY}\left(k_{\perp} = \frac{\ell+1/2}{\chi},k_z=0\right)
\label{eqn:ClXY}
\end{equation}
where we have included the lowest order correction to the Limber approximation, $\ell\to\ell+1/2$, to increase the accuracy to $\mathcal{O}(\ell^{-2})$ \cite{Loverde08}. 

Lensing is sourced by the Weyl
potential, which is related to the total
matter power spectrum (including neutrinos)
by the Poisson equation.
Writing $C_{\ell}$ in terms of the galaxy-matter
and matter-matter power spectra $P_{\rm mg}$ and $P_{\rm gg}$, the weight functions $W(\chi)$
are
\begin{equation}
  W^{\kappa}(\chi) = \frac{3}{2}(\Omega_m + \Omega_{\nu})H_0^2(1+z)
  \ \frac{\chi(\chi_\star-\chi)}{\chi_\star}
  \quad , \quad
  W^{g}(\chi) = b(z) H(z)\,\frac{dN}{dz}
\end{equation}
with $\chi_\star$ the distance to last scattering and $\int dz\ dN/dz=1$.

Besides density-density and density-lensing correlations, there are also
correlations induced by lensing magnification of background sources:
\begin{equation}
C_\ell^{\kappa g} \rightarrow C_\ell^{\kappa g} + C_{\ell}^{\kappa \mu}
\end{equation}
\begin{equation}
C_\ell^{g_1 g_2} \rightarrow C_\ell^{g_1 g_2} + C_{\ell}^{g_1\mu_2} + C_{\ell}^{g_2\mu_1} + C_\ell^{\mu_1 \mu_2}
\end{equation}
where
\begin{equation}
W^{\mu,i}(\chi) =  (5s-2)\,\frac{3}{2} (\Omega_m + \Omega_{\nu}) H_0^2(1+z) g_i(\chi)
\end{equation}
\begin{equation}
g_i(\chi) = \int_{\chi}^{\chi_{\star}} d\chi' \ \frac{\chi(\chi' - \chi)}{\chi'} \ H(z') \ \frac{dN_i}{dz'}
\end{equation}
where $s \equiv d \log_{10}N/dm$ is the response of the number density to a multiplicative
change in brightness.
Given our complex selection function, we measure the response by finite difference, artificially changing each magnitude by the same amount (in analogy to lensing magnification) and measuring the change in number of galaxies satisfying our selection criteria. This procedure is discussed in detail in Appendix \ref{sec:magbias_slope}
for both the unWISE galaxies (necessary for modeling the angular power spectra)
and for the spectroscopic samples (necessary for determining magnification bias contamination to the clustering redshifts).
For the color-selected unWISE samples,
the response $s$ may be significantly different
from the slope of the luminosity function
at the magnitude limit because the color cut
is magnitude dependent.

\subsection{HaloFit model}
\label{sec:halofit}

In order to compute $C_\ell$ we need to model $P_{\rm gg}(k,z)$, $P_{\rm mg}(k,z)$ and $P_{\rm mm}(k,z)$. In this paper, we do not explore the cosmological implications of our measurement, but rather seek to characterize the unWISE samples and their redshift distribution, and present a measurement of the cross-correlations. For this purpose, a phenomenological fit will be sufficient, and we choose to model the auto and cross correlation in terms of a linear bias, multiplied by the ``HaloFit'' fitting function \citep{Takahashi12} to the non-linear matter power spectrum  as implemented in the CLASS code \citep{Blas11}:
\begin{equation}
  P_{\rm mg}(k,z) = b_{\rm lin}(z) P_{\rm mm}(k,z) \ \ , \ \ P_{\rm gg}(k,z) = b^2_{\rm lin}(z) P_{\rm mm}(k,z) + {\rm Shot \ Noise}
\end{equation}
This procedure has been shown to produce fairly reasonable phenomenological fits to the auto and cross correlations.\footnote{For the magnification bias terms, each $\ell$ maps to higher $k$ than for the clustering terms; therefore the linear bias times Halofit model is less adequate for
$C_{\ell}^{\mu g}$. However, the magnification bias terms are subdominant compared to the clustering terms, so inaccuracy in modeling $C_{\ell}^{\mu g}$ is not significant.} While the fit may be good, ref.~\cite{Modi17} has shown that the value of the inferred cosmological parameters can be significantly biased if HaloFit is used, and for this reason we will explore a more sophisticated bias model to better model non-linearities in our cosmological analysis in ref.~\cite{paper2}.

Since the galaxy field responds to dark matter and baryons only \citep{Costanzi13,VN14,Castorina14,Castorina15,Vagnozzi18}, $P_{\rm gg}$ is the power spectrum
of non-neutrino density fluctuations. Although lensing responds to the power spectrum
of total fluctuations, on the scales of interest here the neutrinos cause a scale-independent suppression of power.  Therefore, using the non-neutrino power spectrum throughout and substituting $\Omega_m + \Omega_{\nu} \rightarrow \Omega_m$ greatly simplifies the modelling and makes less than 1\% difference compared to the exact calculation.

\section{Angular clustering}
\label{sec:angular_clustering}

In this section we discuss our method of estimating the auto and cross spectra, as well as their covariance matrix.

\subsection{Angular power spectra estimation}

In order to estimate the binned cross and auto power spectra, we use a pseudo-$C_\ell$ estimator \cite{Hivon02} based on the harmonic coefficients of the galaxy and lensing fields.
The measured pseudo-$C_\ell$ on the cut sky are calculated as 
\begin{equation}
\Tilde{C}_\ell^{XY} = \frac{1}{2\ell + 1}\sum_m X_{\ell m} Y^{\star}_{\ell m}
\end{equation}
where $X,Y\in\{g_1,g_2,g_3,\kappa\}$ are the observed fields on the cut sky.
Because of the mask, these differ from the true $C_\ell$ that are calculated from theory, but their expectation value is related through a mode-coupling matrix, $M_{\ell \ell'}$, such that 
\begin{equation}
\langle \Tilde{C}_\ell \rangle = \sum_{\ell'} M_{\ell \ell'}  C_{\ell'}
\label{eq:mode_coupling}
\end{equation}
The matrix $M_{\ell \ell'}$ is purely geometric and can be computed from the power spectrum of the mask itself. While Eq.~(\ref{eq:mode_coupling}) is not directly invertible for all $\ell$, the MASTER algorithm \cite{Hivon02} provides an efficient method to do so assuming that the power spectrum is piecewise constant in a number of discrete bins, $b$.
Defining a ``binned'' mode-coupling matrix, $\mathcal{M}_{b b'}$ \cite{Alonso18}, we can recover unbiased binned bandpowers 
\begin{equation}
   C_b = \sum_{b'} \mathcal{M}^{-1}_{b b'} \tilde{C}_{b'} \quad .
\label{eq:mode_coupling_binned}
\end{equation}
We use the implementation in the code \texttt{NaMaster}\footnote{\url{https://github.com/LSSTDESC/NaMaster}} \cite{Alonso18}.
Finally, the theory curve must be binned in the same way as the data when comparing theory and measurements. Since the true $C_{\ell}$ are not piecewise constant, this involves the following steps \cite{Alonso18}: First, the theory curve is convolved with $M_{\ell \ell'}$ using Eq.~(\ref{eq:mode_coupling}). Then the convolved theory, $\tilde{C}^{\rm theory}_{\ell}$, is binned in the same bins, $b$, as the data to form bandpowers, $\tilde{C}^{\rm theory}_b$, and finally the bandpowers are decoupled using Eq.~(\ref{eq:mode_coupling_binned}) to obtain $C^{\rm theory}_b$. While for simplicity the plots show unbinned theory curves, all of the calculations are performed with binned quantities.

In short, our pipeline works as follows: first, we mask the Planck lensing map with the mask provided by the Planck team, apodized with a Gaussian smoothing kernel with FWHM 1 deg. For the unWISE galaxies, we use the custom-made mask described in Section \ref{sec:masking}, which includes different apodization schemes for the wide Galactic mask and point sources.  In addition, we have to consider that Galactic stars can mask galaxies behind them or in their vicinity, a problem that becomes more severe closer to the galactic plane. To correct for this, we create an ``area lost'' mask (described in Section \ref{sec:masking}) and divide the observed galaxy number count by the area available in each pixel, to obtain an unbiased estimate of the local number of galaxies. Then a galaxy overdensity field is created, and cross-correlated with the CMB lensing maps using \texttt{NaMaster}.
Finally, we need to correct for the pixel window function, due to the assignment of galaxies to discrete pixels: we divide $C_{\ell}^{\kappa g}$ by the HEALPix pixel window function at the center of each bandpower.
The procedure is more complicated for $C_{\ell}^{gg}$: a shot-noise power spectrum has correlation
length zero and thus does not need to be corrected
for the pixel window function, whereas
the signal part of $C_{\ell}^{gg}$
should be divided by the square of the pixel window function.
Therefore, we first subtract the estimated
shot noise from $C_{\ell}^{gg}$ using $\bar{n}$ from Table 1, then
divide by the squared pixel window function,
and then add the estimated shot noise back.

We tested this pipeline on Gaussian realizations of the CMB lensing and galaxy fields, and noted that the final ``deconvolved'' $C_\ell$ are rather sensitive to the choice of apodization scale, especially for the CMB lensing map, but are not affected by the inclusion of unapodized
components in the galaxy mask. Our choice of smoothing was determined by optimizing the recovered power spectrum in simulations with known input angular correlation.
In particular, we use the above \texttt{NaMaster} pipeline to measure $C_{\ell}$ for 100 simulated Gaussian
lensing and galaxy maps (generated with the correct cross-correlation). We find significant biases of several percent due to power leakage outside the measured range, if the $\ell_{\rm max}^{\rm NaMaster}$ used in \texttt{NaMaster} is close to the $\ell_{\rm max} = 1000$ used in our analysis. To remedy this, we run \texttt{NaMaster} with $\ell_{\rm max}^{\rm NaMaster} = 6000$, before extracting the bandpowers in our analysis range and discarding the higher $\ell$ ones.

\begin{figure}
    \centering
    \resizebox{\columnwidth}{!}{\includegraphics{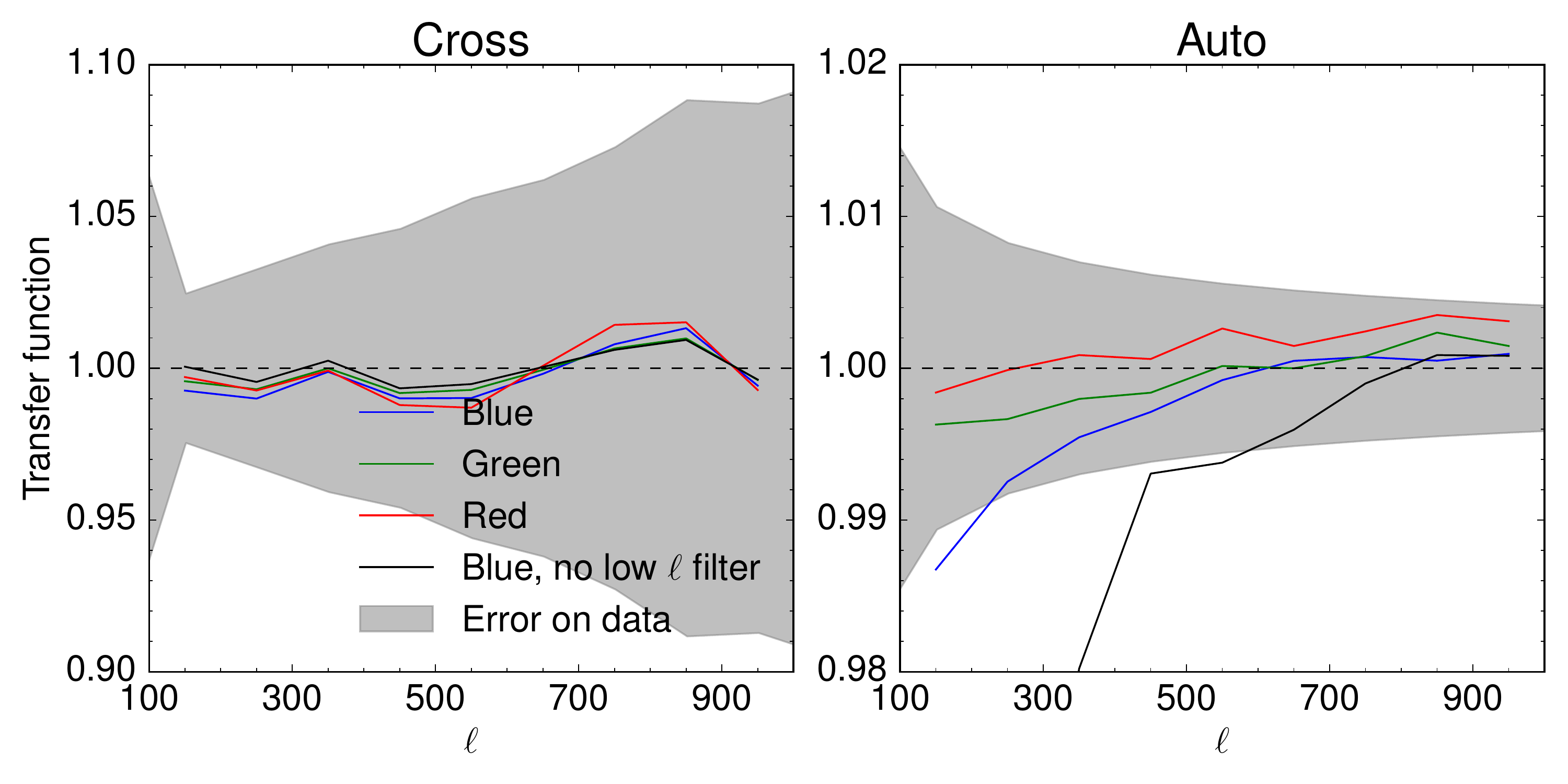}}
    \caption{Mask deconvolution transfer function for the CMB lensing cross-spectra (left) and galaxy auto spectra (right), i.e.\ comparison between
    input power spectrum and output after
    masking, pseudo-$C_{\ell}$ estimation,
    and mask deconvolution. Maps were generated
    from power spectra assuming a Gaussian
    field.  Colored curves are transfer
    functions for different samples
    after filtering $\ell < 20$ modes
    from the unmasked map, whereas no filtering
    was applied to black curves. Recovery
    of the cross-spectrum is unbiased
    even without filtering, but recovery
    of the auto-spectrum requires filtering
    $\ell < 20$ modes. To ensure sub-percent
    accuracy, we additionally
    correct the auto-spectrum by the transfer
    function displayed here.
    }
    \label{fig:transfer}
\end{figure}

With the Gaussian simulations,
we also find biases of several percent
in the recovery of the galaxy
auto-spectrum at $\ell < 300$ (Figure~\ref{fig:transfer}).
Mask-induced mode coupling
causes $\ell < 50$
systematic power
in the auto-spectrum to leak to considerably
higher $\ell$.  We find that if we
turn off the extra low-$\ell$
power by using the theory prediction
rather than the measured $C_{\ell}^{gg}$ as the input power spectrum,
we can recover $C_{\ell}^{gg}$
with no bias.

We therefore filter all modes with
$\ell < 20$ in the unmasked galaxy
map.
To do this, we take the spherical
harmonic transform of the raw galaxy
map (before applying the mask),
apply a sharp cut setting all modes
with $\ell < 20$ to zero,
and apply the inverse
transform to recover the filtered map.
We then use the filtered map
as input for our {\textsc NaMaster} pipeline.
We find this procedure
leads to considerably less
biased recovery of the auto-spectrum
(Figure~\ref{fig:transfer}).
Other approaches
(i.e.\ setting the edge of the smallest-$\ell$ bin to $\ell_{\rm min}^{\rm NaMaster} = 20$ or filtering $\ell < 50$ modes instead)
also recover the unbiased
auto and cross-spectra.
We correct the auto-spectra
for the residual mask-transfer
bias. Since the residual
bias is $\leq 1\%$, 
smaller than the statistical
errors on the cross spectrum
or the statistical errors
from uncertain $dN/dz$, this
correction has only a small impact on our results (compare the ``no transfer function'' row
to the fiducial row in Figure~\ref{fig:systematics}). 




We conclude that with our pipeline we can measure all of the auto and cross-correlations between the different samples with sub-percent accuracy over the whole range of scales considered,
once the input maps have been filtered and the mask-deconvolution transfer
function has been applied.

\subsection{Covariance matrix}
While an exact computation of the covariance matrix after applying the MASTER algorithm for a Gaussian random field is possible, it is computationally very demanding, involving $\mathcal{O}(\ell_{\rm max}^6)$ operations. Refs. \cite{Efstathiou:2003dj, Garcia-Garcia:2019bku} have proposed an approximate method to estimate the Gaussian part of the covariance matrix that makes it as computationally expensive as the power spectrum itself. This procedure has been validated on simulations and shown to work extremely well \cite{Garcia-Garcia:2019bku}. This algorithm is implemented in \texttt{NaMaster}, and takes as input the true auto and cross spectra (for example, from the theory curves with the correct value of parameters including the galaxy bias). Since measuring the bias  requires a covariance matrix to start with, an iterative approach may be used.  For computational simplicity, we adopt a further approximation which will assume the
decoupled covariance matrix to be diagonal, and where the on-diagonal elements for binned bandpowers of width $\Delta \ell$ are given by \cite{Hivon02}:
\begin{equation}
\label{eq:cov}
    {\rm Cov}(C^{XY}_\ell, C^{XY}_{\ell'}) =  \sigma^2(C^{XY}_\ell) \delta_{\ell, \ell'} = \frac{ \left[ C^{XX}_\ell C^{YY}_\ell + \left(C^{XY}_\ell\right)^2 \right]_{\rm measured} }{f_{\rm sky} (2\ell + 1) \Delta \ell} \ \frac{w_4}{ w_2^2 } \delta_{\ell, \ell'}
\end{equation}
Here the weights $w_2$ and $w_4$ are defined in terms of the arbitrary mask weights $W(\hat{\boldsymbol{n}})$ as:
\begin{equation}
    w_i f_{\rm sky} = \frac{1}{4 \pi} \int_{4\pi} d \Omega_{\hat{\boldsymbol{n}}} W^i(\hat{\boldsymbol{n}})
\end{equation}
with $w_1 f_{\rm sky} = f_{\rm sky}$. If $X \neq Y$ and the fields have different masks, we take $w_2$ and $w_4$ to be the geometric means of the ones computed with each of the individual masks.

Using the method  for analytic Gaussian pseudo-$C_{\ell}$ covariance in refs.~\cite{Efstathiou:2003dj, Garcia-Garcia:2019bku}, we have checked that the largest off-diagonal correlation between bandpowers is 4\% for the two lowest $\ell$ bins, and that the on-diagonal elements agree to percent level.  Therefore we conclude that the approximation in Equation \ref{eq:cov} is adequate for our purposes.
Furthermore, we neglect any non-Gaussian contribution to the covariance matrix, since we will only model scales that are in the linear or mildly non-linear regimes, where these corrections are expected to be small.

\section{Galaxy redshift distribution}
\label{sec:dndz}


Since the unWISE galaxy sample is selected from two-band imaging, $dN/dz$ cannot be determined by photometric redshifts.
We instead measure $dN/dz$ using cross-correlations with large-area spectroscopic surveys \cite{Newman08,McQuinn13,Menard13},
supplemented by redshifts from cross-matching to deep multi-band photometry in a small field.
Cross-correlation redshifts measure $b(z) dN/dz$
(in the absence of a small contribution from magnification bias), which is the relevant kernel for modeling $C_{\ell}^{\kappa g}$ and $C_{\ell}^{g g}$ (Section~\ref{sec:model}). Therefore, unlike previous work, we are not concerned with disentangling $dN/dz$ from the bias evolution of the unWISE galaxies.  $C_{\ell}^{\kappa g}$ and $C_{\ell}^{g g}$ do contain a subdominant
contribution from magnification bias, which depends on $dN/dz$ alone; in this context, we use $dN/dz$ measured from cross matches to the COSMOS
 photometric catalog.  
Consistency between the cross-match $dN/dz$ and cross-correlation $b(z) dN/dz$ requires that the
bias increase strongly with redshift.
In Appendix~\ref{sec:xcorr_redshift_details}, we show
that a simple halo occupation distribution of the unWISE galaxies exhibits a similar increase in bias, demonstrating that our approach is self-consistent.

\begin{figure}
    \centering
    \resizebox{1.2\columnwidth}{!}{\includegraphics{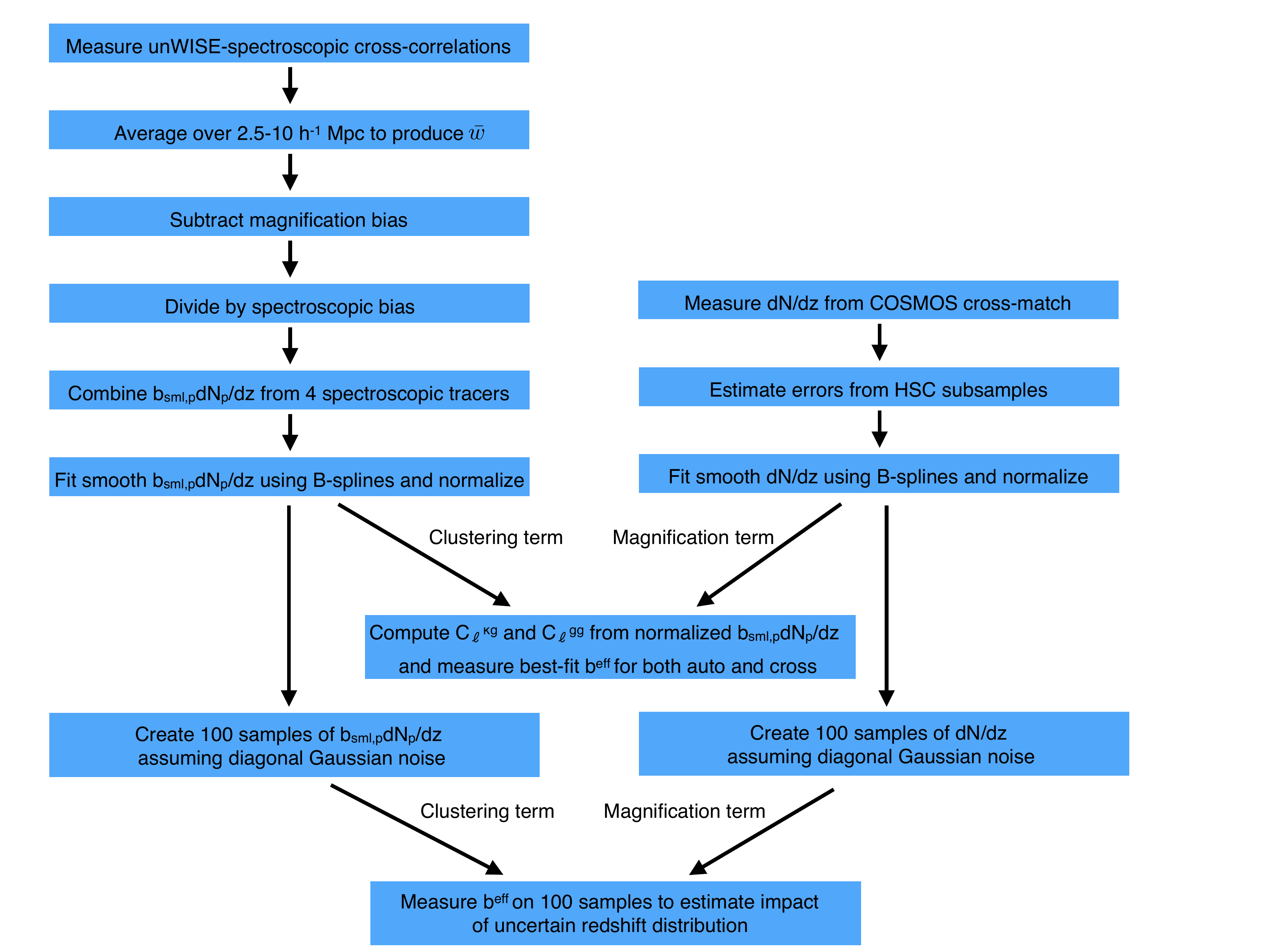}}
    \caption{An outline of the steps required to interpret
    the angular clustering of the unWISE samples:
    estimate
    the redshift distribution, fit a linear bias
    to $C_{\ell}^{\kappa g}$ and $C_{\ell}^{gg}$, and estimate the impact of uncertainty in the redshift distribution
    on the fitted biases.
    }
    \label{fig:flowchart}
\end{figure}

In Sections~\ref{sec:xmatch_dndz} and~\ref{sec:xcorr_dndz} we describe
our methodology for measuring the cross-correlation and cross-match redshifts and estimating their uncertainties, which constitute a substantial portion of the error budget in modeling
the angular power spectra.
 In Figure~\ref{fig:flowchart} we list the steps outlined
in Sections~\ref{sec:dndz} and~\ref{sec:results}
to interpret the clustering of the unWISE samples:
estimate the redshift distribution, determine the best-fit
linear bias of each sample, and estimate the uncertainty on the bias
due to uncertain redshift distribution.
Throughout Sections~\ref{sec:dndz} and~\ref{sec:systematics},
we quantify systematic errors in terms of their impact
on the best-fit bias to the CMB cross ($C_{\ell}^{\kappa g}$) and galaxy auto 
power spectra ($C_{\ell}^{gg}$) of the unWISE samples.

\subsection{Cross-match redshifts}
\label{sec:xmatch_dndz}

One estimate of $dN/dz$ can be obtained by matching the unWISE samples to deep catalogs with photometric redshifts.
The deepest sample of well-measured photometric redshifts comes from
the COSMOS field, where deep photometry in many bands spanning the ultraviolet through infrared allows precise photometric redshifts for all sources detected by WISE
with $\Delta z / ( 1 + z)$ = 0.007
\cite{Laigle16}.  We first trim the COSMOS catalog to include only objects brighter than 20.7 (19.2) Vega mag at 3.6 (4.2) $\mu$m.  These depths are roughly 2.5\,mag fainter than the 50\% completeness limit for the unWISE catalog \cite{Schlafly19}, so excluding fainter objects removes no objects that WISE could conceivably detect.  We then match COSMOS sources to unWISE sources at a radius of $2.75''$ over the 2\,deg$^2$ overlap,
considering the closest COSMOS source within $2.75''$ to be the true match.

The COSMOS catalog marks many bright stars as galaxies, so we additionally edit the COSMOS catalog so that bright objects which Gaia identifies as pointlike are classified as stars, as long as those objects are not X-ray selected.  We find that stellar contamination of the unWISE samples is very low, with 1.8\%, 1.6\%, and likely $< 1$\% \footnote{The red sample only has 188 matches to the COSMOS photometric catalog and none of them are stars; the error on the stellar contamination fraction may therefore be quite large.  For the fainter red samples
reaching to W2 = 16.5 or 16.7, we find stellar contamination of 0.6\% and 0.8\%, respectively.} of the blue, green and red samples classified as stars.

For each source, we use the redshift corresponding
to the median of the likelihood distribution (``photoz'' in
the COSMOS catalog).  If the SED is better fit by an AGN
template than a galaxy template, we instead use the redshift
from the AGN template fit (``zq'' in the catalog); we find 19\%, 30\% and 41\% of the blue, green and red sample are classified as AGN by this criterion.  However, 
for these objects ``zq'' and ``photoz'' are very similar.

Due to the small area of the COSMOS field, sample variance can be larger than the Poisson variance on $dN/dz$.
We therefore estimate uncertainty on $dN/dz$ by constructing 44 subsamples, each of ${\sim}2$ deg$^2$, from the HSC SSP survey \cite{HSCDR1,Tanaka18}.
Compared to COSMOS, HSC is slightly shallower but covers a much larger area (${\sim}120$ deg$^2$).
However, the HSC photometric redshifts are 
less accurate than COSMOS and become biased at $z > 1.5$,\footnote{See page A5 in \url{https://hsc-release.mtk.nao.ac.jp/doc/wp-content/uploads/2017/02/s16a_demp_median.pdf}, which plots bias, scatter, and outlier fraction as a function of reference redshift for the training set used by HSC.}
where a substantial fraction of galaxies scatter to $z_{\rm \, HSC} \sim 1$, biasing $dN/dz_{\rm HSC}$ at $z \geq 1$
compared to $dN/dz_{\rm \, COSMOS}$.
As a result,  we restrict
the HSC comparison to $z < 1$.
 We use the DEmP photometric redshifts, as these are the most accurate
redshifts available for all ``primary'' HSC objects \citep{Tanaka18}.
 We require that the HSC objects have clean photometry:
we only use ``primary'' sources, and remove sources with pixel flags indicating
saturated or interpolated pixels,
bad pixels, cosmic ray hits, suspect and clipped pixels, and  poor centroid measurements.  We also require that the objects are classified as extended sources or have $i > 23$, where the star-galaxy classification performs poorly.  As with COSMOS, we use the closest match within $2.75''$.

We find that the $dN/dz$ errors are larger than
Poisson statistics would indicate, by roughly redshift-independent factors of 3.8, 1.9 and 1.1 for the blue, green and red samples at $z < 1$.  Since we cannot use HSC
to determine $dN/dz$ errors at $z > 1$, where the DEmP photometric redshifts
become significantly biased, we extrapolate the $dN/dz$ uncertainty to higher redshift by multiplying
the Poisson error bars by a constant factor of 3.8, 1.9 and 1.1 for blue, green and red samples.
This extrapolation yields error bars
appropriate for the cosmic variance
contribution alone;
in Section~\ref{sec:systematics_dndz}
we discuss the impact of photometric
redshift errors on $dN/dz$ and on our bias
results.

We give summary statistics for the cross-match $dN/dz$ in Table~\ref{tab:dndz_statistics} and plot
the cross-match $dN/dz$ in Figure~\ref{fig:dNdz_cross}.
Even at low redshift, there is a
systematic shift between the COSMOS and HSC
$dN/dz$ for the red sample; this may
be due to errors in the $z < 1$ HSC
redshifts. The impact of this shift
is limited because the cross-match
$dN/dz$ is only used to model the magnification
bias term. Therefore, even the $\Delta z = 0.3$
shift required to reconcile the COSMOS and HSC
$dN/dz$ for the red sample makes $\lesssim 0.7\sigma$ difference on the bias fitted to the auto and cross-correlation.

\begin{table}[]
    \centering
    \begin{tabular}{c | c c c c c | c c c}
Color & Median $z$ & Std & Percentile & Median $z$  & Median $z$  & Median $z$ & Std & Percentile \\
   & ($b$-weighted) & & (5\%-95\%) & $\ell < 155^{\circ}$ & $\ell > 155^{\circ}$ & & & (5\%-95\%) \\
\hline
Blue & 0.72 & 0.031 & 0.11 & 0.75 & 0.72 & 0.63 & 0.022 & 0.06 \\
Green & 1.38 & 0.026 & 0.09 & 1.35 & 1.38 & 1.09 & 0.019 & 0.06 \\
Red & 1.70 & 0.064 & 0.23 & 1.71 & 1.68 & 1.46 & 0.030 & 0.09 
    \end{tabular}
    \caption{Summary statistics of the redshift distribution of the WISE sample. Statistics on the left are computed from samples
    of $b_{\rm sml,p}dN_{\rm p}/dz$ (i.e.\ clustering redshifts), while statistics on the right are computed from samples of the cross-match $dN/dz$. Std gives the standard deviation of the medians of the 100
    $dN/dz$ samples. Galactic longitude $155^{\circ}$
    approximately splits the sky in half, so the $\ell = 155^{\circ}$ split
    provides another estimate of the uncertainty on the clustering redshifts.
    \label{tab:dndz_statistics}}
\end{table}

\subsection{Cross-correlation redshifts}
\label{sec:xcorr_dndz}

Another method for determining $dN/dz$ is through cross-correlation with a spectroscopic sample.
This is an old method has been revived in several recent works
\cite{Newman08,Menard13,McQuinn13,Schmidt13,Schmidt15,Rahman15,Rahman16,Rahman16b,Scottez16,Scottez18,Johnson17,Davis18,Cawthon18,Gatti18,Bates19}
(including validation against spectroscopic redshifts in ref.~\cite{Rahman15}),
but here we present one of its first applications
to modeling galaxy power spectra.  We therefore discuss and quantify several sources of systematic error, including
nonlinear clustering and nonlinear bias evolution; magnification bias contribution to the photometric-spectroscopic cross-correlation; and bias evolution of the various spectroscopic
samples as required to combine cross-correlations with multiple spectroscopic samples.

In the Limber approximation the cross-correlation of a photometric survey with scale-independent bias $b_{\rm sml, p}(z)$ and redshift distribution $dN_{\rm p}/dz$,
and a spectroscopic survey with bias $b_{\rm sml, s}(z)$,
in a narrow bin between $z_{\rm min}$ and $z_{\rm max}$ is
\begin{equation}
C_{\ell}^{\rm p-s \ cross} = b_{\rm sml, s}(z) \, b_{\rm sml, p}(z) \, H(z) \, \frac{dN_{\rm p}}{dz} \, \int_{z_{\rm min}}^{z_{\rm max}} \, dz \, \frac{dN_{\rm s}}{dz} \, \frac{P_{\rm mm}\left( k = (\ell + \frac12) /\chi , z \right)}{\chi^2} + C_{\ell}^{\rm mag}
\label{eqn:cl_dndz}
\end{equation}
where $C_{\ell}^{\rm mag}$ includes the contributions from the three lensing magnification bias terms; we assume the bin is sufficiently narrow that  the biases and $dN_{\rm p}/d\chi$ are constant across the bin; and both biases are assumed to be scale-independent.

We refer to the biases here as $b_{\rm sml}$
to emphasize that they are defined on relatively small scales ($2.5-10$ $h^{-1}$ Mpc) 
on which we measure the spectroscopic-photometric cross-correlations.
This is in contrast to the large-scale bias, $b_{\rm lin}$, relevant to the modeling of the angular power spectrum in Section~\ref{sec:model}.  
However, as discussed below and in Section~\ref{sec:systematics},
$b_{\rm sml}$ and $b_{\rm lin}$ are within 15\% of each
other for all of the unWISE samples,
and the systematic error from the discrepancy
between $b_{\rm sml}$ and $b_{\rm lin}$ is subdominant
to the statistical error from uncertainty in $b\, dN/dz$.

We have implemented a cross-correlation $dN/dz$ estimate in both harmonic and configuration space, obtaining consistent results.  It is convenient, and consistent with past results, to first present the spectroscopic-photometric cross-correlations in configuration space.  We use the estimator of Ref.~\cite{Menard13}, in which the correlation function is weighted by $r^{-1}$ to increase the signal-to-noise ratio
 and integrated over a range of scales:
\begin{equation}
  \bar{w}_{\rm{sp}}(z) = \sum_i \, r_i^{-1} \, \Delta r_i \, w_{\rm sp, binned}(r_i,z)
\label{eqn:wbar}
\end{equation}
where we use three log-spaced bins in $r$ between
2.5 and 10 $h^{-1}$ Mpc.  The binned correlation function is
given by
\begin{equation}
w_{\rm sp, binned}(r_i,z) = \int \, \frac{\ell d\ell}{2\pi} \, C_{\ell}^{\rm{p-s \ cross}} \, \frac{1}{\pi(r_{\rm max,i}^2 - r_{\rm min,i}^2)} \int_{r_{\rm min,i}}^{r_{\rm max,i}} \, 2 \pi r dr \, J_0(\ell \theta)
\label{eqn:binned_corrfunc}
\end{equation}
Noticing that $\theta = r/\chi$\footnote{
When measuring $w(\theta)$, we count pairs in angular
bins with $\theta_i = r_i/\chi_{\rm central}$, where
$\chi_{\rm central}$ is the comoving distance
to the center of the sub-bin.  Therefore, within each sub-bin, $r = \theta \chi_{\rm central}$,
which is well approximated by $r = \theta \chi$ because the sub-bins are narrow.
This simplifies the triple integral in Equation~\ref{eqn:iz}
by allowing the $r$ integral to be redshift-independent.}
and switching integration variables
from $\ell$ to $k$ we can write
\begin{equation}
\bar{w}_{\rm{sp}}(z) = b_{\rm sml, s}(z) \, b_{\rm sml, p}(z) \, H(z) \, \frac{dN_{\rm p}}{dz} \, I(z) + \bar{w}_{\rm mag}(z)
\label{eqn:wbar_final}
\end{equation}
with
\begin{equation}
    I(z) = \int \frac{ k \ dk}{2 \pi} \, \int_{z_{\rm min}}^{z_{\rm max}} \, dz \frac{dN_{\rm s}}{dz} \, P_{\rm mm}(k, z) \, \sum_{r_i} \, \frac{ r_i^{-1} \, \Delta r_i}{\pi r_{\rm{max},i}^2 - \pi r_{\rm{min},i}^2} \int_{r_{\rm{min},i}}^{r_{\rm{max},i}} 2 \pi r \, J_0(kr) \,  dr
\label{eqn:iz}
\end{equation}
In the linear regime $I(z)$ is equal to $D^2(z)$ times a redshift-independent integral, which is degenerate with the normalization of $dN_{\rm p}/dz$.  On our scales of interest,
$I(z)$ deviates only slightly from $D^2(z)$ ($\sim 5\%$ at $z =2$).
To compute $I(z)$ we use the HaloFit nonlinear matter power spectrum from Ref.~\cite{Mead15} for $P_{\rm mm}(k,z)$ and continue to assume scale-independent bias.
While the $I(z)$ term introduces a cosmology dependence
into the clustering redshifts, normalizing the clustering redshifts
eliminates the relationship between redshift distribution
and power spectrum amplitude, allowing us to constrain
the power spectrum amplitude (i.e. $\sigma_8(z)$) averaged
over the redshift distribution of each unWISE sample.

Lensing magnification can correlate samples widely separated in redshift
and therefore
 bias clustering redshifts in the tails of the distribution \cite{McQuinn13,Gatti18}.
We estimate the contribution of magnification bias
$\bar{w}_{\rm mag}(z)$ using the COSMOS cross-match $dN/dz$ and the measured $s$ for unWISE and the spectroscopic samples (Appendix \ref{sec:magbias_slope}), and assuming a scale-independent bias times the HaloFit power spectrum. We use the following form for the bias evolution of each sample:
 \begin{subequations}
\begin{align}
   b_{\rm sml,p}(z) &= 0.8 + 1.2 z &&\text{Blue} \\
   b_{\rm sml,p}(z) &= \max{(1.6z^{2},1)} &&\text{Green} \\
   b_{\rm sml,p}(z) &= \max{(2z^{1.5},1)} &&\text{Red}
\end{align}
\end{subequations}
with $\max{(a,b)}$ meaning the larger of $a$
and $b$. This form
is roughly consistent both with the observed
 clustering given the cross-match $dN/dz$ and the expected bias evolution
 from a simple HOD of the unWISE samples (Figure~\ref{fig:implied_bias}).
 Since the unWISE bias evolution is only required
to model the magnification bias correction to the cross-correlation redshifts,
more quantitative agreement with the observed clustering is not needed.
If we instead use $b_{\rm sml,p}(z)$
from the cross-correlation redshifts
after the initial magnification bias correction,
the bias fitted to $C_{\ell}^{\kappa g}$
and $C_{\ell}^{gg}$
changes by $<0.4\sigma$.

We show $\bar{w}$ and the magnification bias correction in Figure~\ref{fig:magbias_subtract}.  Magnification bias has the largest
impact on the blue sample, with CMASS galaxies at $z > 0.6$ showing the largest impacts.

We can invert Equation~\ref{eqn:wbar_final} to derive
$b_{\rm p, sml} dN_{\rm p}/dz$ given a measurement of $\bar{w}_{\rm sp}$.
We measure the binned correlation function using the estimator of ref.~\cite{Davis+Peebles83} 
\begin{equation}
    \hat{w}_{\rm{sp, binned}}(\theta) = \frac{D_{\rm{s}}D_{\rm{p}}}{D_{\rm{s}}R_{\rm{p}}} \frac{N_{\rm R}}{N_{\rm D}} - 1
\label{eqn:dp}
\end{equation}
using three log-spaced bins 
between\footnote{At all redshifts, the lower limit corresponds
to much larger angular scales than those affected by the WISE PSF
and suppression by nearby bright sources; see 
Fig.~25 in \url{http://wise2.ipac.caltech.edu/docs/release/allsky/expsup/sec6\_2.html\#brt_stars}
for estimation of this scale.} 2.5 and $10\,h^{-1}$ Mpc,
with the inner radius set to reduce the contributions from scale-dependent bias and ``1-halo'' effects.
Since the unWISE galaxy density varies across the sky (slightly decreasing towards the Galactic center),  we measure the normalization $N_{\rm R}/N_{\rm D}(\theta)$ in
\texttt{NSIDE} = 8 HEALPix pixels.
If the annulus in which
we count pair straddles two \texttt{NSIDE} = 8
pixels, we average the normalization
in the two pixels.
 Our correlation function
code is publicly available at 
\url{https://github.com/akrolewski/BallTreeXcorrZ} and has been tested to ensure
that the correlation function as measured on the curved
sky is correct.

\begin{figure}
    \centering
    \resizebox{\columnwidth}{!}{\includegraphics{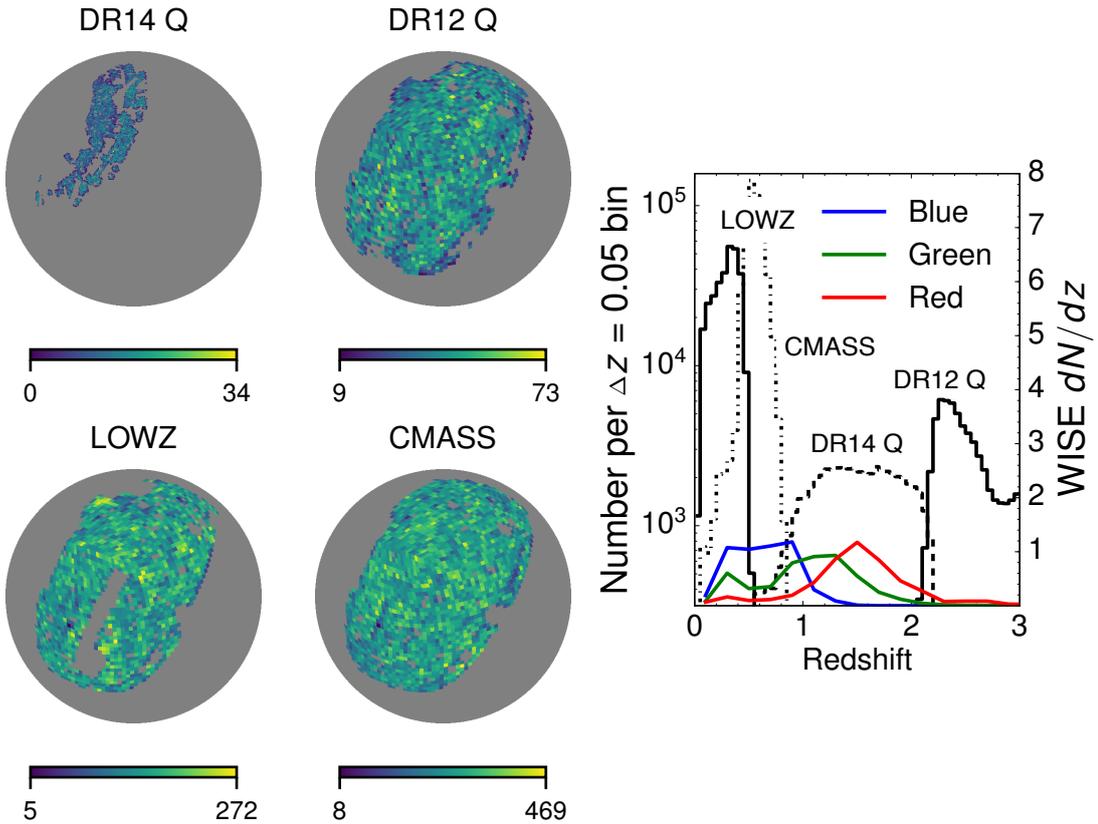}}
    \caption{\textit{Left:}
     Sky maps of spectroscopic samples used for cross-correlation redshifts in Galactic coordinates.
    \textit{Right:}
    Redshift distributions of spectroscopic samples (black curves) used for cross-correlation redshifts.  unWISE galaxy distributions (colored curves) are overplotted with arbitrary amplitude. 
    }
    \label{fig:spec_zdist}
\end{figure}

We cross-correlate the unWISE photometric galaxies with spectroscopic quasars from BOSS DR12 \cite{Paris12} and eBOSS DR14 \cite{Ata18}\footnote{While the BOSS and SDSS quasar catalogs are independent, eBOSS includes previously observed quasars.  We remove these quasars to create an independent sample; they comprise 45\% of the northern eBOSS catalog.} and galaxies from BOSS CMASS and LOWZ \cite{Reid2016}.
We plot the redshift and sky distributions of the spectroscopic samples used for the clustering redshifts in Figure~\ref{fig:spec_zdist}.

We split the spectroscopic
samples into bins of width $\Delta z$ = 0.05 at $z < 0.8$ and $\Delta z = 0.2$ at $z > 0.8$ where the errorbars become much larger due to the sparser quasar samples.
As $\Delta z$ becomes narrower, the signal-to-noise in each individual
bin decreases, but the total signal-to-noise of $dN/dz$
increases modestly 
(by $\sim35\%$ as $\Delta z$ changes from 0.1 to 0.02).  We 
prefer having a relatively high signal-to-noise in the individual bins, particularly at high redshift where $dN/dz$
is nearly zero but our errorbars are also relatively large.

We restrict the BOSS quasars to the ``{\sc core}''-like sample from ref.~\cite{Eftekharzadeh15} (similar to the ``{\sc qso\_core\_main}'' targeting flag, but only including quasars
that would have been selected by the {\sc xdqso} method \citep{Bovy11},\footnote{\url{https://xdqso.readthedocs.io/en/latest/}} and removing
quasars lying in regions where the targeting completeness is $<75\%$), as was done in previous quasar clustering analyses \cite{White12,Eftekharzadeh15}.
We remove all objects in the southern galactic cap (SGC), which are a small fraction of the total 
spectroscopic sample. Differences in photometric calibration between the SGC and NGC lead to slightly different galaxy samples \cite{Alam17,Ross17}, unexplained differences in quasar clustering between NGC and SGC \cite{White12,Eftekharzadeh15,Myers15}, and differences in $\hat{w}_{\rm sp}$, possilby resulting
from the different spectroscopic bias evolution.
We remove DR12 quasars at $z < 2$, DR14 quasars at $z > 2.2$, and DR14
quasars at $z < 0.8$, since these objects are outliers with redshifts different from each survey's target redshift range and thus may have different clustering properties than the sample as a whole
(indeed, we find somewhat discrepant measurements of $\bar{w}_{\rm sp}$ when comparing to BOSS and eBOSS quasars outside their primary redshift ranges).
We also remove CMASS galaxies at $z > 0.8$ and $ z < 0.1$ and LOWZ
galaxies at $z > 0.5$, where 
the spectroscopic samples become too sparse to measure
$b_{\rm sml,s}$ (see below).
We summarize the key properties of these samples in Table \ref{tab:spec_samples}.

We apply the corresponding spectroscopic mask to each sample. For eBOSS we use the BOSS veto masks \cite{Reid2016}, pixelized to \texttt{NSIDE} = 256 HEALPix pixels, and we also mask \texttt{NSIDE} = 128 pixels where more than 80\% of eBOSS quasars are in DR7 or DR12.
For DR12 quasars, we apply the BOSS veto masks \cite{Reid2016} and remove \texttt{NSIDE} = 256 pixels with $<75\%$ completeness as computed from the {\sc bossqsomask} software\footnote{\url{http://faraday.uwyo.edu/~admyers/bossqsomask/}} \cite{White12,Eftekharzadeh15}.
For CMASS and LOWZ we use the corresponding BOSS DR12 LSS catalog masks.\footnote{\url{https://data.sdss.org/sas/dr12/boss/lss/}}
We also apply the same WISE masks that we use for the cross-correlation analysis.
For the spectroscopic cross-correlations, we threshold all masks by setting
pixels with mask value $<0.9$ to zero and $>0.9$ to one.

\begin{table}[]
    \centering
        \begin{tabular}{c|ccccc}
        \multirow{2}{*}{Sample} & \multirow{2}{*}{$z_{\rm min}$} & \multirow{2}{*}{$z_{\rm max}$} & \multirow{2}{*}{$N$} & Jackknife & Area \\
        & & & & regions & (deg$^2$) \\
        \hline
        DR14 Q  & 0.8 & 2.2  & 54708 & 29 & 1178 \\
         DR12 Q  & 2.0 & 4.0  & 67175 & 34 & 6030 \\
         LOWZ  & 0.0 & 0.5 & 273549 & 31 & 5656 \\
         CMASS  & 0.1 & 0.8 & 544308 & 37 & 6670
        \end{tabular}
    \caption{Properties of the spectroscopic samples used for cross-correlation redshifts.
    \label{tab:spec_samples}}
\end{table}

We use jackknife\footnote{We also investigated bootstrap resampling to estimate the covariance matrix of $w(\theta)$, and found that on scales smaller
than the resampling pixel size (which is always larger than $\theta_{\rm max}$), jackknife errors agree well with errors from
the ``marked bootstrap'' of Refs.~\cite{LohStein04,Loh08}.  We prefer
jackknife errors to bootstrap errors because sampling a pixel more than once
double-counts all pairs in that pixel and is not a reasonable physical situation. Moreover, it leads to ambiguities in the situation where a spectroscopic
source lies in pixel $i$ and a photometric source lies in pixel $j$. In a naive implementation
of the bootstrap, intra-pixel pairs are resampled $0,1,2...N$ times
while cross-pixel pairs are resampled $0,1,4...N \times M$ times, leading
to larger variance on all scales than jackknife or marked bootstrap.  The marked
bootstrap avoids this issue by resampling only one of the tracers.
We also find very little difference between leave-one-out jackknife
resampling and leave-two-out jackknife, so we opt for leave-one-out jackknife
in the interest of simplicity.
} resampling to estimate errors on $\bar{w}_{\rm {sp}}$. We start by splitting the sky into \texttt{NSIDE} = 4 HEALPix pixels\footnote{We use \texttt{NSIDE} = 8 pixels for the smaller-area eBOSS samples.
We find that using too small pixels can underestimate the errorbars (as argued in ref.~\cite{Norberg09}), so we set the size of the resampling pixels so that $\sim30$ are available for each sample.} and then combine neighboring pixels until the unmasked area within each region reaches a threshold, which we vary between 80 and 120\% of the maximum pixel area, choosing the threshold that minimizes the difference between the largest and smallest regions. We list the number of regions used for each sample in Table \ref{tab:spec_samples}.  
The error on $\bar{w}_{\rm sp}$ is then
\begin{equation}
    \sigma_{\bar{w}}^2(z) = \sum_{L=1}^N \frac{R_{[L]}}{R} \left(\bar{w}_{[L]}(z) - \langle \bar{w}(z)\rangle \right)^2
\label{eqn:sigma_bdndz}
\end{equation}
where $R$ refers to the number of randoms, the subscript $[L]$ indicates that we exclude the $L$th region,
and $\langle \bar{w}(z) \rangle$ is the average over all
$N$ jackknife resamples
$\bar{w}_{[L]}(z)$.
The replacement of the conventional factor $(N-1)/N$ with $R_{[L]}/R$ is an empirical correction for the fact that the regions have different areas (equation 5 in ref.~\cite{Myers05}).

\begin{figure}
    \centering
    \resizebox{\columnwidth}{!}{\includegraphics{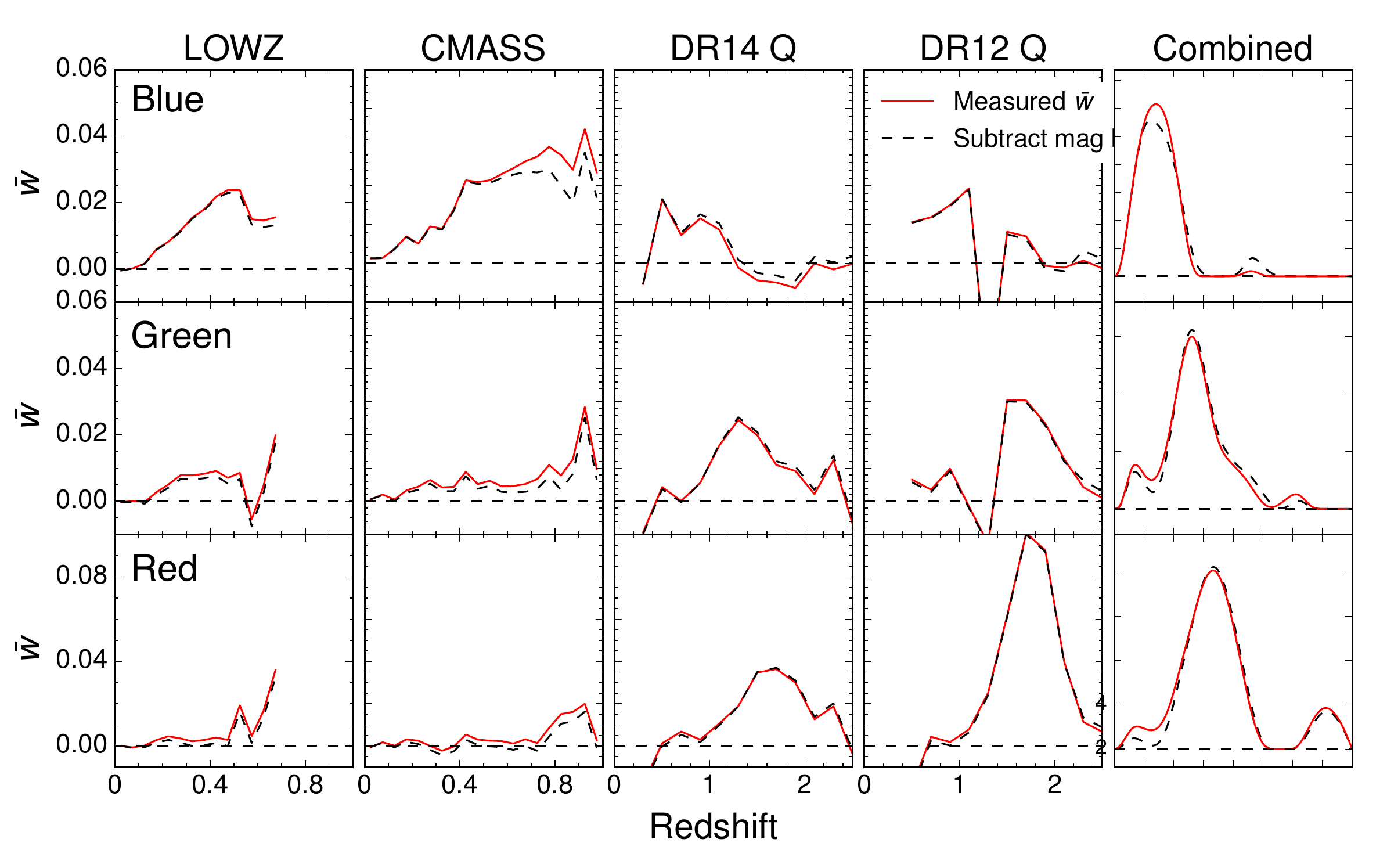}}
    \caption{Measured $\bar{w}$ without (black dashed) and with (red solid) magnification bias subtracted for unWISE galaxies crossed with our spectroscopic samples (labeled above each column), and for all samples combined (right-most panel).
    }
    \label{fig:magbias_subtract}
\end{figure}

Combining multiple spectroscopic tracers
(as is necessary in our case, due to the broad $dN/dz$ of the unWISE
samples) requires a measurement
 of $b_{\rm sml,s}(z)$.
We measure $b_{\rm sml,s}(z)$ by fitting a scale-independent bias
times Halofit to the measured $w(\theta)$
between 2.5 and 10 $h^{-1}$ Mpc:
\begin{equation}
w_{\rm auto,s}(\theta, z) = b^2_{\rm sml,s}(z) \int_{0}^{\infty} \, k \, dk \, P_{\rm mm}(k) \int_{\chi_{\rm min}}^{\chi_{\rm max}} \frac{d\chi}{2\pi} \, J_{0}(k \chi \theta) \left (\frac{dN_{\rm s}}{d\chi}\right)^2
\label{eqn:wtheta_for_spec_bias}
\end{equation}
where the integral over $\chi$ ranges between the lower
and upper boundaries of each redshift bin.
 We omit SDSS DR7 quasars
 from our spectroscopic samples due to their poorly measured
 autocorrelation \cite{Ross09}.
For BOSS galaxies
and eBOSS quasars we use publicly available galaxy and random catalogs,\footnote{\url{https://data.sdss.org/sas/dr12/boss/lss/} for BOSS and \url{https://data.sdss.org/sas/dr14/eboss/lss/catalogs/} for eBOSS.} and
for DR12 quasars we generate randoms using {\sc bossqsomask}.
For BOSS
galaxies and eBOSS quasars we weight each object by the combined angular systematics,
fiber collision and redshift failure weight (Equation 50 in ref.~\cite{Reid2016}),
and for BOSS quasars we weight by the inverse of the 
targeting completeness \cite{White12,Eftekharzadeh15}.
Note that the $55''$ SDSS fiber collision radius is smaller than our inner bin of $2.5\,h^{-1}$Mpc at all redshifts that we consider. 
Previous work has shown that application
of these systematics weights allows unbiased
recovery of the correct cosmological clustering \citep{Ross12,Ross14,Eftekharzadeh15,Ross17,Laurent17}.

\begin{figure}
    \centering
    \resizebox{\columnwidth}{!}{\includegraphics{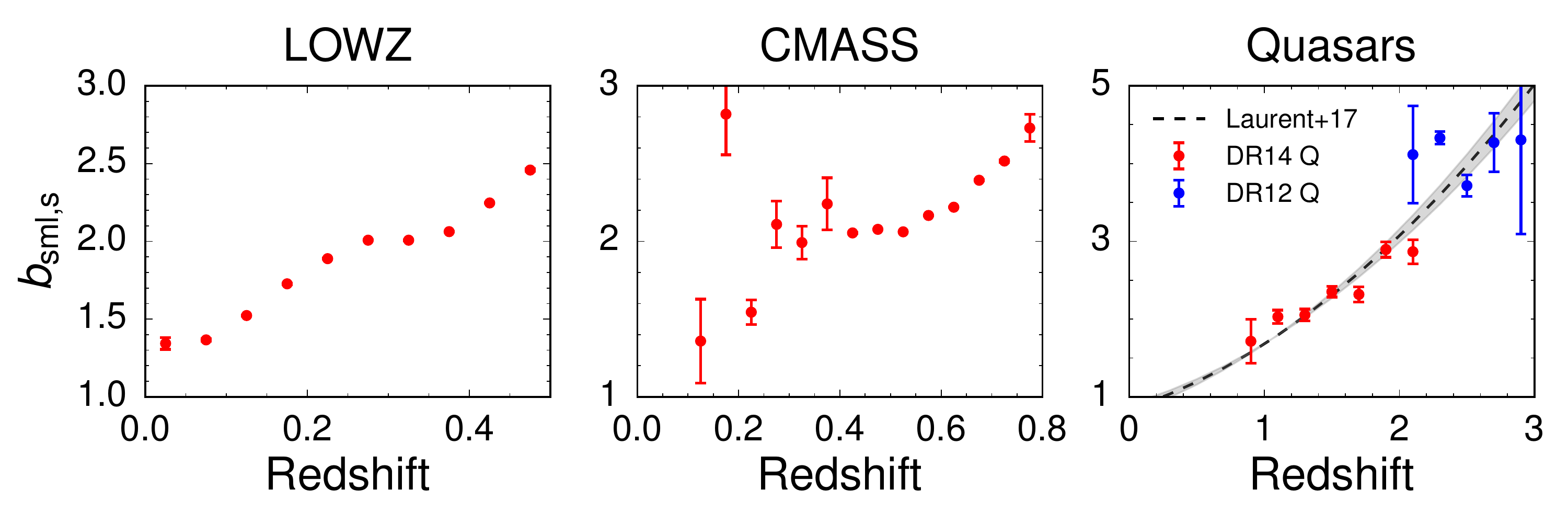}}
    \caption{Measured bias, $b_{\rm sml,s}$, of spectroscopic samples as a function of redshift. The dashed line in the right panel corresponds to the fit from ref.~\cite{Laurent17} (with the gray band giving uncertainty in the fit). See Table~\ref{tab:spec_bias} for a tabulated compilation.}
    \label{fig:spec_bias}
\end{figure}

While previous measurements of the clustering exist for all spectroscopic
samples,  measurements for BOSS galaxies
have generally been made in coarse redshift
bins over a somewhat restricted redshift range
($0.2 < z < 0.6$) \citep{Chuang17,Tojeiro12,Zhai17}, so we use
our measured bias values 
(Figure~\ref{fig:spec_bias} and Table~\ref{tab:spec_bias})
to ensure that the scales
used and redshift bins are consistent with
the clustering redshifts.
We check our results by replacing our BOSS galaxy spectroscopic
bias measurements with those from Figure 12 in ref.~\cite{ChiangMenard18} (who measure the bias
in similar redshift bins, to measure clustering
redshift distributions for Galactic dust maps),
and find this makes $\lesssim 0.1\sigma$
difference in the bias fitted to $C_{\ell}^{gg}$ and $C_{\ell}^{\kappa g}$.
We also propagate the fitting error on the spectroscopic bias to our clustering $dN/dz$
measurement, although this is almost always
subdominant to the statistical error on the cross-correlation.

For 
quasars, we find that the fitting function of ref.~\cite{Laurent17} provides a very good
approximation to the measured bias evolution:
\begin{equation}
    b_{\rm sml,s}(z) = (0.278 \pm 0.018) \left[ \left(1+z\right)^2-6.565\right]+(2.393 \pm 0.042)
    \label{eqn:laurent_b}
\end{equation}

\begin{figure}
    \centering
    \resizebox{\columnwidth}{!}{\includegraphics{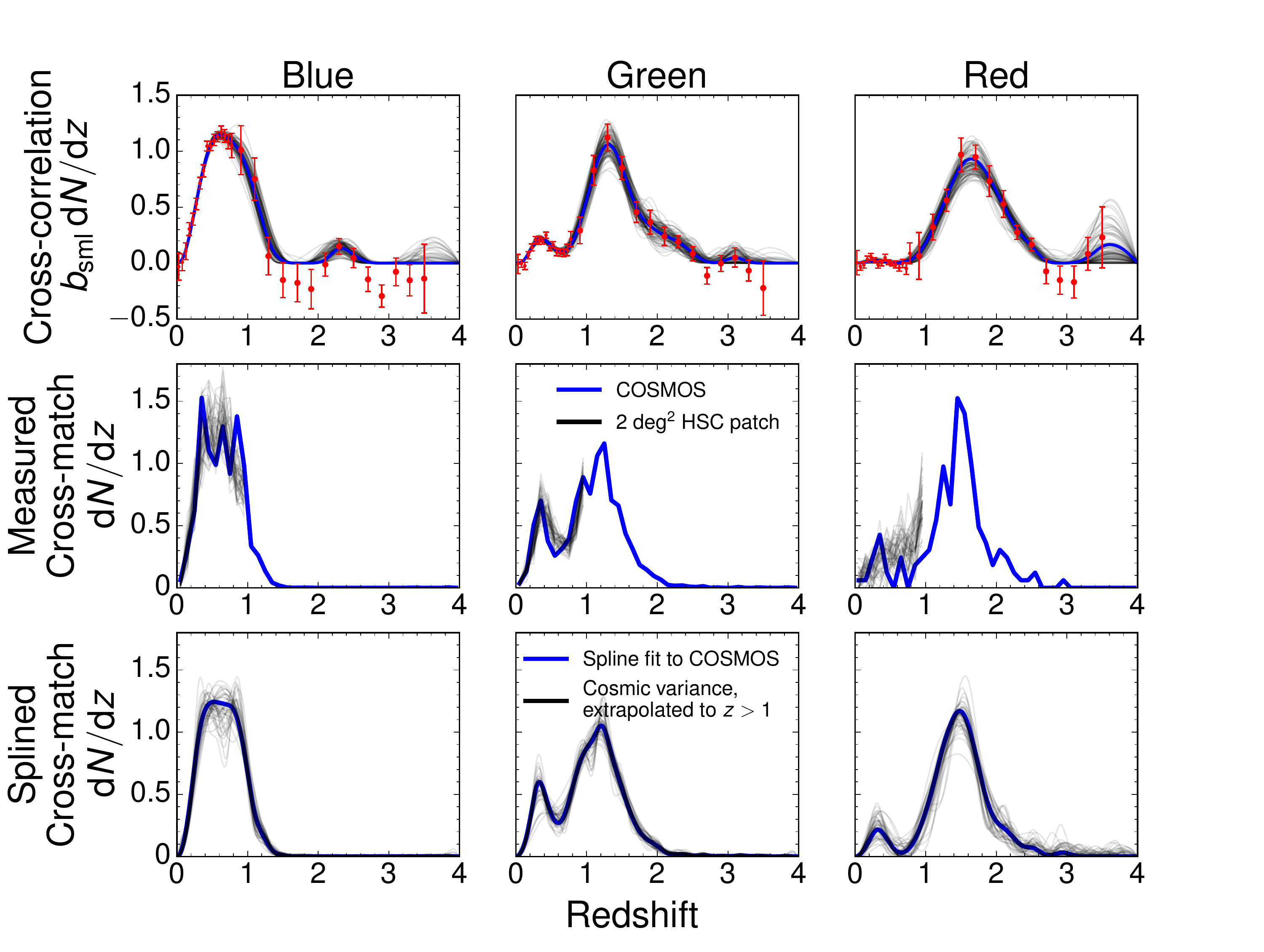}}
    \caption{\textit{Top}: $b_{\rm sml,p} dN_{\rm p}/dz$ combined for all tracers, with best-fit spline plotted in blue and 100 splined samples drawn from diagonal Gaussian realizations
    of the noise overplotted in gray.
    The difference in $dN/dz$ between the two halves of the sky
    (split at $\ell = 155^{\circ}$; not shown) is comparable to the range
    of the 100 spline realizations.
   \textit{Middle}: $dN/dz$ from the COSMOS cross-matches
    (thick solid lines)
    compared to $dN/dz$
    from 44 COSMOS-like ${\sim}2$deg$^2$ patches of HSC (gray lines).
     While HSC is deep enough to contain nearly
    all of the WISE objects, its photometric
    redshifts become biased at $z > 1$ and therefore
    we do not display them in this range.
    \textit{Bottom}: Spline fit to the 
    COSMOS $dN/dz$ (thick solid line) compared to
    44 realizations of the noise
    assuming diagonal Gaussian errors (thin gray lines). To account for cosmic variance, the standard deviation is a fixed
    multiple of the Poisson error set to match the observed
    scatter from the 44 HSC patches at $z < 1$, 
    as described in Section~\ref{sec:xmatch_dndz}.
    }
    \label{fig:dNdz_cross}
\end{figure}

Once we have measured the spectroscopic
bias, we can divide $\bar{w}$ by $b_{\rm sml,s}$ (as in equation~\ref{eqn:wbar_final})
to obtain
$dN_p/dz$ for each spectroscopic tracer
(quasars, CMASS and LOWZ galaxies).  We then combine $dN_p/dz$ for the different tracers
by inverse-variance weighting
in each redshift bin.
We find good agreement between the clustering redshift measurements
from CMASS and LOWZ in the redshift range where they overlap ($0 < z < 0.5$) with $\chi^2 = 12.1$ over 10 dof (7.7/10, 2.8/10) for the blue
(green, red) samples.
 We plot the final redshift distributions
in Fig.~\ref{fig:dNdz_cross}, along with samples
drawn from the uncertainty in the redshift distribution.

To model the angular power spectra, the redshift distribution
must satisfy
physical constraints 
($b_{\rm sml,p}dN_p/dz > 0$ and  $b_{\rm sml,p}dN_p/dz |_{z=0}$ = 0$)$)\footnote{The number of galaxies per area per comoving distance, $dN/d\chi$,
is related to the comoving number density
$\bar{n}$ as
    $dN/d\chi(\chi) = \bar{n} 4 \pi \chi^2$
so at $\chi = 0$, $dN/d\chi$
and therefore $dN/dz$ must go to zero.}
and have well-characterized uncertainties.
To create a smooth and physical $dN_p/dz$,
we therefore model  $b_{\rm sml,p}dN_p/dz$ using cubic B-splines
with the spline coefficients required to be positive,
satisfying the positivity constraint on $b_{\rm sml,p}dN_p/dz$.
Considering the penalized $\chi^2$:
\begin{equation}
\chi^2 = \sum_i \left(\frac{\hat{y_i}-y_i}{\sigma_i}\right)^2 + \lambda \int \, dx \, [\hat{y}''(x)]^2
\end{equation}
we determine $\lambda$ by minimizing $\chi^2$ using cross-validation \citep{CravenWahba1978}.
This method is sufficiently
flexible to fit almost any shape of  $b_{\rm sml,p}dN_p/dz$,
while satisfying our constraints.
We use knots evenly spaced between $z = 0.1$ and $z = 3.5$
with $\Delta z = 0.2$.  

By requiring $\hat{y}$ to be positive, this procedure
introduces a bias into the theory predictions for $C_{\ell}$,
since in regions of nearly zero $dN/dz$, we
will fit to positive noise fluctuations but
not negative noise fluctuations.
Moreover, the magnitude of this bias is different
for $C_{\ell}^{\kappa g}$ and $C_{\ell}^{gg}$.
We find that the differential bias is generally
small ($< 5\%$) and therefore do not consider it further in this paper. However, cosmological parameter
constraints from these data will require a more careful approach \cite{paper2}, such as simulating
the $C_{\ell}$ and $dN/dz$ measurement given some
known input cosmology and $dN/dz$,
and subtracting the 
contribution to $C_{\ell}$ from the 
bias in $dN/dz$.

We create smooth $dN/dz$ in a similar fashion for the cross-match redshifts.  Here we use bins of $\Delta z = 0.06$
instead, as we find that more knots are required to accurately represent the shape of the cross-match $dN/dz$.

We propagate errors on $dN/dz$ by drawing 100 samples
from the data (assuming uncorrelated Gaussian errors between redshift bins), finding the best-fit $dN/dz$, and using it to model the auto and cross-power spectra.
Additionally, we test the assumptions behind
the jackknife $dN/dz$ errors by splitting the sky in half
at Galactic $\ell = 155^{\circ}$ and measuring
$dN/dz$ separately for each half.
Summary statistics for both the cross-correlation and cross-match redshifts are given in Table~\ref{tab:dndz_statistics}.

\subsection{Systematic errors in the cross-correlation $dN/dz$}

One source of systematic error in the measurement
of $dN/dz$ is the discrepancy between $b_{\rm sml,p}$, as measured by the configuration space photometric-spectroscopic
cross-correlations, and $b_{\rm lin,p}$, as required
to model the autospectra and CMB cross-spectra.
Exactly matching the scales used for the cross-correlation redshifts and the CMB cross-correlations is undesirable because it would push the cross-correlation redshifts to large scales where the signal-to-noise is lower and the potential impact of observational systematics is larger \cite{Menard13}; indeed, previous work uses scales of several Mpc at most \cite{Menard13,Schmidt13,Rahman15,Rahman16,Rahman16b,Davis18,Cawthon18,Gatti18}.  Conversely working on very small scales can be problematic, as the cross-correlation could depend upon galaxy formation physics in addition to the redshift distribution.

To study potential deviations between $b_{\rm sml,p}$
and $b_{\rm lin,p}$,
we populate an $N$-body simulation with a simple HOD model
for the WISE galaxies (Appendix~\ref{sec:xcorr_redshift_details}), which is roughly consistent
with the spectroscopic cross-clustering given the cross-match
$dN/dz$ (Figure~\ref{fig:implied_bias}).  We then measure
$b_{\rm sml}$ and $b_{\rm lin}$ from the autocorrelation of halos
in the simulation at four representative redshifts ($z = 0.41$, 1.00, 1.27
and 1.78; Figure~\ref{fig:choosing_xcorrz_scales}).  At $z = 0.41$, $b_{\rm sml}$ is 0.7\% (1.7\%, 2.5\%) greater
than $b_{\rm lin}$ for halos representative of the blue (green, red)
samples.  At $z = 1.78$, $b_{\rm sml}$ is 7.2\% (15.3\%) greater
than $b_{\rm lin}$ for green (red) halos; and at $z = 1.00$, $b_{\rm sml}$
is 1.6\% greater than $b_{\rm lin}$ for blue halos.
We discuss the implications of these discrepancies between
$b_{\rm sml,p}$ and $b_{\rm lin,p}$ in Section~\ref{sec:systematics};
we find that their impact is subdominant to the statistical uncertainty on $dN/dz$.

\begin{figure}
    \centering
    \resizebox{\columnwidth}{!}{\includegraphics{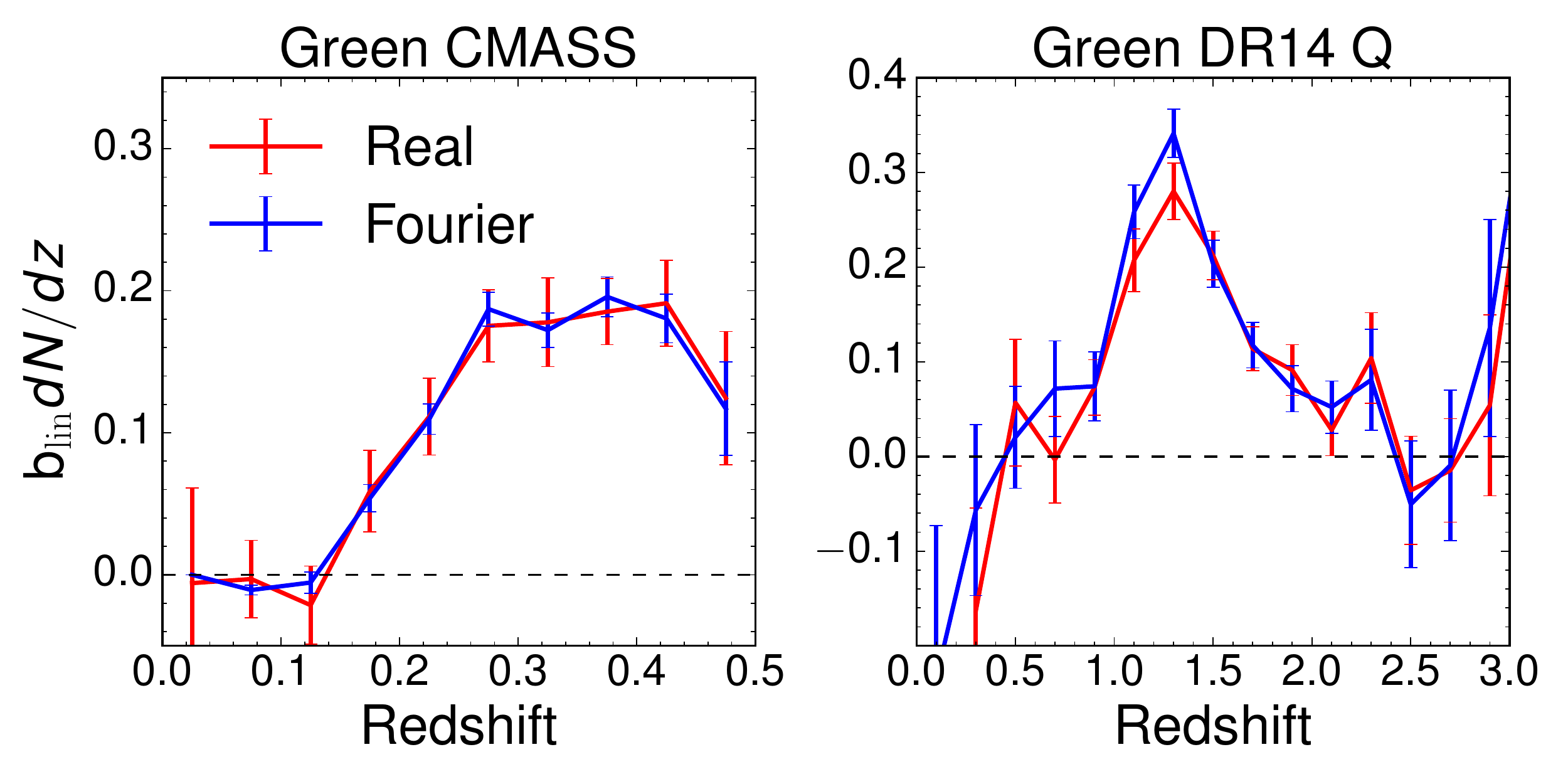}}
    \caption{Comparison between the fiducial $b_{\rm sml,p}dN_{\rm p}/dz$ measured
    in configuration space (Equation~\ref{eqn:wbar_final}) and $b_{\rm lin,p} dN_{\rm p}/dz$ measured in harmonic space using the pipeline described in Section~\ref{sec:angular_clustering}, with $100 < \ell < 1000$.
    }
    \label{fig:real_vs_fourier}
\end{figure}

We also compare the fiducial real-space $b_{\rm sml,p}dN_{\rm p}/{dz}$
to $b_{\rm lin,p}dN_{\rm p}/{dz}$ measured in Fourier
space using Eq.~\ref{eqn:cl_dndz} on the same angular scales ($\ell = 100$ to 1000) as 
the CMB lensing cross-correlation.\footnote{At $z < 0.3$, $\ell_{\rm max} = 1000$ corresponds to $k_{\rm max} > 1$ $h$ Mpc$^{-1}$, and the scale-independent bias assumption begins to break down.  As a result, we
set $\ell_{\rm max} = \rm{min}(1000, k_{\rm max}\chi)$,
where $\chi$ is the comoving distance to the redshift bin center
and $k_{\rm max} = 2.5$ $h$ Mpc$^{-1}$.}
We find good agreement for $b dN/dz$ in both configuration and harmonic space
(Figure~\ref{fig:real_vs_fourier}),
suggesting that discrepancies between $b_{\rm sml,p} dN_{\rm p}/dz$ and $b_{\rm lin,p} dN_{\rm p}/dz$ are minor.

We test the sensitivity of the $dN/dz$ results
to the presence of angular systematics in the spectroscopic
data by measuring the weighted cross-correlation,
using the combined angular systematics, fiber collision
and redshift failure weights for BOSS galaxies and eBOSS; and the
inverse of the targeting completeness for BOSS quasars.
We also create and apply
systematics weights for the unWISE
samples, following the methodology
of ref.~\cite{Ross17} for the BOSS and eBOSS samples. We construct
\texttt{nside}=128 maps
of the unWISE density (correcting
for sub-pixel masking) and of several
systematics: stellar density from Gaia \citep{Gaia18},
5$\sigma$ limiting magnitude in W2,
and $N_{\rm HI}$ column density
from the HI4PI 21 cm survey \citep{hi4pi}.  We fit piecewise
linear functions to the relationship
between density and systematic,
using stellar density and W2 limiting
magnitude for the green and blue samples, and stellar density and $N_{\rm HI}$
for the red sample (other maps are not significantly correlated with galaxy density).  We then define the weights
for each systematic as the inverse of the predicted density.

We found $< 0.8\sigma$ change between the weighted
and unweighted cross-correlations among all bins in redshift,
spectroscopic and photometric tracers; this suggests
that angular systematics correlated with unWISE fluctuations
do not significantly affect our results.

We also compare the observed cross-correlation in a given 
redshift bin to the cross-correlation between Gaia stars and the spectroscopic sample.  Since $\sim 2\%$ of all unWISE
samples are stars, star-driven fluctuations in the spectroscopic
sample may lead to spurious correlations between unWISE
and spectroscopic samples.  We find that the LOWZ-Gaia cross-correlation, times a fiducial stellar contamination fraction of 2\%, is $<5\%$ of the LOWZ-unWISE cross-correlation at $z > 0.15$, but comparable to the LOWZ-unWISE cross-correlation at $z < 0.15$ (for all three colors), although the error bars on the LOWZ-Gaia cross-correlation
are comparable to the measured cross-correlation at these redshifts.

\section{Galaxy-lensing auto and cross-spectra}
\label{sec:results}

In this section we present our measurements. We parameterize the amplitude of the correlations by a single effective linear bias 
\begin{equation}
b^{\rm eff} = \int \, dz \, b_{\rm lin,p}(z) \frac{dN_{\rm p}}{dz} 
\end{equation}
where we follow our convention of $\int dz \ dN/dz = 1$. 
For our theory model, we use the cross-correlation redshifts, $b_{\rm sml,p}dN_p/dz$, to approximate $b^{\rm eff}$
\begin{equation}
b^{\rm eff} \approx \int \, dz \, b_{\rm sml,p}(z) \frac{dN_{\rm p}}{dz} 
\end{equation}
We insert the normalized cross-correlation redshifts,
$f(z) dN_p/dz$
\begin{equation}
f(z) \frac{dN_{\rm p}}{dz} \equiv \frac{b_{\rm sml,p}(z) \frac{dN_{\rm p}}{dz}}{b^{\rm eff}}
\end{equation}
into equation~\ref{eqn:ClXY},
and allow the amplitude $b^{\rm eff}$ to vary to match the data:
\begin{multline}
    C_{\ell}^{\kappa g} = b^{\rm eff} \int d\chi \frac{W^{\kappa}(\chi)}{\chi^2} H(z) \left[f(z) \frac{dN_p}{dz}\right] 
    P(k \chi = \ell \, + \, 1/2) \\
    + \int d\chi \frac{W^{\kappa}(\chi) W^{\mu}(\chi)}{\chi^2} P(k \chi = \ell \, + \, 1/2)
\end{multline}
\begin{multline}
    C_{\ell}^{g g} = (b^{\rm eff})^2 \int d\chi \frac{1}{\chi^2} H(z)^2 \left[f(z) \frac{dN_p}{dz}\right]^2
    P(k \chi = \ell \, + \, 1/2) \\
    + b^{\rm eff} \int d\chi \frac{W^{\mu}(\chi)}{\chi^2} H(z) \left[f(z) \frac{dN_p}{dz}\right]  P(k \chi = \ell \, + \, 1/2) \\
    + \int d\chi \frac{W^{\mu}(\chi) W^{\mu}(\chi)}{\chi^2} P(k \chi = \ell \, + \, 1/2)  
\end{multline}
For the magnification bias term $W^{\mu}(\chi)$, we take the cross-matched $dN/dz$, and the values of $s$ from Appendix \ref{sec:magbias_slope}. 

\begin{figure}[ht]
    \centering
    \includegraphics[width=15cm]{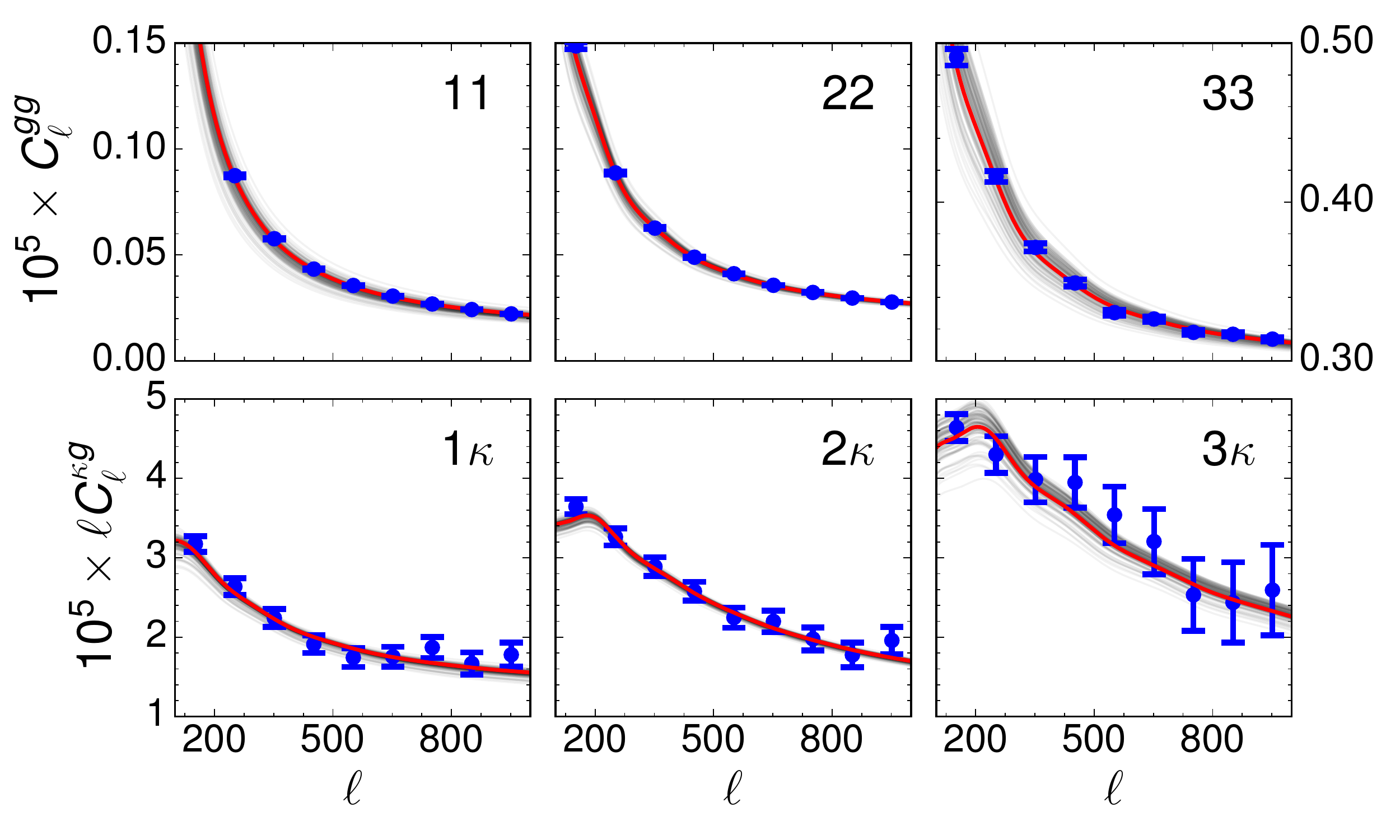}
\caption{Auto correlation (top) and cross correlation between the unWISE catalog and Planck CMB lensing (bottom);  numbers label the samples (1: blue, 2: green, 3: red). The best-fit theory curve assuming a constant bias times HaloFit
is shown as a solid red line and the uncertainty in the model
from the uncertainty in $dN/dz$ is given by the gray lines.
We fit angular scales to the right of the dashed line.
Magnification bias is a few times larger than the errorbars in the auto-spectra,
and $\sim 50\%$ of the errorbars in the CMB cross-spectra.
}
\label{fig:results_cross}
\end{figure}

Figure \ref{fig:results_cross} shows the auto correlation of our three galaxy samples as well as their cross-correlation with the CMB lensing convergence, $\kappa$.  Table \ref{tab:halofit_xcorr} summarizes the results.
We quote both statistical and $dN/dz$ error bars on $b^{\rm eff}$; the statistical errors
are from the errors on $C_{\ell}^{gg}$ and $C_{\ell}^{\kappa g}$ using the fiducial $dN/dz$,
whereas the $dN/dz$ error bars are the standard deviation of $b^{\rm eff}$ from fitting
$C_{\ell}^{gg}$ and $C_{\ell}^{\kappa g}$ to 100 samples of $dN/dz$ with uncorrelated
Gaussian error added (as described earlier).

Over the range of scales that we model ($100 < \ell < 1000$), we obtain cross-correlation $S/N = \sqrt{\chi^2_{\rm null} - \chi^2_{\rm cross}}$ of 59.2, 68.5 and 41.4 for the blue, green and red samples, respectively.  The combined cross-correlation $S/N$ for the sample as a whole (taking into account the covariance between the three galaxy samples) is 79.3.

\begin{table}[h!]
    \centering
    \small
    \begin{tabular}{ccccc||ccc}
	WISE & \multirow{2}{*}{$b^{\rm eff}_{\rm auto}$}  & Shot Noise & $\sigma_b$ from & \multirow{2}{*}{$\chi^2_{\rm auto}$/dof} & \multirow{2}{*}{$b^{\rm eff}_{\rm cross}$}  & $\sigma_b$ from & \multirow{2}{*}{$\chi^2_{\rm cross}$/dof}  \\
	sample & & $(\times 10^7)$ & $dN/dz$ & & & $dN/dz$ & \\
	        \hline
	       Blue & $1.74 \pm 0.0052$ & $0.92 \pm 0.012$ 
	       & 0.0865 & 24.3/4 & $1.56 \pm 0.0276$ & 0.0355 & 6.11/5 \\ 
	       Green & $2.44 \pm 0.0083$ & $1.81 \pm 0.012$ 
	        & 0.0793 & 8.69/4 & $2.23 \pm 0.0352$ & 0.0308 & 2.93/5 \\
	       Red & $3.47 \pm 0.0383$ & $29.6 \pm 0.09$ 
	       & 0.2435 & 8.21/4 & $3.29 \pm 0.090$ & 0.1541 & 4.56/5 \\
    \end{tabular}
    \caption{Results from fitting a constant bias times HaloFit power spectrum (using cross-correlation $dN/dz$). Note that the value of $\chi^2$ here is for a \textit{fixed} fiducial $dN/dz$ and for a linear bias model with HaloFit power spectrum. A high $\chi^2$ value indicates the need to marginalize over the redshift distribution for any cosmological interpretation, and highlights the importance of going beyond linear bias. In the follow up paper \cite{paper2}, we fully marginalize over the uncertainty in $dN/dz$ and a non-linear model for galaxy biasing, obtaining a good fit.}
    \label{tab:halofit_xcorr}
\end{table}

\section{Systematics in the cross-correlation and null tests}
\label{sec:systematics}

In this section we explore the impact of stellar contamination, foregrounds in the CMB maps and the galactic latitude dependence of the signal.

\subsection{Stellar Contamination}

Due to the photometric nature of the catalog, with only two broad-band filters available, some fraction of the objects in our catalog will be stars or other non-cosmological sources such as nebulae or artifacts in the images. For simplicity, below we shall refer to any non-cosmological source that is uncorrelated with the true galaxies in our samples as ``stars.''
On scales where stars can be considered unclustered, i.e.\ where their clustering power is negligible compared to the galaxies, their effect is to lower both the auto and cross correlations in a way that is completely degenerate with the galaxy bias, and hence can be marginalized over in a cosmological analysis.  To see this, let's assume that average number density of objects in our catalog, $\Bar{n}$, is the sum of galaxies, $\Bar{n}_g$, and ``stars'', $\Bar{n}_s$. The observed ``galaxy overdensity'' necessarily includes both
\begin{equation}
    g^{\rm obs} = \frac{\delta n_g}{\Bar{n}_g + \Bar{n}_s} + \frac{\delta n_s}{\Bar{n}_g + \Bar{n}_s}
\end{equation}
We expect the second term to be uncorrelated with CMB lensing, given its non-cosmological origin. This is an important assumption that can be violated, for example, if galactic dust emission affects CMB lensing reconstruction, and at the same time modulates the number density of galaxies observed in WISE.  We test this in the next section by applying different galactic cuts. Assuming the second term is uncorrelated with $\kappa$, we can write 
\begin{equation}
    \langle \kappa \ g^{\rm obs} \rangle = \langle \kappa \ g^{\rm true} \rangle \frac{\Bar{n}_g}{\Bar{n}_g + \Bar{n}_s} = \langle \kappa \ g^{\rm true} \rangle \frac{1}{1 + \epsilon_s}
\end{equation}
where we have defined $\epsilon_s = \Bar{n}_s / \Bar{n}_g$ to be the stellar contamination fraction.
Similarly, on scales where stars are approximately unclustered (see below), 
\begin{equation}
    \langle g^{\rm obs} \ g^{\rm obs} \rangle = \langle g^{\rm true} \ g^{\rm true} \rangle \left (\frac{1}{1 + \epsilon_s} \right)^2
\end{equation}

From the argument above, we can see that the effect of stellar contamination is to lower the auto and cross correlations in a scale-independent way. Since $\langle \kappa \ g^{\rm true} \rangle \propto b_g$ and $\langle g^{\rm true} \ g^{\rm true} \rangle \propto b_g^2$, we conclude that unclustered stellar contamination is completely degenerate with a scale-independent galaxy bias and that our analysis actually measures the ``effective bias''
\begin{equation}
    b^{\rm eff} = b^{\rm true} \frac{1}{1 + \epsilon_s}
\end{equation}
so that marginalization over galaxy bias will automatically also marginalize over the (in general unknown) amount of stellar contamination. We further note that the ratio 
\begin{equation}
    \frac{\left(C_\ell^{\kappa g}\right)^2}{C_\ell^{g g}} \sim \frac{(b_g^{\rm eff})^2 \sigma_8^4 }{(b_g^{\rm eff})^2 \sigma_8^2} \sim \sigma_8^2
\end{equation}
is proportional to $\sigma_8^2$ in linear theory, and is therefore independent of $b^{\rm eff}$ on linear scales.

The power spectrum of galactic contaminants such as stars is typically very large on large scales, falling off steeply with increasing $\ell$ (faster than the typical galaxy power spectrum). For example, we have checked that if the stellar contamination in unWISE traces a Gaia stellar map\footnote{We also find a similar power spectrum for Gaia stars that meet our 
blue or green WISE color selection.} with stellar contamination fraction $\sim 1\%$ (as expected from the cross-match to COSMOS), the stellar power in the lowest-$\ell$ bin used in the analysis is $< 0.5\%$ of the galaxy clustering power on the same scale, ensuring that the argument above holds.

\subsection{Foreground contamination to CMB lensing cross-correlations}
\label{subsec:foregrounds}

The Minimum Variance (MV) reconstruction we use in the fiducial analysis is dominated by CMB temperature (rather than polarization), and is therefore subject to possible contamination by both galactic and extragalactic foregrounds.  When these foregrounds are correlated with the galaxy sample of interest, they can lead to biases in the cross-correlation \cite{Schaan:2018tup,Madhavacheril:2018bxi,Ferraro:2017fac,vanEngelen:2013rla,Osborne:2013nna}.

Regarding galactic foregrounds, we expect the largest contaminant to be galactic dust, seen in emission in the CMB maps, and causing reddening and/or extinction on most galaxy catalogs. Imperfect foreground separation can impact the CMB lensing maps. While we expect the IR-selected unWISE sources not to be directly affected by galactic dust,
 nonetheless their local density can be dependent on (for example) stellar density, which is itself correlated with galactic dust.
The Planck team \cite{PlanckLens18} performed a large number of null tests regarding the reconstructed map, and find general stability of the baseline reconstruction on the \texttt{SMICA} component separated temperature map.  Importantly, the reconstruction is stable with respect to choice of galactic mask, with variations consistent with those expected from the change in area. Most of the null test tensions come from the reconstruction on the 217 GHz frequency map, which may contain non-negligible galactic contamination. We caution that the tSZ-free map has larger weight given to the 217 GHz channel and may therefore have a larger dust contamination.

Regarding extragalactic foregrounds, the effect on the lensing auto-power spectrum has been thoroughly investigated in Section 4.5 of ref.~\cite{PlanckLens18}. The Planck team has found that at lensing $\ell < 1000$, both tSZ and CIB biases are expected to be a small fraction of 1\%, significantly below the statistical errors. The effect of kSZ biases for the Planck SMICA map has been calculated in ref.~\cite{Ferraro:2017fac} and shown to be negligible. Calculating the bias to the cross-correlation with galaxies is more difficult, since it depends on the particular sample, its redshift distribution, HOD and IR luminosity function.  Using realistic correlated CMB and large-scale structure simulations, refs.~\cite{Schaan:2018tup, Ferraro:2017fac} have found that for a galaxy sample with median redshift $\approx 0.8$ and $b(z) \approx 1 + 0.84z$, the size of the biases in auto and cross-correlations are very comparable. This sample is rather similar in redshift distribution and bias (and hence mass) to the WISE blue and green samples, and therefore we expect that the biases in cross be the same order of magnitude of the ones in the CMB lensing auto-spectrum, and hence safely sub-percent. While this argument only provides a rough estimate, it appears very likely that any extragalactic source of bias will be well below the statistical significance of our cross-correlations.



As a further test, we repeat the cross-correlation with the Planck 2018 lensing reconstructed from tSZ-deprojected temperature maps, shown in Table~\ref{tab:halofit_xcorr_tsz}. Apart from removing the possible tSZ contamination, the CIB contribution will be significantly different due to the different weighting of the single-frequency channels. While the absence of tSZ bias could in principle be partly compensated by a larger CIB-induced bias, the consistency between the fiducial and tSZ-free cross-correlations provides further confidence that foreground contamination is subdominant to our other sources of uncertainty. 

\begin{table}[h!]
    \centering
    \small
    \begin{tabular}{ccc||cc}
	WISE & \multirow{2}{*}{$b^{\rm eff}_{\rm cross}$}   & \multirow{2}{*}{$\chi^2_{\rm cross}$/dof}  & \multirow{2}{*}{$b^{\rm eff}_{\rm cross, tSZ-free}$} &
	\multirow{2}{*}{$\chi^2_{\rm cross, tSZ-free}$/dof} \\
	sample & & & & \\
	        \hline
	       Blue & $1.56 \pm 0.0276$ & 6.11/5 & $1.54 \pm 0.0305$ &  9.34/5 \\ 
	       Green & $2.23 \pm 0.0352$  & 2.93/5 & $2.19 \pm 0.0389$ &  3.87/5 \\ 
	       Red 16.2 & $3.29 \pm 0.090$  & 4.56/5 & $3.25 \pm 0.102$ &  6.03/5 \\ 
    \end{tabular}
    \caption{Comparison between fiducial $b^{\rm eff}_{\rm cross}$ (reproducing Table~\ref{tab:halofit_xcorr})
    and $b^{\rm eff}_{\rm cross}$ for the tSZ free sample.
    }
    \label{tab:halofit_xcorr_tsz}
\end{table}

\subsection{Galactic mask dependence of the sample properties}
\label{subsec:gal_mask}

If the redshift distribution varies across the sky, the clustering $dN/dz$ measured in the SDSS region
could be unrepresentative of the true $dN/dz$ across the entire WISE mask.
We test this possibility by restricting the $C_{\ell}^{gg}$ and $C_{\ell}^{\kappa g}$
measurement to the SDSS footprint used to measure $dN/dz$ and repeating our measurements.  We find good agreement between the biases measured in the SDSS region and the biases measured across the full sky (Table~\ref{tab:halofit_xcorr_sdss}).  We also find that the galaxy-galaxy cross-spectra (i.e. Fig.~\ref{fig:red_check}) are quite similar in the SDSS region as in the full unWISE footprint, changing by $<10\%$.

\begin{table}[h!]
    \centering
    \small
    \begin{tabular}{cccc||cc}
	WISE & \multirow{2}{*}{$b^{\rm eff}_{\rm auto}$}  & \multirow{2}{*}{$10^7 \times$ Shot Noise} & \multirow{2}{*}{$\chi^2_{\rm auto}$/dof}  & \multirow{2}{*}{$b^{\rm eff}_{\rm cross}$} &
	\multirow{2}{*}{$\chi^2_{\rm cross}$/dof} \\
	sample & & & & & \\
	        \hline
	       Blue & $1.76 \pm 0.0117$ & $0.87 \pm 0.027$  & 10.3/4 & $1.55 \pm 0.0632$ &  12.7/5 \\ 
	       Green & $2.42 \pm 0.0188$ & $1.78 \pm 0.027$  & 10.1/4 & $2.19 \pm 0.0797$ &  11.8/5 \\ 
	       Red 16.2 & $3.60 \pm 0.0845$ & $28.9 \pm 0.21$  & 7.27/4 & $3.27 \pm 0.206$  &  5.73/5 \\ 
    \end{tabular}
    \caption{Results from fitting a constant bias times HaloFit power spectrum (using cross-correlation $dN/dz$), restricting measurements
    to the CMASS area. 
    }
    \label{tab:halofit_xcorr_sdss}
\end{table}

\begin{figure}
    \centering
    \resizebox{\columnwidth}{!}{\includegraphics{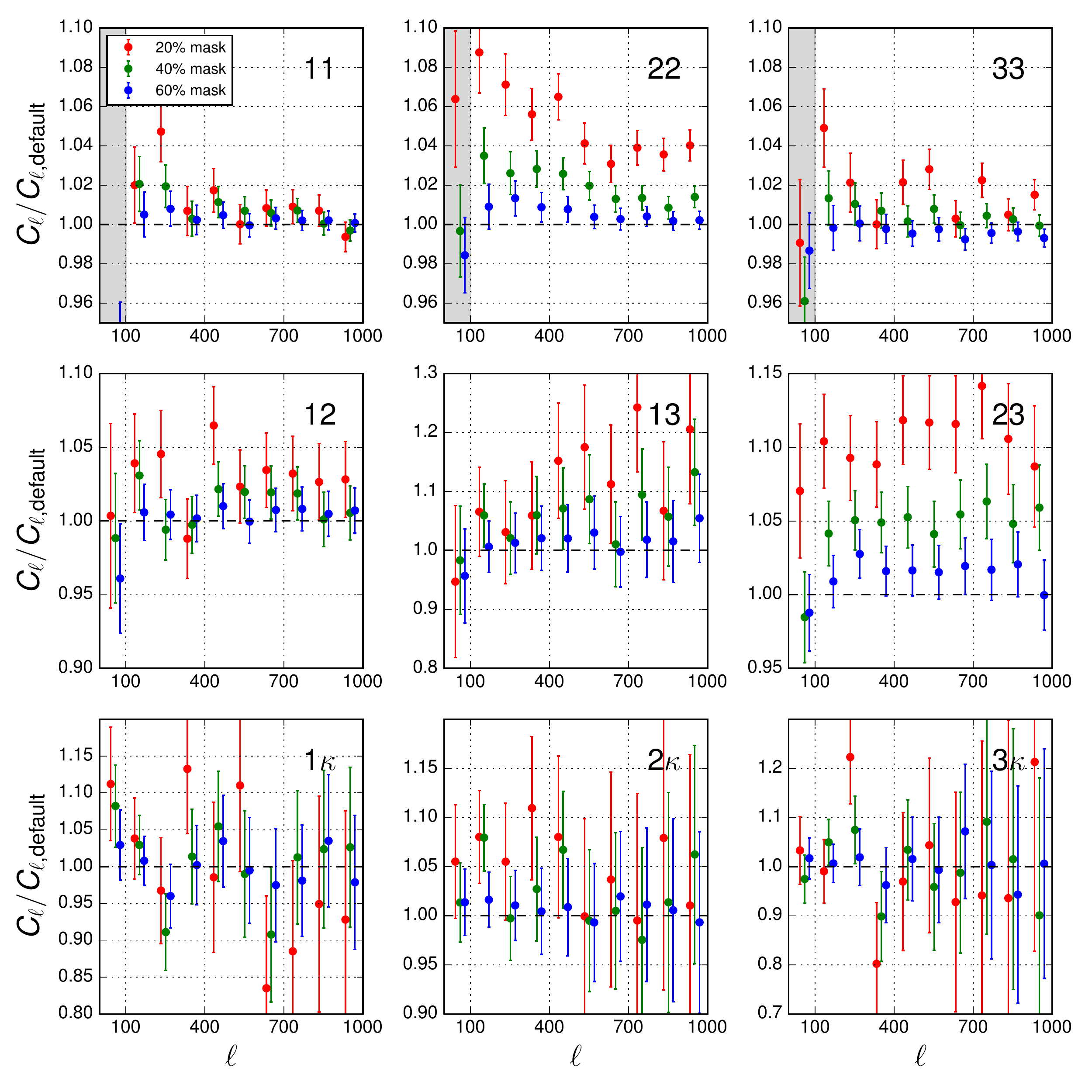}}
    \caption{Change in clustering when masking is changed from the default Planck and WISE masks
    to the Planck 20\%, 40\% and 60\% Galactic masks. Top row shows galaxy auto-spectra, middle
    row shows galaxy-galaxy cross-spectra, and bottom row gives galaxy-CMB cross spectra.  
    Gray regions indicate scales excluded ($\ell < 100$)
    because the power depends too strongly on the Galactic mask.}
    \label{fig:gal_masking}
\end{figure}

\begin{figure}
    \centering
    \resizebox{\columnwidth}{!}{\includegraphics{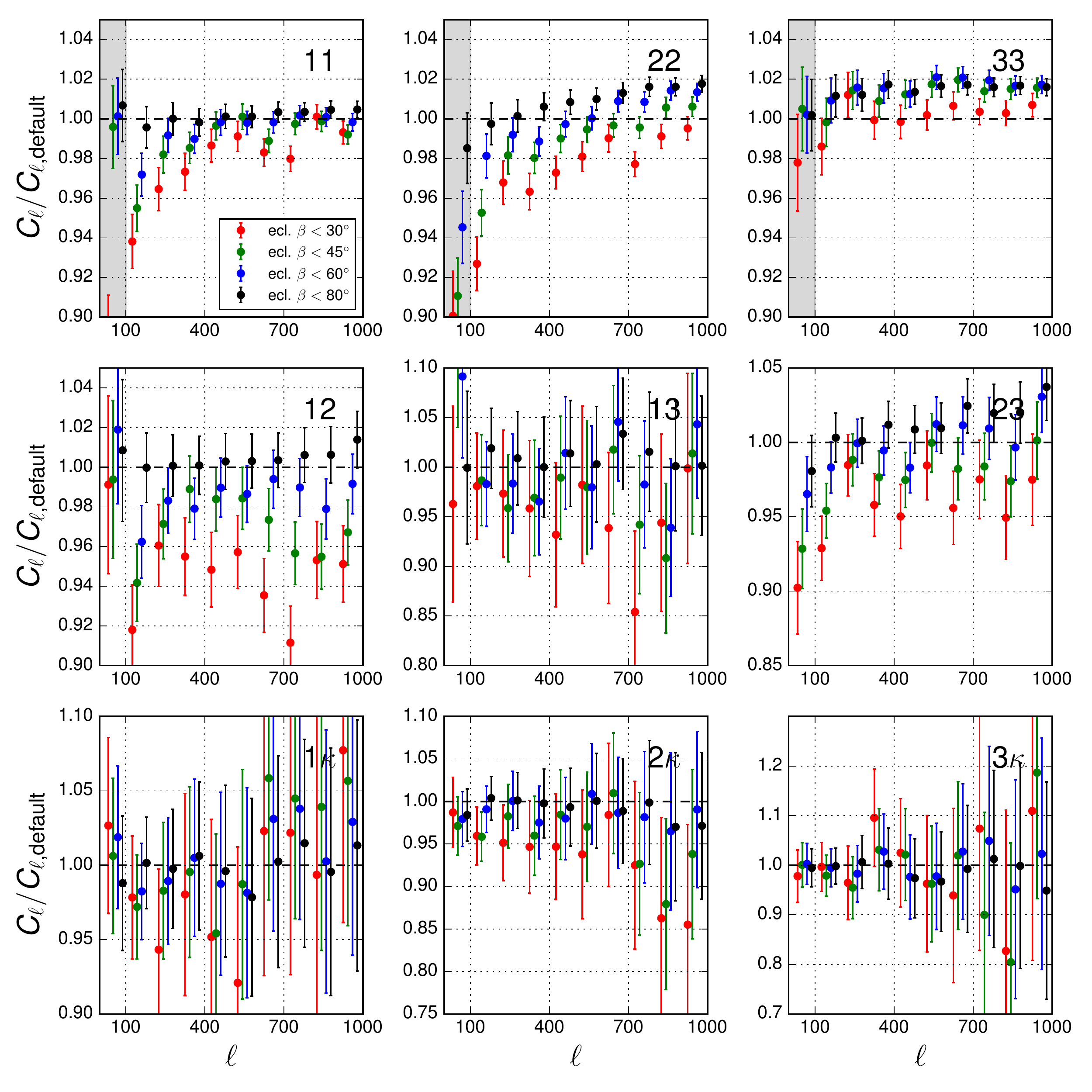}}
    \caption{Change in clustering when masking is changed from the default Planck and WISE masks to cover more
    of the ecliptic.
    We sequentially exclude the sky
    at ecliptic latitude $\beta < 30^{\circ}$, $40^{\circ}$,
    $60^{\circ}$, and $80^{\circ}$
    (i.e.\ black points include the smallest fraction
    of the sky, near the
    Ecliptic poles).  Since
    the WISE depth of coverage
    is highest at the ecliptic
    poles, the black points are
    the deepest and the
    red points are the shallowest.
    Top row shows galaxy auto-spectra, middle
    row shows galaxy-galaxy cross-spectra, and bottom row gives galaxy-CMB cross spectra.
    Gray regions indicate scales excluded ($\ell < 100$)
    because the power depends too strongly on the Galactic mask.}
    \label{fig:ecl_masking}
\end{figure}

We further test the impact of restricting our sample to higher Galactic latitudes by sequentially applying
the Planck 60\%, 40\% and 20\% Galactic masks (retaining the ``cleanest'' 60, 40 and 20\% of the extragalactic sky)\footnote{Available at \url{https://irsa.ipac.caltech.edu/data/Planck/release_2/ancillary-data/masks/HFI_Mask_GalPlane-apo0_2048_R2.00.fits}} in addition to the standard WISE masks described in Section~\ref{sec:masking}.  We find no significant change in $C_{\ell}^{\kappa g}$ as the Galactic
masking is changed.  In contrast,
we find changes of several percent
with differing Galactic masks in the $\ell < 100$
bin; consequently, we only fit the bias to $\ell > 100$.
At $\ell > 100$, we find a mild scale-independent trend in the amplitude of $C_{\ell}^{gg}$ with Galactic latitude, which may be caused by changes in the galaxy population due to changing selection function at higher Galactic latitudes (modifying both the bias and $dN/dz$). 
This should not affect our cosmological constraints as long as the area over which we measure $dN/dz$ and the auto and cross correlations are the same.  In practice, we prefer not to restrict
the $dN/dz$ measurement to the footprint of the spectroscopic samples ($f_{\rm sky} = 0.15$)
and find that measuring $b^{\rm eff}$ over the spectroscopic footprint leads to variations $< 1 \sigma$, suggesting $dN/dz$ varying on the sky is not a major systematic.

Changes in the galaxy-galaxy cross-spectra with Galactic latitude
suggest that $dN/dz$ does vary on the sky in addition to the bias. If only the bias were
changing as we changed the Galactic masks, the increase in the cross-spectrum would go as the geometric mean of the increase
in the individual galaxy auto-spectra, but the red-blue and red-green
cross-correlations increase by $\sim5-10\%$ more than the geometric mean,
implying a varying $dN/dz$\footnote{Using fainter red samples
(e.g.\ with a faint cut at W2 = 16.5 or 16.7) leads to even larger
variation in red cross blue compared to red and blue separately,
suggesting that $dN/dz$ variations are more severe for the fainter
red samples.} or more complex bias.

 We perform a similar test for masking the ecliptic plane below $\beta = 30^{\circ}$, $45^{\circ}$, $60^{\circ}$ and $80^{\circ}$
(Figure~\ref{fig:ecl_masking}). Similarly, we find only
mildly scale-dependent trends, with deviations of $\sim$5\% in the auto-spectrum at $\ell > 100$.

We also find that doubling the WISE stellar masking radius changes $b^{\rm eff}$ by less than 1 $\sigma$.
 Similarly, applying the more conservative masks of ref.~\cite{Kitanidis19} (both the QSO and ELG masks,
equation 20, which mask around considerably
fainter stars than our mask)
changes $b^{\rm eff}$ by less than 0.5 $\sigma$.

\subsection{Systematic uncertainties in the redshift distribution}
\label{sec:systematics_dndz}

Due to the $6''$ WISE PSF, blending is a source of systematic error in the cross-match
redshifts, as it could lead to spurious matches with COSMOS.  Many of our sources are blended, and it is possible that the low-redshift tails
in the red and green samples result from source blending rather than from the presence of low-redshift sources.  However, because
we only use the cross-match redshifts
in the magnification bias term,
we expect blending to have a negligible impact on $b^{\rm eff}_{\rm auto}$ and $b^{\rm eff}_{\rm cross}$. If we replace all unWISE
magnitudes in the cross-match with Spitzer
3.6 and 4.5 $\mu$m magnitudes
(reducing the possibility of source confusion
due to the 2'' resolution of Spitzer)
and replace the cross-match $dN/dz$
with the Spitzer $dN/dz$, we find shifts of $< 0.5\sigma$ in the fitted biases.

Systematic errors in the COSMOS
$dN/dz$ may also impact our results,
and are an important
systematic in
 cosmic shear \citep{Wright19,Hildebrandt20}.
Again, because we only use the COSMOS $dN/dz$
in the magnification term, the impact
of this systematic shift on our results is limited.
Because our redshift bins are broad,
the impact of scatter in the photometric
redshifts should be minimal, as the scatter is
$\Delta z < 0.1$ for the three samples,
compared to the $\Delta z \sim 0.5$
redshift bins.
Catastrophic errors may have a more
significant impact; from Figures 11 and 12 in
ref.~\cite{Laigle16}, we conservatively estimate
10\% catastrophic errors for the higher-redshift
$i \lesssim 24$ red and green samples, and 1\% catastrophic errors for the $i \lesssim 22$
blue samples.
The maximum impact of these errors
would be to create a population of galaxies
with $\Delta z \sim 1$ redshift errors. We therefore
create two $dN/dz$ for the green and red samples where 10\% of the galaxies
are scattered uniformly into a $0 < z < 1$
tail or a $2 < z < 3$ tail. We find that using
these $dN/dz$ instead of the fiducial  $dN/dz$ makes $\lesssim 0.5\sigma$ difference in our results.

We study the impact of a discrepancy
between $b_{\rm sml,p}$ and $b_{\rm lin,p}$
using $b_{\rm sml}$ and $b_{\rm lin}$
from the autocorrelation of WISE-like
samples in an $N$-body simulation (Appendix~\ref{sec:xcorr_redshift_details}).
We parameterize $b_{\rm sml}/b_{\rm lin} = 1 + Az^2$
(smooth curves in Figure~\ref{fig:bsml_blin_ratio}),
allowing $A$ to vary between zero and twice its fiducial value, $A_{\rm fid}$.
We pick $1+Az^2$ because it has roughly the right functional
form. Since the HODs are approximate anyway, quantitative
agreement with the $N$-body results is not required. Indeed,
the $1+Az^2$ fitting function is somewhat conservative,
as it predicts a slightly larger increase in $b_{\rm sml}/b_{\rm lin}$
than indicated by the $N$-body simulations (compare the $N$-body simulations
to the fitting function at $z\sim0.4$).
We show the impact of using both the fiducial
value, $A_{\rm fid}$ = 0.025, 0.025, 0.05 for blue, green and red, and the maximal value, $A_{\rm max}$ = 0.1, 0.15, 0.2 for blue, green and red, for the most extreme bias evolution
allowed by the data.


\begin{figure}
    \centering
    \resizebox{\columnwidth}{!}{\includegraphics{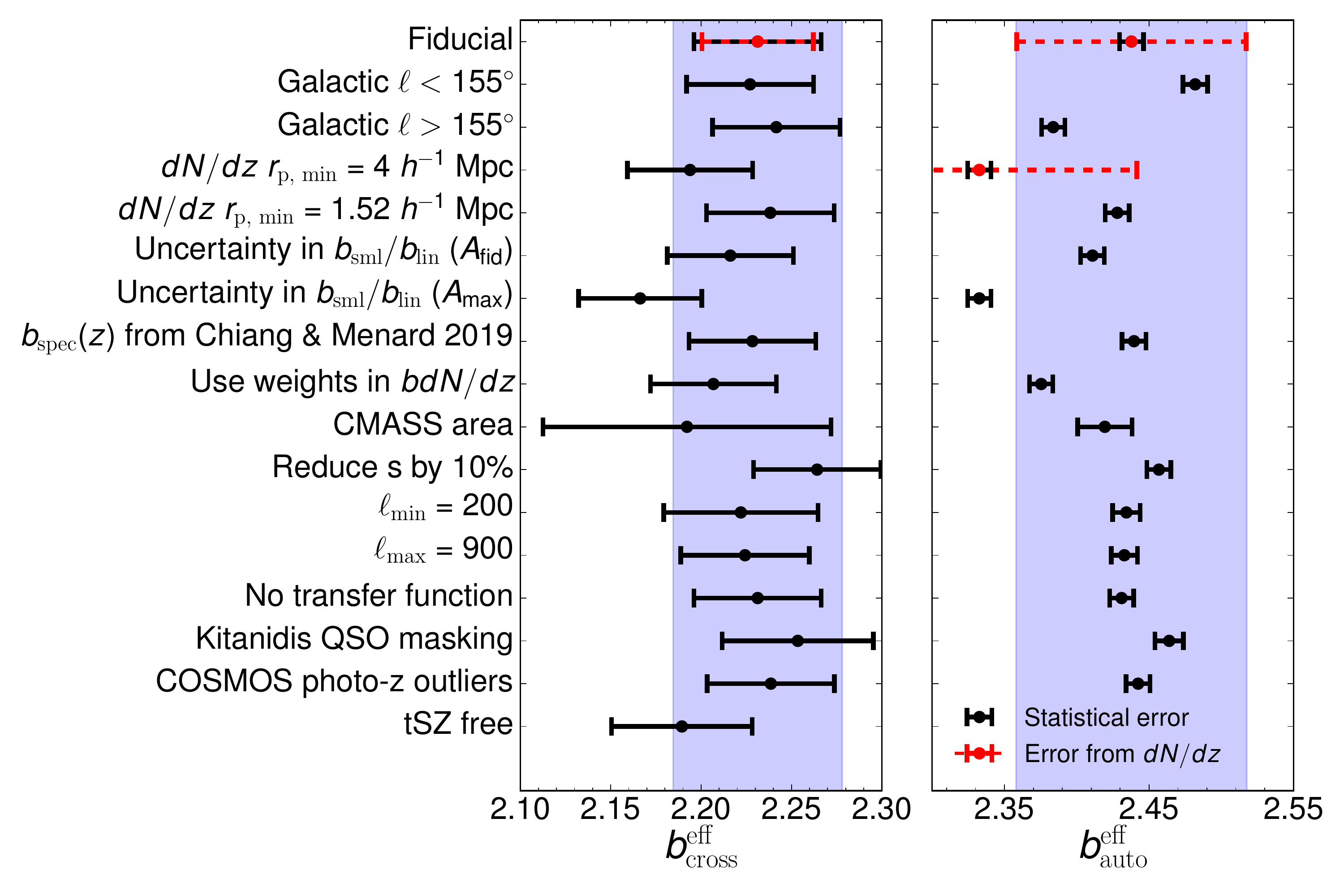}}
    \caption{Impact of systematic
    errors on $b^{\rm eff}$ for the green sample.
    Black errorbars give the statistical uncertainty, red dashed errorbars give the systematic uncertainty
    from errors in $dN/dz$ (only plotted for the fiducial values), and the blue band displays their quadrature sum.  We also plot the uncertainty
    from $dN/dz$ error for the $r_{\rm p,min} = 4$ $h^{-1}$ Mpc $dN/dz$, to emphasize that this point
    is only $1.3\sigma$ discrepant if we define $\sigma$ using the $dN/dz$ errors.
    Top row gives fiducial value matching Table~\ref{tab:halofit_xcorr}.
    The next two rows show $b^{\rm eff}$
    for the split-sky sample, giving
    an estimate of uncertainty due to uncertain $dN/dz$.  The next four rows
    are concerned with nonlinear bias evolution; either by increasing/decreasing $r_{\rm p,min}$ to be more/less robust against nonlinear bias;
    or using the parameterized
    function from
    Appendix~\ref{sec:xcorr_redshift_details},
    $b_{\rm sml,p}/b_{\rm lin,p} = (1+Az^2)$
    with $A_{\rm fid}$ ($A_{\rm max}$) = 0.025 (0.1), 0.025, (0.15), 0.05 (0.2)
    for blue, green and red.
    The next row shows the impact of changing
    the CMASS and LOWZ spectroscopic bias evolution used in the clustering redshifts from the measured values of Table~\ref{tab:spec_bias} to values from Fig. 12 of Ref.~\cite{ChiangMenard18}.
    The next rows show the impact of using
    weights
    in the cross-correlation redshifts,
    restricting to the CMASS footprint (Table~\ref{tab:halofit_xcorr_sdss}),
    reducing magnification bias response $s$ by 10\%,
    changing scale cuts for the auto- and CMB-cross power spectra,
    using the tSZ-free lensing map,
    using stricter stellar masking
    from ref.~\cite{Kitanidis19},
    and the impact of COSMOS photo-$z$
    catastrophic errors adding a spurious low-$z$ tail
    to $dN/dz$.
    \label{fig:systematics}}
\end{figure}

We summarize the impact of different systematics
on $b^{\rm eff}_{\rm auto}$
and $b^{\rm eff}_{\rm cross}$ for the green sample in Figure~\ref{fig:systematics}.
The analogous plots for the blue and red samples
look similar; in fact, green is the only sample
to have $> 1\sigma$ discrepancies from the fiducial value (measured using the quadrature sum of the $dN/dz$ error and the statistical error),
with a $1.3\sigma$ discrepancy when using $A_{\rm max}$ to correct for nonlinear bias evolution and a 1.3$\sigma$ discrepancy
when the cross-match $dN/dz$ uses $r_{\rm p,min} = 4$ $h^{-1}$ Mpc
(where $r_{\rm p,min}$ indicates the minimum separation
    bin used in the clustering redshifts, fiducially 2.5 $h^{-1}$ Mpc).  For the $r_{\rm p,min} = 4$ $h^{-1}$ Mpc $dN/dz$, we use the error for the $r_{\rm p,min} = 4$ $h^{-1}$ Mpc $dN/dz$ for $\sigma$ rather
than the quadrature sum of this error and the error from the fiducial $r_{\rm p,min} = 2.5$ $h^{-1}$ Mpc $dN/dz$,
because the two $dN/dz$ errors are highly correlated.

\section{Conclusions and lessons learned}
\label{sec:conclusions}

We have presented a tomographic measurement of the cross-correlation of the unWISE galaxies and CMB lensing. We report a combined detection significance of 55.1, which is the highest-significance detection of lensing by large-scale structure to date.

One of the greatest challenges was the characterization of the redshift distribution for the three samples. Since for most galaxies only the W1 and W2 magnitudes were available, we did not attempt to assign individual photometric redshifts, but just split the full catalog into three samples with different mean redshifts, but with non-negligible overlap between them. We use two techniques to measure the redshift distribution.

First, we cross-match our objects with the COSMOS catalog, obtaining a direct measurement of the redshift distribution $dN/dz$. A direct cross-match is insensitive to modeling assumptions and measures $dN/dz$, required to calculate the magnification bias contribution. If used to predict clustering, assumptions on the redshift evolution of the bias evolution are necessary. One disadvantage is the high completeness required of the survey, which limits the area available.  
Another disadvantage is source blending,
which could lead to spurious cross-matches
and thus modify $dN/dz$.
The small overlap area (2 square degrees) not only limits the measurement statistically, but given the inhomogeneous depth of the WISE survey and possible spatial dependence of the selection function, the results may not be representative of the full WISE footprint. While we take steps to ensure that our catalogs are magnitude limited over the whole footprint by applying appropriate magnitude cuts, residual effects such as blending and background subtraction can potentially lead to inhomogeneity in the selection function. The mild trends in bias with respect to Galactic mask observed in Figure \ref{fig:gal_masking} may be an indication of this.

Second, we cross-correlate the unWISE samples with a number of overlapping spectroscopic samples, thus determining the product of the bias and redshift distribution. This can be advantageous when calculating the clustering signal, since it is this product that enters the auto-correlation and the cross-correlation with CMB lensing. Another advantage is the typically large overlap area (important for when the selection function is inhomogeneous), and the fact that there are no completeness requirements on the spectroscopic sample.  However, assumptions on the redshift evolution of the bias are necessary when calculating the magnification bias contribution, and the impact of magnification needs to be taken into account in the spectroscopic-photometric cross-correlation. Moreover, this measurement is subject to the usual modeling challenges such as non-linearities in clustering and bias.

As discussed in Appendix \ref{sec:xcorr_redshift_details}, the two measurements of $dN/dz$ are consistent with each other when assuming a simple model for bias evolution.
Further, the consistency of the unWISE bias measured on the CMASS  overlap region (Table \ref{tab:halofit_xcorr_sdss}) compared to the whole unWISE footprint (Table \ref{tab:halofit_xcorr}) indicates that the cross-correlation redshifts should be unaffected by spatial variations in the selection function.
In the fiducial analysis, we use the cross-correlation result to predict the clustering and the cross-matched distribution to predict magnification bias and therefore we don't need to assume a redshift evolution of the bias.

Once the redshift distribution is known (or the uncertainties appropriately marginalized over), theoretical modeling of the signal on intermediate to small scales is the next challenge. Non-linear corrections to both clustering and bias become important at $\ell$ of few hundred, where the statistical $S/N$ is still large in each bandpower. This implies that even if $dN/dz$ were known perfectly, our ability to extract cosmological information could still be limited by our theoretical models.  We defer consideration of modeling the signal to future work \cite{paper2}.

In conclusion, we believe that the cross-correlations presented here are an extremely sensitive probe of late-time cosmology. A spectroscopic followup of a subsample of the sources as well as improved modeling of intermediate and small scales can lead to sub-percent measurement, with important possible applications for tests of gravity,  measurement of neutrino masses and the properties of Dark Energy.

\section*{Acknowledgments}

We thank David Alonso, Enea Di Dio, Yu Feng, Julien Guy, Colin Hill, Ellie Kitanidis, Alexie Leauthaud, Thibaut Louis, Emmanuel Schaan, David Schlegel, Uros Seljak, Blake Sherwin, Zachary Slepian, David Spergel, Katherine Suess and Michael Wilson for very useful discussions.  S.F. is supported by the Physics Division at Lawrence Berkeley National Laboratory.
M.W.~is supported by the U.S.~Department of Energy and by NSF grant number 1713791.
This research used resources of the National Energy Research Scientific Computing Center (NERSC), a U.S. Department of Energy Office of Science User Facility operated under Contract No. DE-AC02-05CH11231.
This work made extensive use of the NASA Astrophysics Data System and of the {\tt astro-ph} preprint archive at {\tt arXiv.org}.

\newpage 

\appendix

\section{Optical properties of unWISE samples and prospects for spectroscopic followup}

In this section, we describe the optical
properties of the unWISE galaxies from archival
photometric and spectroscopic data.
We also discuss the prospects
and requirements for spectroscopic followup
of the unWISE samples to better determine $dN/dz$.

\begin{figure}
    \centering
    \resizebox{\textwidth}{!}{ \includegraphics{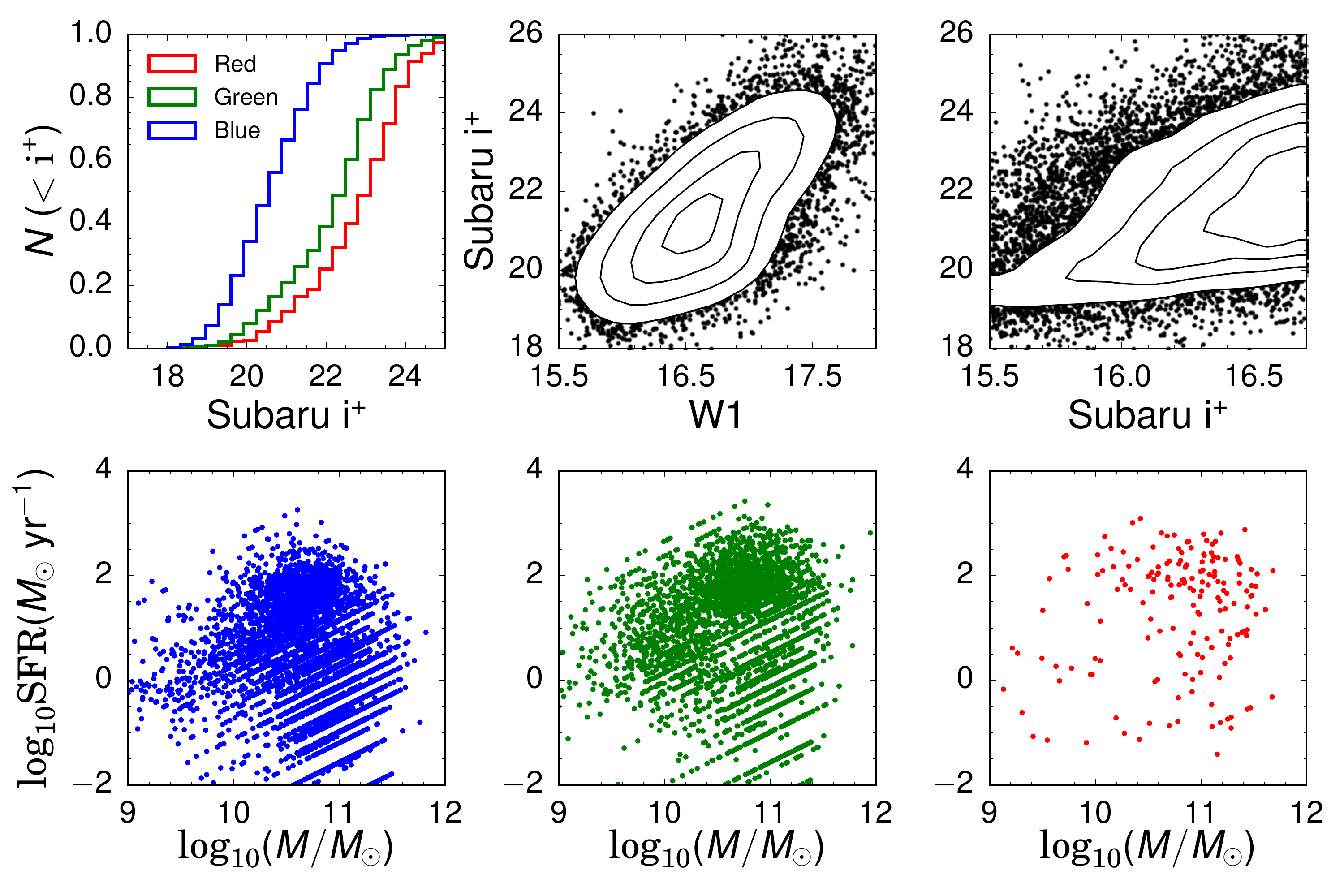}} 
    \caption{
    \textit{Top left:} Distribution of Subaru $i^{+}$ magnitudes (``mag\_auto,'' measured in flexible elliptical apertures,
    as in ref.~\cite{Kron80}) from unWISE matched to COSMOS.
    \textit{Top center and right:} Subaru $i^{+}$
    versus WISE W1 and W2 magnitudes for the combined red, green and blue samples in COSMOS.
    We show 20\%, 40\%, 60\% and 80\% iso-contours of the cumulative distribution function (Gaussian smoothed with $\sigma = 1$ mag) of galaxies
    in $i^{+}$ and W1/W2.
    \textit{Bottom:} Distribution of stellar 
    mass and star formation rate
    for each of the three samples,
    from COSMOS broad-band photometry.
    }
    \label{fig:wise_colors}
\end{figure}

While the full unWISE sample only has infrared fluxes, by cross-matching to COSMOS we can determine
the optical colors and properties of unWISE. In Figure~\ref{fig:wise_colors}, we show the distribution
of Subaru $i^{+}$ for the unWISE galaxies, and the relationship between $i^{+}$ and the WISE bands.
For the blue sample a 90\% completeness is achieved at $i^{+}\simeq 22$ while for the green and red samples 90\% completeness occurs at $i^{+}\simeq 24$.

We also show the stellar mass and star formation rates of unWISE galaxies from COSMOS broad-band photometry. All three unWISE samples have similar stellar masses (with $\log_{10} (M/M_{\odot}) = 10.80$, 10.78, and 10.87 for the blue, green and red samples) but the star-formation rates of the green and red samples ($\log_{10} \textrm{SFR} / M_{\odot} \textrm{yr}^{-1}$ = 1.03, 1.61) are significantly higher than the star-formation rate
of the blue sample ($\log_{10} \textrm{SFR} / M_{\odot} \textrm{yr}^{-1}$ = 0.12).

Some of the galaxies in the unWISE samples have been observed in the
VVDS survey \citep{LeFevre13} using VIMOS on VLT, allowing us to both better characterize
the galaxy samples and understand the feasibility of spectroscopic
followup.
The VVDS-Deep survey has a simple selection function, uniformly targeting galaxies
at $17.5 < I < 24$.  Not every $17.5 < I < 24$ galaxy
is targeted; to determine the completeness of the unWISE galaxies in VVDS,
we must divide the number of matches by the VVDS targeting selection rate
(typically 20-30\%) and then compare to the number of unWISE galaxies
lying within the VVDS spectroscopic mask.\footnote{As with the COSMOS matches, we additionally
remove VVDS galaxies with Spitzer-SWIRE \citep{Lonsdale03} 4.5 $\mu$m magnitude $> 19.2$ (18.7 for red sample), although relatively few VVDS galaxies have SWIRE matches so this cut makes little difference.}  We find 444, 261 and 13
VVDS matches to the blue, green and red samples (418, 191 and 10 with high confidence redshifts, $\textsc{ZFLAGS} >= 2$), implying 101.3\%,\footnote{Completeness higher than 100\% likely indicates that the targeting selection rate is somewhat underestimated.} 89.6\% and 70.0\% of the blue, green and red samples yield a VVDS spectrum.
This agrees well with the fraction of galaxies with I $< 24$, which is 99.6\% (91.8\%, 84.5\%) for the three samples, implying that within the range of galaxies
that could have been targeted, 97.8\%, 69.0\% and 59.4\% of blue, green and red galaxies received
a high-confidence VVDS redshift.  VVDS-Deep exposure times are 4.5 hr on a $R \sim 230$
spectrograph, suggesting that spectroscopic followup of the unWISE samples is feasible
on 8-10 m class telescopes (and perhaps smaller telescopes for the brighter blue sample).

\begin{figure}
    \centering
    \resizebox{\textwidth}{!}{ \includegraphics{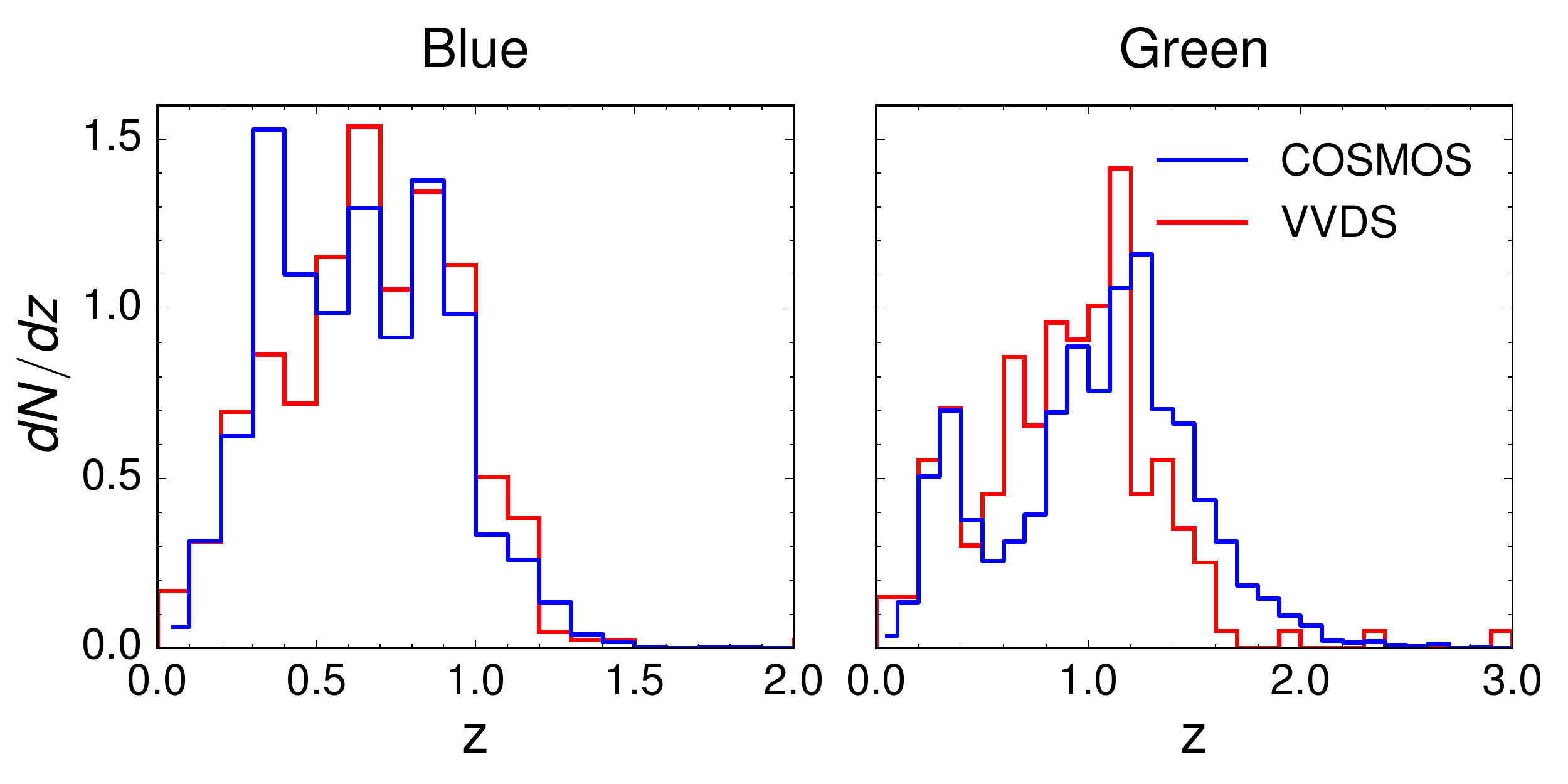}} 
    \caption{
   Comparison of $dN/dz$ for COSMOS photometric (blue) and VVDS spectroscopic (red) matches for the blue and green samples (including spectra from VVDS-Deep and VVDS-UltraDeep).  The red sample has too few VVDS matches to compare the two $dN/dz$.  While VVDS is largely complete for the blue sample, incompleteness
   at the faint, high redshift end of the green sample may bias $dN/dz$ relative to COSMOS.
    }
    \label{fig:cosmos_vvds_zdist}
\end{figure}

We also incorporate galaxies from the VVDS UltraDeep survey, which includes
12, 19, and 1 galaxies from the blue, green and red samples (10, 11, and 1 with $\textsc{ZFLAGS} >= 2$).
The redshift distribution of blue galaxies with redshifts from VVDS
is quite similar to the COSMOS $dN/dz$ (Figure~\ref{fig:cosmos_vvds_zdist}); for the green sample,
the VVDS $dN/dz$ is suppressed relative to COSMOS at $z > 1.6$, possibly
because of increased redshift failures at high redshifts where the [OII]
line redshifts beyond the red end of the spectrograph.

\begin{figure}
    \centering
    \resizebox{\textwidth}{!}{ \includegraphics{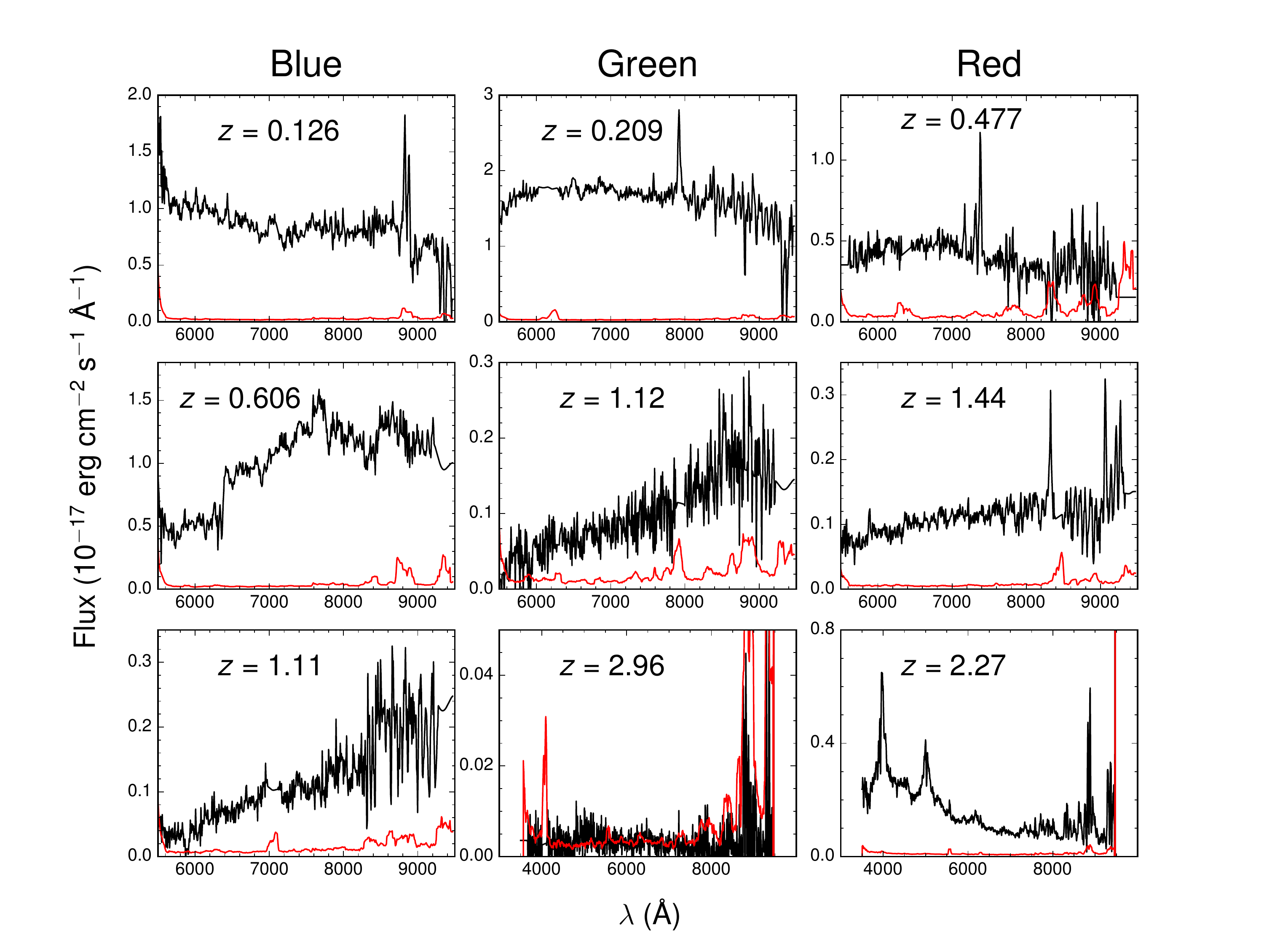}} 
    \caption{
    VVDS spectra from each of the three unWISE samples, with spectra in black and noise in red.  For each of the three
    samples we display galaxies representative of the low, medium, and high portion of the redshift distribution.  The highest redshift green and red galaxies are from VVDS-UltraDeep, which took spectra
    in both the blue and red grisms of
    VIMOS; all other galaxies are from VVDS-Deep, which only observed using
    the red grism.}
    \label{fig:vvds_spectra}
\end{figure}

We display three spectra from each sample in Figure~\ref{fig:vvds_spectra},
representing each sample at low, medium and high redshift.
Blending presents a similar challenge for the VVDS cross-match as for COSMOS,
so we only display galaxies that are well-isolated in optical imaging.
Due to the paucity of red spectra, the red sources at $z = 0.477$
and $z = 2.27$ are blends, although in both cases only one of the two potential
optical matches has a VVDS spectrum.

\begin{figure}
    \centering
    \resizebox{\textwidth}{!}{ \includegraphics{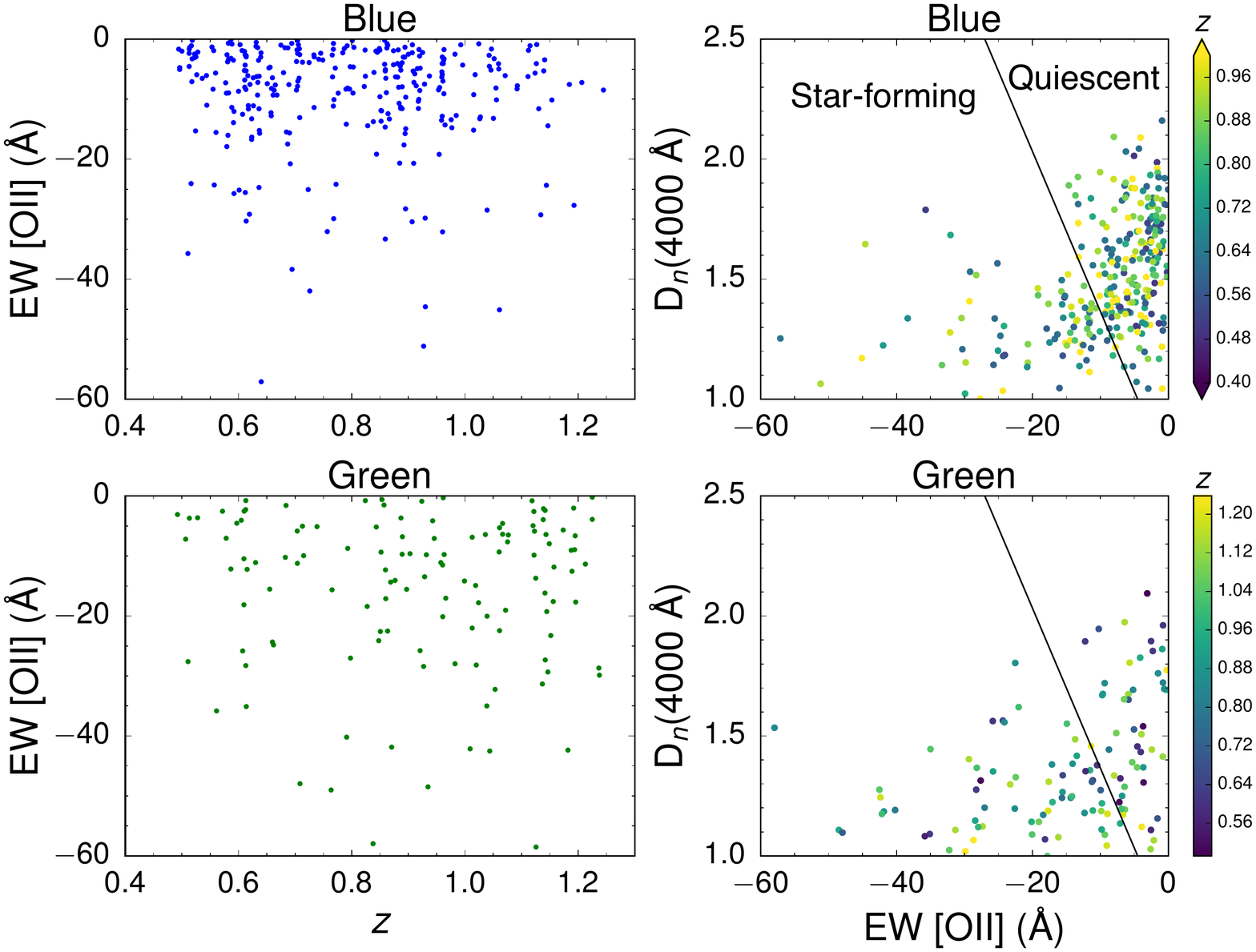}} 
    \caption{
   \textit{Left:} From VVDS spectra, distribution of rest frame [OII] 3727 \AA\ EW versus redshift for the blue sample (top)
   and green sample (bottom). \textit{Right:}
   Distribution of [OII] 3727 \AA\ EW
   versus $D_n(4000 \AA )$ to separate
   star forming and quiescent galaxies
   as in ref.~\cite{Franzetti07}.
   Points are color-coded to represent redshift.
    }
    \label{fig:vvds_summary}
\end{figure}

In Figure~\ref{fig:vvds_summary},
we plot the distribution of redshift
and rest frame [OII] 3727 \AA\ EW for the green
and blue samples
from VVDS spectra, as well
as $D_n(4000 \AA )$ versus [OII] 3727 \AA\ EW,
to separate star-forming from quiescent galaxies as in ref.~\cite{Franzetti07}.
We find median rest frame [OII] 3727 \AA\ EW of 6.4 (11.8) \AA\ in emission for the blue (green) sample,
and from the star-forming versus quiescent
cut from ref.~\cite{Franzetti07},
29.0\% (60.3\%) of blue (green)
galaxies are star-forming.

A direct measurement of $dN/dz$ with smaller errors than the COSMOS cross-match
$dN/dz$ would allow for improved modeling of the unWISE samples and better control
of systematic errors.
Even with improved cross-match $dN/dz$, we would still require the photometric-spectroscopic
cross-correlations to determine $b_{\rm lin}(z)$, but we could greatly improve the simple
HOD modeling in Appendix~\ref{sec:xcorr_redshift_details} and Fig.~\ref{fig:implied_bias},
allowing for better understanding of $b_{\rm lin}(z)$ and potentially better control
of systematics such as nonlinear bias evolution.
With observations in several fields, we could also better understand
the variation in $dN/dz$ on the sky.
Finally, we could better understand the impact of blending in our sample by re-targeting
both (or all) galaxies blended together by the $6''$ WISE PSF.

If the errors on $dN/dz$ were much smaller than the errors on the photometric-spectroscopic
clustering measurement, we could neglect $dN/dz$ errors and better model the unWISE
galaxy population.  This is not the case in Fig.~\ref{fig:implied_bias}; at $z < 0.2$
 and $z \sim 0.5$ in the blue sample, the error from uncertain $dN/dz$ (gray band)
 is larger than the statistical error on the clustering (blue errorbars).
However, this is driven by the HSC-derived cosmic variance correction,
which is a factor of 3.8 for the blue sample (Section~\ref{sec:xmatch_dndz}). If instead of measuring $dN/dz$
on a single field, we measured $dN/dz$ in multiple fields spread across the sky,
the errors would be dominated by Poisson rather than cosmic variance,
and no such correction would be necessary.
Indeed, if we divide out this correction in the $dN/dz$ errors in Figure~\ref{fig:implied_bias},
we find that $dN/dz$ errors are at most 80\% of the statistical errors for blue and green (peaking
at $z \sim 0.3-0.5$); for red, the $dN/dz$ errors are larger at low redshift, 150\% of the statistical
errors at $z \sim 0.5$.  Scaling from the number of galaxies with secure COSMOS redshifts
(5557, 3024, and 164 for blue, green and red), we estimate that achieving $dN/dz$
errors that are at most 50\% of the statistical errors will require 14000, 7500, and 1500 spectra
for the blue, green and red samples.  However, a smaller effort focused solely at low redshift
could be just as effective for the red sample, since the low redshift tail is much more uncertain
than the higher redshift $dN/dz$.

By measuring $dN/dz$ across multiple fields,
a spectroscopic followup program could constrain
 variations in $dN/dz$ on the sky.
Using the standard deviation of the COSMOS
cross-match $dN/dz$, we estimate that we could
measure a 5\% shift in the mean $dN/dz$ at 3$\sigma$
with 1000 spectra per field for both the blue and green samples.
For the red sample, with 400 spectra per field we could
measure a 10\% shift at 3$\sigma$.

\section{Simple HOD model for unWISE samples}
\label{sec:xcorr_redshift_details}

\begin{table}[]
    \centering
        \begin{tabular}{cc | cc | cc || cc | cc | cc}
        & & \multicolumn{2}{c}{LOWZ} & \multicolumn{2}{c}{CMASS} & & & \multicolumn{2}{c}{eBOSS Q}  & \multicolumn{2}{c}{BOSS Q}  \\
        $z_{\rm min}$ & $z_{\rm max}$ & $b_{\rm sml,s}$ & $\sigma_{b}$ &  $b_{\rm sml,s}$ & $\sigma_b$ & $z_{\rm min}$ & $z_{\rm max}$ & $b_{\rm sml,s}$ & $\sigma_{b}$ &  $b_{\rm sml,s}$ & $\sigma_b$\\
        \hline
 0.00 & 0.05 & 1.34 & 0.0381 & & & 0.00 & 0.20\\
 0.05 & 0.10 & 1.37 & 0.0077 & & & 0.20 & 0.40\\
 0.10 & 0.15 & 1.52 & 0.0037 & 1.36 & 0.2698 & 0.40 & 0.60\\
 0.15 & 0.20 & 1.73 & 0.0045 & 2.82 & 0.2615 & 0.60 & 0.80\\
0.20 & 0.25 & 1.89 & 0.0039 & 1.54 & 0.0796 & 0.80 & 1.00 & 1.72 & 0.2803 \\
0.25 & 0.30 & 2.01 & 0.0018 & 2.11 & 0.1489 & 1.00 & 1.20 & 2.03 & 0.0851 \\
0.30 & 0.35 & 2.01 & 0.0021 & 1.99 & 0.1054 & 1.20 & 1.40 & 2.05 & 0.0759 \\
0.35 & 0.40 & 2.06 & 0.0019 & 2.24 & 0.1674 & 1.40 & 1.60 & 2.35 & 0.0683 \\
0.40 & 0.45 & 2.25 & 0.0017 & 2.05 & 0.0020 & 1.60 & 1.80 & 2.32 & 0.0992 \\
0.45 & 0.50 & 2.46 & 0.0079 & 2.08 & 0.0006 & 1.80 & 2.00 & 2.89 & 0.1004 \\
 0.50 & 0.55 & & & 2.06 & 0.0008 & 2.00 & 2.20 & 2.87 & 0.1548 \\
 0.55 & 0.60 & & & 2.17 & 0.0007 & 2.20 & 2.40 & & & 4.33 & 0.0808 \\
 0.60 & 0.65 & & & 2.22 & 0.0010 & 2.40 & 2.60 & & & 3.72 & 0.1394 \\
 0.65 & 0.70 & & & 2.39 & 0.0022 & 2.60 & 2.80 & & & 4.27 & 0.3772 \\
 0.70 & 0.75 & & & 2.52 & 0.0090 & 2.80 & 3.00 & & & 4.30 & 1.2109 \\
 0.75 & 0.80 & & & 2.73 & 0.0872 & 3.00 & 3.20 & & & 4.30 & 1.1164 \\
  &  & & &  &  & 3.20 & 3.40 & & & 5.33 & 1.9984 \\
\end{tabular}
\caption{Bias of the spectroscopic samples and the 1-$\sigma$ error bar, as defined in Equation~\ref{eqn:wtheta_for_spec_bias}.
\label{tab:spec_bias}}
\end{table}

When computing $dN/dz$ using cross-correlations we assumed a scale-independent bias, and we found that in order for the cross-match and cross-correlation $dN/dz$ to match, the biases needed to evolve relatively rapidly with redshift.  In this Appendix we check whether this assumption and its implications are consistent with expectations from simple models of the manner in which galaxies populate dark matter halos.

A scale-independent bias is likely to be true on large, linear scales, but the extent to which this approximation is valid on the scales used in the $dN/dz$ analysis is unclear.  If the bias is scale-dependent, the redshift evolution of $b_{\rm sml,p}$ may not match the redshift evolution of $b_{\rm lin,p}$ (relevant for $C_{\ell}^{\kappa g}$ and $C_{\ell}^{gg}$), potentially introducing a systematic bias.  
To investigate this issue we model the unWISE galaxies using a simple HOD applied to dark matter halos in $N$-body simulations, allowing us to study the scale and redshift dependence of the unWISE galaxy bias.  Since our goal is modest, we simply use a 1-parameter family\footnote{There is some evidence that HOD parameters scale approximately universally with number density, e.g.~ref.~\cite{Brown08}.  A similar assumption is at the root of the `SHAM' approximation \cite{SHAM}.} of HODs based on ref.~\cite{Zheng05} with 
\begin{equation}
  \langle N_{\rm cen} \rangle = \frac{1}{2}\left[1 + {\rm erf}\left(\frac{{\rm log_{10}} M - {\rm log_{10}} M_{\rm cut}}{\sqrt{2}\sigma_{{\rm log_{10}}M}}\right)\right]
  \qquad ; \quad \sigma_{{\rm log_{10}}M}=0.25
\label{eqn:hod_cen}
\end{equation}
and
\begin{equation}
  \langle N_{\rm sat} \rangle = \left[ \frac{M - 0.1 M_{\rm cut}}{15 M_{\rm cut}}\right]^{0.8}
  \quad .
\label{eqn:hod_sat}
\end{equation}
The values of $\sigma_{{\rm log_{10}} M}$, and the power-law index and denominator in $\langle N_{\rm sat}\rangle$ are typical of magnitude-selected galaxy samples and our final results are not very sensitive to them.  The number density and large- and small-scale biases $b_{\rm HOD}(z)$ can then be computed as a function of ${\rm log_{10}}M_{\rm cut}$.  We compute the comoving number density of unWISE galaxies from the COSMOS cross-match $dN/dz$ (Section~\ref{sec:xmatch_dndz}) and choose the cutoff mass $M_c$ to match the abundance of each sample at all redshifts.  The results are given in Table \ref{tab:unwise_lgmc}.

\begin{table}[h!]
    \centering
    \small
    \begin{tabular}{c|cc|cc|cc}
    & \multicolumn{2}{c}{Blue} & \multicolumn{2}{c}{Green} & \multicolumn{2}{c}{Red} \\
    \hline
    $z$ & \multirow{2}{*}{$\log_{10}(M_{\rm cut})$} & Abundance  & \multirow{2}{*}{$\log_{10}(M_{\rm cut})$} & Abundance & \multirow{2}{*}{$\log_{10}(M_{\rm cut})$} & Abundance  \\
    & & ($h^{3}$ Mpc$^{-3}$) & & ($h^{3}$ Mpc$^{-3}$) & & ($h^{3}$ Mpc$^{-3}$) \\
    \hline
    0.41 & 12.25 & $3.44 \times 10^{-3}$ & 13.00 &  $5.57 \times 10^{-4}$ & 13.50 & $1.39 \times 10^{-4}$ \\
    1.00 & 12.50 & $1.43 \times 10^{-3}$ & 12.75 &  $7.19 \times 10^{-4}$ & 13.50 & $6.55 \times 10^{-5}$ \\    
    1.27 & 13.25 & $1.11 \times 10^{-4}$ & 12.75 &  $5.71 \times 10^{-4}$ & 13.50 & $4.26 \times 10^{-5}$ \\  
    1.78 &  &  & 13.00 &  $5.57 \times 10^{-4}$ & 13.50 & $1.39 \times 10^{-4}$ \\
    \end{tabular}
    \caption{Halos populated with the HOD of Eq.~\ref{eqn:hod_cen} and Eq.~\ref{eqn:hod_sat}, at four output times.  All HODs use $\sigma_{\log_{10}M} = 0.25$ decades, and $\log_{10}{M}_{\rm cut}$ is then selected to roughly match the abundance of each unWISE sample at the specified redshift.
    }
\label{tab:unwise_lgmc}
\end{table}

\begin{figure}
    \centering
     \includegraphics[width=15cm]{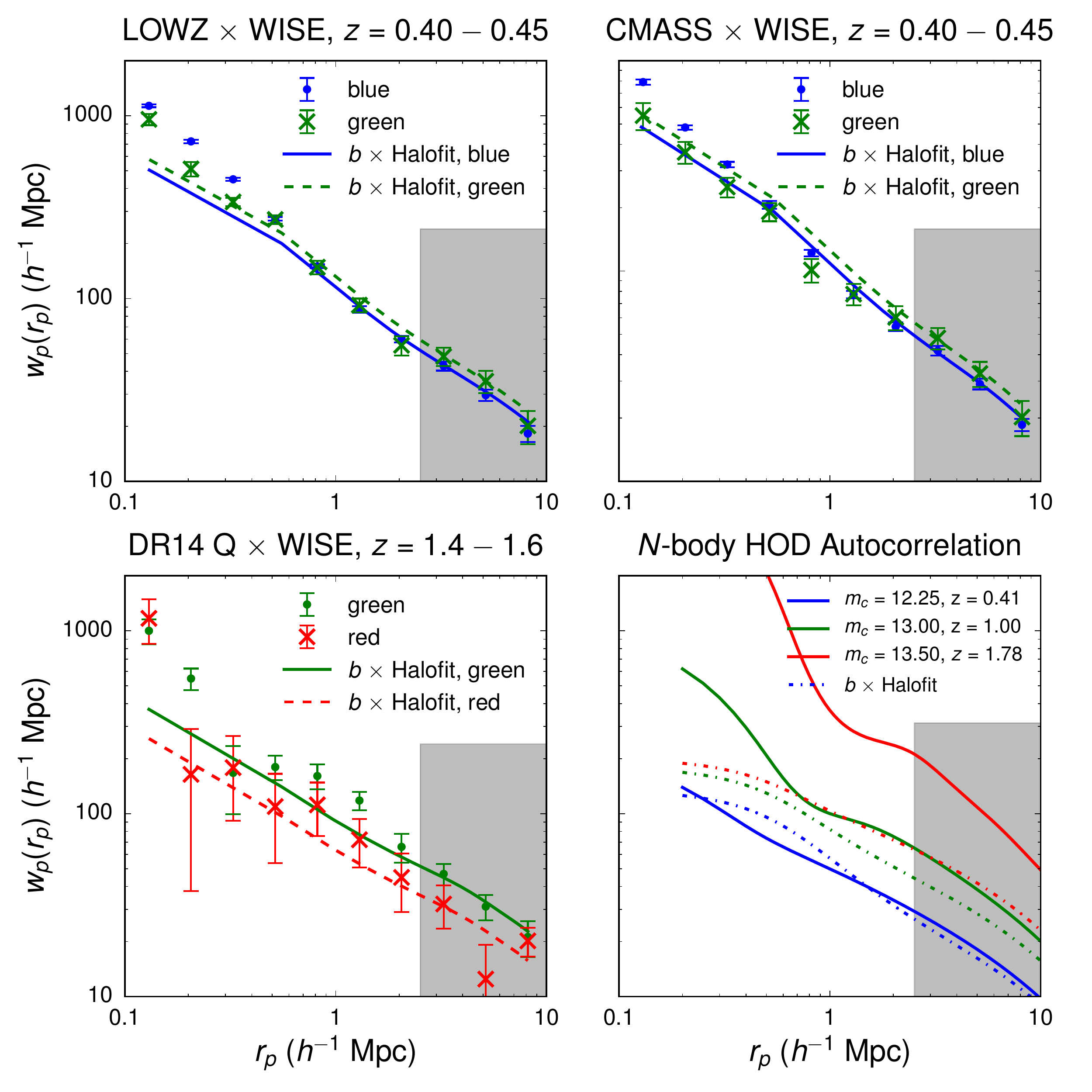} 
    \caption{
    Cross-correlations between spectroscopic tracers and the unWISE galaxy samples,
    compared to a scale-independent bias times nonlinear correlation function fit to the points
    with $2.5 < r_p < 10$ $h^{-1}$ Mpc (shaded region).  
    Deviations from scale-independent
    bias are seen at $r_p < 2.5\,h^{-1}$Mpc, justifying our decision to use $2.5 < r_p < 10\,h^{-1}$Mpc for the cross-correlation redshifts.
    \textit{Lower right}: autocorrelation of galaxies
    populating halos in an $N$-body simulation according to Eq.~\ref{eqn:hod_cen} and~\ref{eqn:hod_sat} with $\sigma_{ \log_{10} M}$ = 0.25 decades.  Redshifts and number
    densities are chosen to be roughly representative of the three unWISE samples.  Since the lower right panel is an autocorrelation $(\propto b_{\rm unWISE}^2)$ while the other panels are cross-correlations $(\propto b_{\rm unWISE})$, it has a much stronger scale-dependent bias at $z \sim 1.5$ (since the unWISE galaxies
    have larger and thus more scale-dependent bias than the quasars).
    }
    \label{fig:choosing_xcorrz_scales}
\end{figure}

To reduce scatter, we averaged the results from halo catalogs generated from 4 simulations, each with $1280^3$ particles in a $640\,h^{-1}$Mpc box, assuming $\Lambda$CDM with
Planck 2014 cosmological parameters \citep{Planck14}.
The simulations use the TreePM code of ref.~\cite{White02}, are described
in section 2.1 of ref.~\cite{Stark15},
and are validated in ref.~\cite{Heitmann08}.
We consider friends-of-friends halos with linking length $0.168$ of the mean interparticle spacing at four representative redshifts $z = 0.41$, $1.00$, $1.27$ and $1.78$.
At each redshift we adjusted ${\log_{10}}M_{\rm cut}$ as in Table \ref{tab:unwise_lgmc} and measured the real-space correlation function by direct pair counting of the halos, hence obtaining the projected correlation function, $w_p(R)$.
We define the real-space bias as a function of scale
\begin{equation}
  b(R) = \sqrt{w_p(R)/w_{p\rm{, HF}}(R)}
\label{eqn:b_rp}
\end{equation}
define $b_{\rm sml}$ with the same
$R^{-1}$ weighting as in Eq.~\ref{eqn:wbar}:
\begin{equation}
    b_{\rm sml} = \int_{r_{\rm min}}^{r_{\rm max}} \, dR \, R^{-1} \, b(R)\bigg/\int_{r_{\rm min}}^{r_{\rm max}} \, dR \, R^{-1}
\label{eqn:b_sml}
\end{equation}
and define $b_{\rm lin} \equiv b(r=40\,h^{-1}{\rm Mpc})$. 

\begin{figure}
    \centering
     \includegraphics[width=8cm]{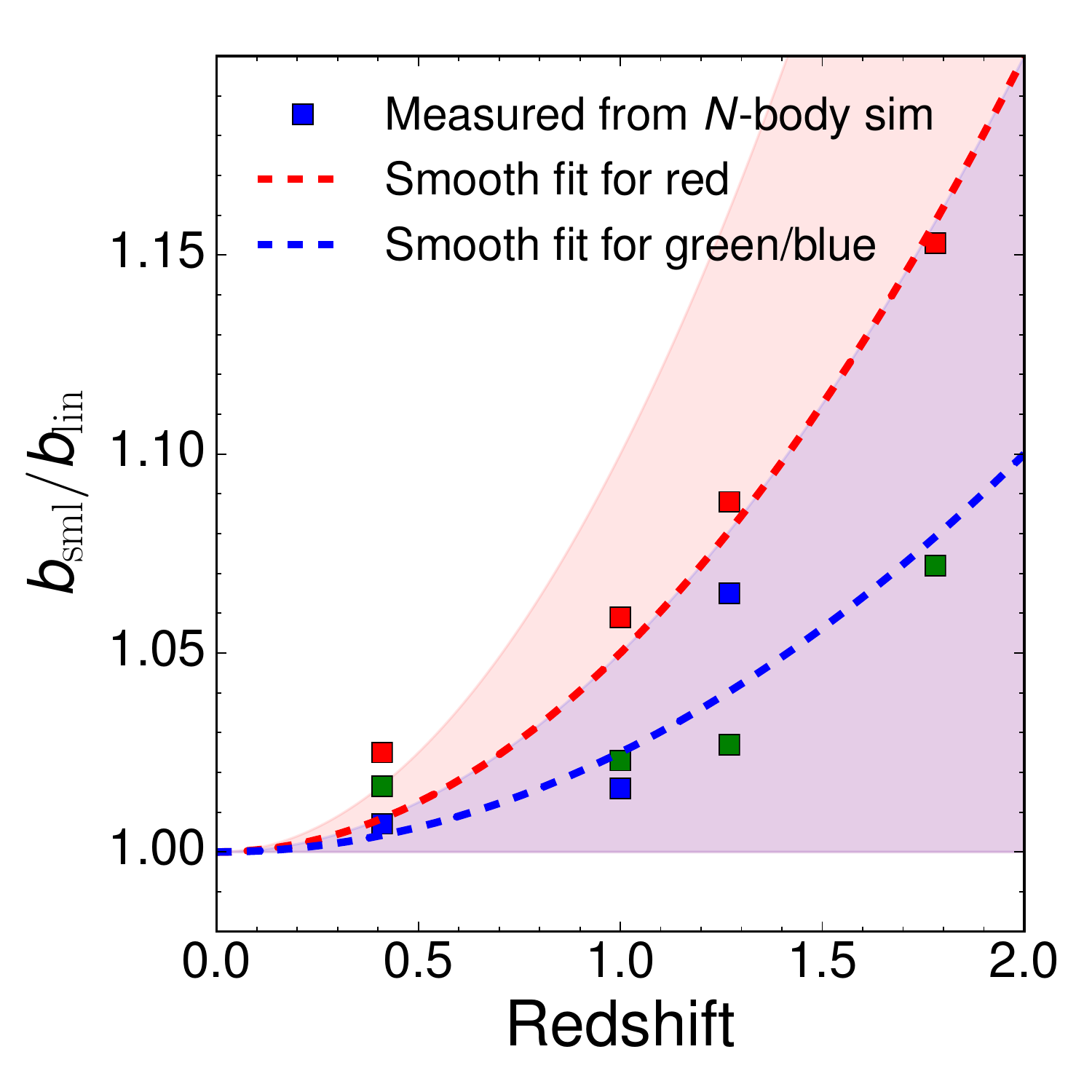} 
    \caption{
    Ratio between $b_{\rm sml}$ and $b_{\rm lin}$ for HODs matched to the three unWISE
    samples at four representative redshifts in the $N$-body simulation.  Dashed lines
    give interpolating functions $b_{\rm sml}/b_{\rm lin} = 1 + Az^2$ with separate
    values of $A$ for red and blue/green; shaded regions represent uncertainty on $A$,
    accounting for uncertainty in the halo occupation of the unWISE galaxies.
    }
    \label{fig:bsml_blin_ratio}
\end{figure}

\begin{figure}
    \centering
    \resizebox{\columnwidth}{!}{\includegraphics{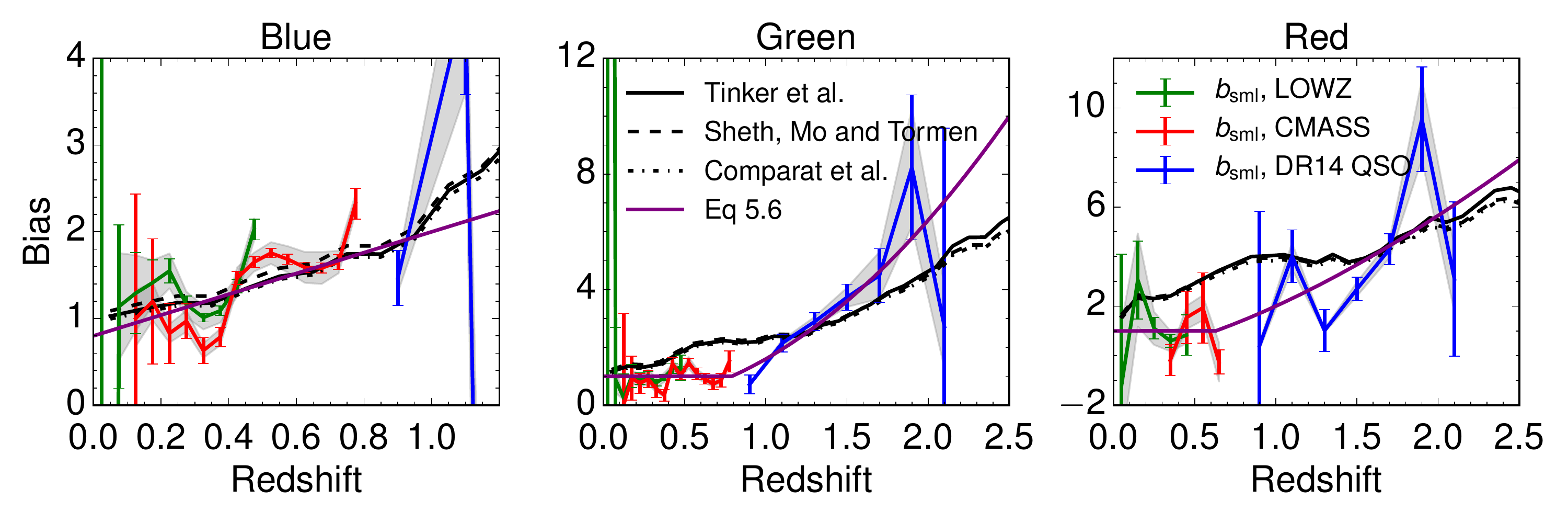}}
    \caption{Bias, derived from $w(\theta)$ at $2.5 < r < 10\,h^{-1}$Mpc
    using Equation~\ref{eqn:wbar_final} and the COSMOS cross-match redshift distribution.
    Colored lines give the measured bias; black lines
    give the bias evolution for an HOD abundance-matched to the density
    of the WISE samples, with different line styles corresponding to different
    bias-number density prescriptions from the literature.  Errorbars on the colored lines are from measurement errors on $w(\theta)$; the gray
    bands give the additional uncertainty from uncertain $dN/dz$, quantified
    by the 16th-84th percentile range from 100 samples of $dN/dz$.
    All fits to clustering include magnification bias using the fiducial values
    in Table~\ref{tab:galaxyselection} and Fig.~\ref{fig:specz_s}.
    }
    \label{fig:implied_bias}
\end{figure}

We find only mild departures from scale-independent bias at $2.5 < r_p < 10$
$h^{-1}$ Mpc, but more significant deviations at smaller scales, in qualitative
agreement with the spectroscopic cross-correlations (Figure~\ref{fig:choosing_xcorrz_scales}).  
From the HOD-populated $N$-body autocorrelations, we find $b_{\rm sml}/b_{\rm lin} = 1.153$
for red at $z = 1.78$, the most massive and highest-redshift sample (Figure~\ref{fig:bsml_blin_ratio}).
To interpolate between the four measured points, we use a function
of the form $b_{\rm sml}/b_{\rm lin} = 1 + Az^2$, with $A = 0.05$ for red and $A = 0.025$ for green and blue.
The HOD has a milder bias evolution
than the data for the green and red samples (Figure~\ref{fig:implied_bias}). If we match the $z=1.9$
clustering with a free number density, we
find somewhat larger $b_{\rm sml}/b_{\rm lin}$
at $z = 1.9$, corresponding to $A \sim 0.15$ and 0.2
for green and red, respectively.
When estimating the impact of this systematic in
Section~\ref{sec:systematics} and Figure~\ref{fig:systematics}, we therefore 
test both the fiducial value of $A$, $A_{\rm fid} = 0.025$, 0.025, 0.1 for blue, green and red; and the maximal value of $A$, $A_{\rm max} = 0.1$, 0.15, 0.2 for blue, green and red.

To assess the compatiblity of the cross-correlation
and cross-match $dN/dz$,
we compare the bias evolution of galaxies in the HOD, $b_{\rm HOD}(z)$, to the observed bias evolution of the unWISE galaxies, $b_{\rm sml,p}(z)$, in Figure~\ref{fig:implied_bias}.  Using Equation \ref{eqn:wbar_final}, the COSMOS $dN/dz$, and $s$ from Appendix~\ref{sec:magbias_slope}, we fit $b_{\rm sml,p}(z)$ to $w(\theta)$ between 2.5 and $10\,h^{-1}$Mpc.
Consistency between the cross-match $dN/dz$ and
photometric-spectroscopic clustering (from which the cross-correlation $dN/dz$ is derived)
requires a steeply evolving galaxy bias (colored
lines in Figure~\ref{fig:implied_bias}).
In fact, the simple abundance-matched HOD yields a galaxy bias that is nearly as steep (Figure~\ref{fig:implied_bias}).
We compare $b_{\rm sml,p}(z)$ to $b_{\rm HOD}(z)$ using the HOD above and one of three different
mass function/mass-bias relationships (Tinker et al., \citep{Tinker08,Tinker10}, Sheth, Mo and Tormen \citep{ST99,Sheth01}, and Comparat et al. \citep{Comparat17}).
We consider both statistical errors on $b_{\rm sml,p}$ from errors
on the cross-correlation (errorbars in Figure~\ref{fig:implied_bias}), and errors on $b_{\rm sml,p}$ from the uncertain
$dN/dz$ (gray band, giving 16th-84th percentile range from 100 draws from $dN/dz$).  While the uncertainty in $dN/dz$ will also affect the bias evolution of the abundance-matched
halos by changing their comoving number density, this effect is smaller than
the impact of uncertain $dN/dz$ on $b_{\rm sml,p}(z)$ because the bias
is a shallow function of halo mass and thus number density.

For the blue sample, and for the red and green samples at $z > 1$, the measured
bias evolution roughly agrees with the HOD prediction within the uncertainty from $dN/dz$.
At $z < 1$, the bias of the red and green samples is significantly lower than the expectation
from the HOD.  However, both the red and green samples are bimodal, and it is possible
that their low-redshift tails are not well-described by the HOD above.
For instance, the low-redshift tails could consist of star-forming galaxies
occupying halos with a duty cycle well below unity, such that at fixed abundance,
the cutoff halo mass is much lower than the HOD above would predict, thus lowering
the bias.  

The rough agreement between $b_{\rm sml,p}(z)$ and $b_{\rm HOD}(z)$
for the abundance-matched halos shows that the combination of cross-correlation
and cross-match redshifts yields a reasonable bias evolution.  This result justifies
our use of both the cross-correlation and cross-match redshifts in modelling
$C_{\ell}^{\kappa g}$ and $C_{\ell}^{gg}$, as it suggests they are consistent
with each other.

\section{Response of the number density to magnification bias}
\label{sec:magbias_slope}

The amplitude of the magnification bias term
depends on the response of the galaxy density to magnification bias, $s \equiv d\log_{10} N / dm $, at the limiting magnitude of the survey.  Since the completeness
of WISE drops over a relatively large range,
measurements of $s$ are affected by incompleteness in WISE.  This can be mitigated by restricting the sample to
high ecliptic latitude, where the greater
depth of coverage results in a fainter limiting magnitude.

Since the WISE galaxies are selected via a magnitude-dependent color cut, one cannot simply histogram them in W2 to determine $s$.  Instead, we compute $s$ by shifting the magnitudes
of all WISE objects by 0.02 magnitudes and re-applying our selection criteria.

\begin{figure}
    \centering
    \resizebox{\columnwidth}{!}{\includegraphics{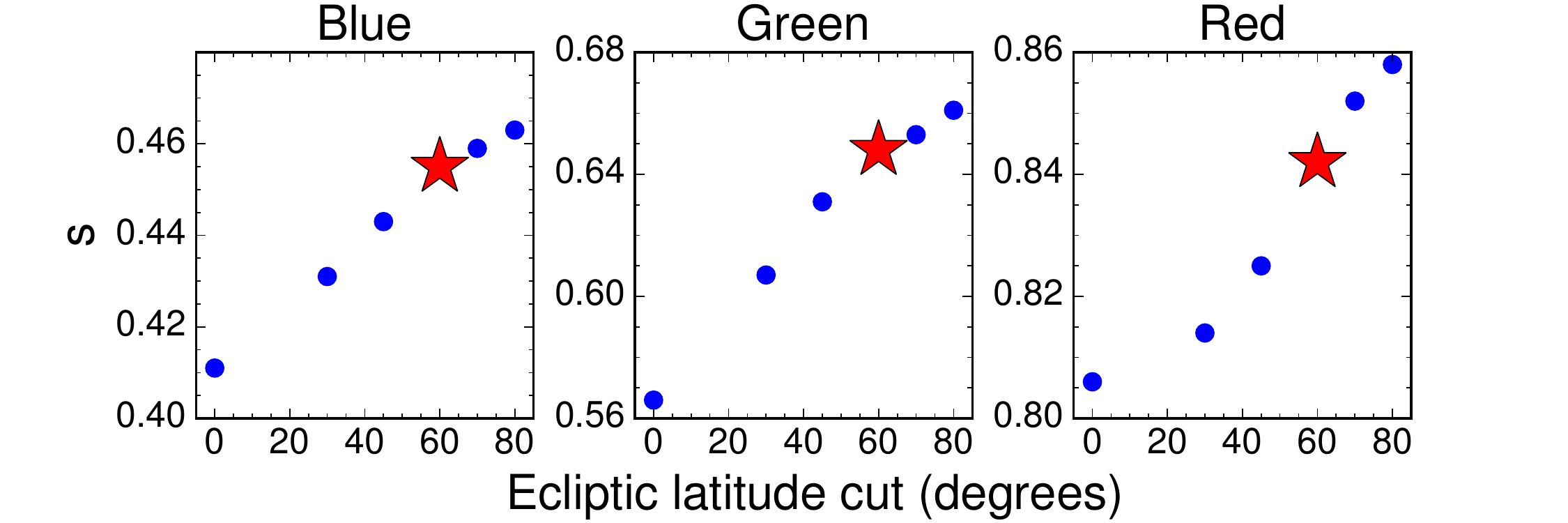}}
    \caption{Dependence of $s$ for the unWISE samples on ecliptic latitude. Each point shows $s(\lambda_{\rm min})$ measured using galaxies with ecliptic $| \lambda | > \lambda_{\rm min}$. The starred point gives the fiducial value
    of $s$, using $\lambda_{\rm min} = 60^{\circ}$.
    }
    \label{fig:magbias_lcut}
\end{figure}

In Figure~\ref{fig:magbias_lcut}, we show
$s$ as a function of $\lambda_{\rm{min}}$, where we sequentially remove all galaxies
with $| \lambda | < \lambda_{\rm{min}}$.
We set the fiducial value of $s$ at $\lambda_{\rm min} = 60^{\circ}$.

We also require $s$ for each of the spectroscopic samples in order
to subtract the magnification bias contribution from $\bar{w}_{\rm sp}$. We measure $s$ by making all galaxies or quasars in the sample
fainter by 0.1 magnitudes, applying the relevant selection criteria and measuring the change
in number counts.  

For LOWZ and CMASS, we use the color cuts described in Ref.~\cite{Reid2016}.
This procedure assumes that every galaxy in the spectroscopic sample with perturbed
photometry was also in the original sample; this is true for both CMASS and LOWZ
(see Figures 3 and 4 in Ref.~\cite{Reid2016} for color-magnitude plots for LOWZ and CMASS, respectively).  

DR12 quasars are selected as point sources with $g < 22$ or $r < 21.85$, $i > 17.8$, and {\sc XDQSO} ``mid-z'' quasar probability (i.e.\ probability the object is a $2.2 < z < 3.5$ quasar) $> 0.424$ \citep{Bovy10,Ross11}.
When we make the quasars fainter by 0.1 magnitudes, we estimate the number
of quasars that would be spuriously categorized as extended using the completeness of SDSS star-galaxy separation as a function of $r$ band magnitude\citep{Strauss02}.\footnote{\url{https://classic.sdss.org/dr7/products/general/stargalsep.html}}  Unlike the color cuts used for the BOSS galaxies,
with the more complicated {\sc XDQSO} color cut it is possible that quasars
could be excluded from the original targeting but included when the photometry is made fainter by 0.1 magnitudes.
To estimate the occurrence of such objects, we use the BOSS {\sc BONUS} sample of non-uniformly-selected quasars, which are not suitable for quasar clustering
analyses but are $\sim 2 \times$ as abundant as the {\sc CORE} sample that we do use.  Since {\sc BONUS} quasars are not selected using {\sc XDQSO}, 
they may have mid-z quasar probability $< 0.424$ but ``scatter into'' our fainter sample.

For DR14 quasars, we follow a similar procedure as for DR12 quasars, applying the selection criteria of Ref.~\cite{Myers15}. 
However, we lack a similarly deep quasar sample (like {\sc BONUS} in DR12) to determine the number of quasars that scatter into the DR14 selection criteria
when the quasar photometry becomes fainter.  Based on the number of quasars that scattered into the DR12 quasar selection, we estimate an additional
systematic error of $\Delta s \sim 0.1-0.2$ for the DR14 quasars.

\begin{figure}
    \centering
    \includegraphics[width=15cm]{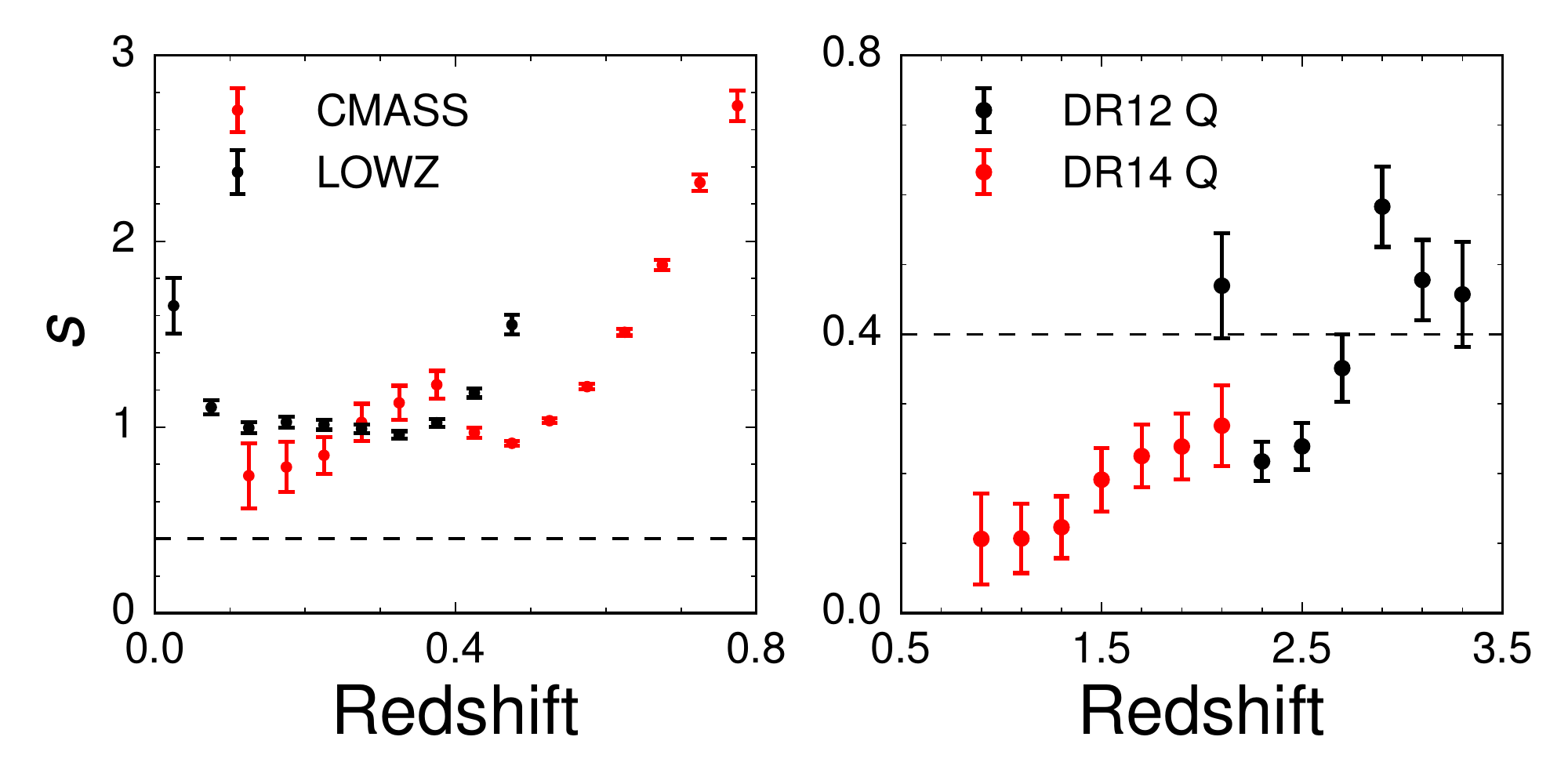}
    \caption{Response of galaxy number density to magnification for spectroscopic samples. Error bars
    are computed as $\Delta s = (\log_{10}N - \log_{10}(N - \sqrt{N}))/\Delta m$.
    Dashed line indicates $s = 0.4$, where magnification bias makes no contribution
    to the observed clustering.
    }
    \label{fig:specz_s}
\end{figure}

We plot the resulting $s$ in Figure~\ref{fig:specz_s}, and use them to remove magnification bias
from $\bar{w}$.
For $z$ beyond the range shown in Figure~\ref{fig:specz_s}, we assume $s$
is a constant function, using the nearest
point for which we have a measurement
of $s$. 

\section{Galaxy-galaxy cross spectra}
\label{sec:galaxy-galaxy}

\begin{figure}[ht]
    \centering
    \includegraphics[width=16cm]{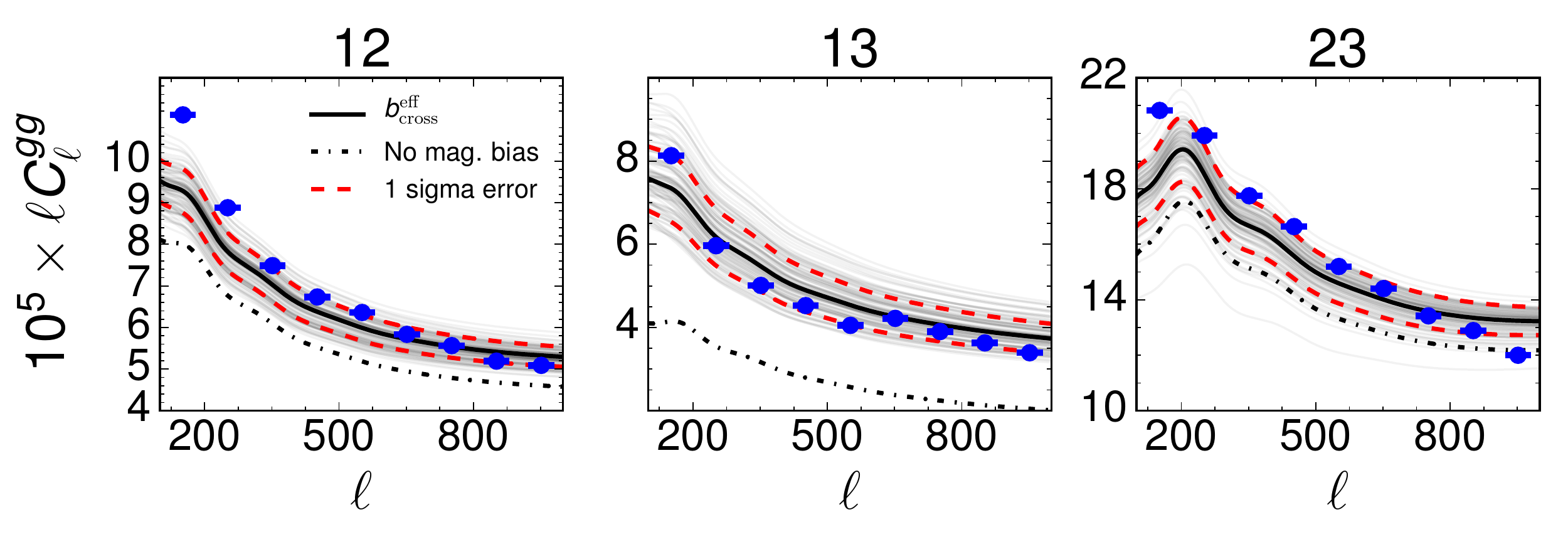}
\caption{Galaxy-galaxy cross-spectra between the different samples.  
The solid black line gives
the predicted theory
curve using $b_{\rm cross}^{\rm eff}$,
and the dot-dashed black line gives the predicted theory curve
with no magnification bias included.
Gray lines show  $C_{\ell}^{gg}$ from 100 realizations of $dN/dz$ (with bias modified from the best-fit $b^{\rm eff}_{\rm cross}$ in Table~\ref{tab:halofit_xcorr} to fit $C_{\ell}^{\kappa g}$ for a given $dN/dz$);
and
dashed red lines show 1 $\sigma$ uncertainty in $C_{\ell}^{gg}$, with contributions
both from uncertain $dN/dz$ and from statistical
errors on $b_{\rm cross}^{\rm eff}$.
Uncertainty on $C_{\ell}^{gg}$
for $dN/dz$ from the two halves of the sky
is comparable to the 1 $\sigma$ uncertainty.
In all cases, a cross shot-noise term is fit to the data and also included in the theory curves.}
\label{fig:red_check}
\end{figure}

In Figure~\ref{fig:red_check} we show the cross-spectra between the different galaxy samples. We use the fiducial cross-correlation $dN/dz$ and $b^{\rm eff}_{\rm cross}$ for the theory calculation.  
Additionally, we fit a shot-noise term, and find good agreement between the expectation and the data once the uncertainty on $dN/dz$ is taken into account.
For red cross blue the shot noise term is negligible, but for blue cross
green and blue cross red, we find shot noise values of $6.22 \times 10^{-9}$ and $4.67 \times 10^{-8}$, respectively.  Shot noise can arise in a cross-correlation if some of the objects
in the two samples occupy the same halo, with density $\bar{n}_{\rm common}$.  The cross shot-noise is then given by
\begin{equation}
    \textrm{Shot Noise} = \frac{\bar{n}_{\rm common}}{\bar{n}_1 \bar{n}_2}
\end{equation}
Using the fitted shot noise for each sample from Table~\ref{tab:halofit_xcorr}, we find $\bar{n}_{\rm common}$ = 130 deg$^{-2}$
for blue cross green and 41 deg$^{-2}$ for green cross red.  
This implies that
 3.8\% (7.0\%) of the blue (green) sample lives in the same halo
as a  green (blue) object, and 2.2\% (28.5\%) of the green (red) sample lives in the same halo as a red (green) object.

We create a simple ``joint HOD'' to understand
the cross shot-noise.  Rather than assume
that every halo well above $M_{\rm cut}$
hosts a central galaxy,
we instead assume that some halos host
red centrals and other halos host green centrals; i.e.\ we multiply $N_{\rm cen}$ by $f_{\rm green}$ or $f_{\rm red}$ where $f_{\rm green} + f_{\rm red} = 1$,
and do not modify $N_{\rm sat}$.
We then ask what fraction of red galaxies
host a green satellite.
If $f_{\rm green} = f_{\rm red} = 0.5$,
we find that 26.3\% of red galaxies
host at least one green satellite.
The common fraction remains similar at 15-25\% if we change some aspects of this toy model (i.e.\ increase $f_{\rm green}$ to 0.9; add a linear ramp where halos transition from hosting green galaxies at low redshift to red galaxies at high redshift; or multiply $N_{\rm sat}$
by 0.5 for both green and red to preserve the total number of satellites).

\bibliographystyle{JHEP}
\bibliography{main}
\end{document}